\documentclass[prb,twocolumn,floatfix]{revtex4}

\usepackage{amsfonts}
\usepackage{amsmath}
\usepackage{verbatim}
\usepackage[dvips]{graphicx}
\usepackage{subfigure}

\def\Tr{\textrm}
\def\etal{\textit{et~al.}}     
\def\dd{\textrm{d}}

\def\vv{\Tr{v}}
\def\pp{\Tr{p}}

\begin{document}

\title{An effective theory of fractional topological insulators in two spatial dimensions}
\author{Predrag Nikoli\'c$^{1,2}$}
\affiliation{$^1$School of Physics, Astronomy and Computational Sciences, George Mason University, Fairfax, VA 22030, USA}
\affiliation{$^2$Institute for Quantum Matter at Johns Hopkins University, Baltimore, MD 21218, USA}
\date{\today}


\begin{abstract}

Electrons subjected to a strong spin-orbit coupling in two spatial dimensions could form fractional incompressible quantum liquids without violating the time-reversal symmetry. Here we construct a Lagrangian description of such fractional topological insulators by combining the available experimental information on potential host materials and the fundamental principles of quantum field theory. This Lagrangian is a Landau-Ginzburg theory of spinor fields, enhanced by a topological term that implements a state-dependent fractional statistics of excitations whenever both particles and vortices are incompressible. The spin-orbit coupling is captured by an external static SU(2) gauge field. The presence of spin conservation or emergent U(1) symmetries would reduce the topological term to the Chern-Simons effective theory tailored to the ensuing quantum Hall state. However, the Rashba spin-orbit coupling in solid-state materials does not conserve spin. We predict that it can nevertheless produce incompressible quantum liquids with topological order but without a quantized Hall conductivity. We discuss two examples of such liquids whose description requires a generalization of the Chern-Simons theory. One is an Abelian Laughlin-like state, while the other has a new kind of non-Abelian many-body entanglement. Their quasiparticles exhibit fractional spin-dependent exchange statistics, and have fractional quantum numbers derived from the electron's charge and spin according to their transformations under time-reversal. In addition to conventional phases of matter, the proposed topological Lagrangian can capture a broad class of hierarchical Abelian and non-Abelian topological states, involving particles with arbitrary spin or general emergent SU(N) charges.

\end{abstract}

\maketitle

\section{Introduction}

Quantum Hall effects are the best known experimentally observed manifestations of electron fractionalization above one spatial dimension \cite{Tsui1982, Stormer1983, Goldman1995, Saminadayar1997, Goldman2001, Camino2005, Camino2007, de-Picciotto1997, Venkatachalam2011}. It is believed that similar fractionalization is also possible in materials with strong spin-orbit coupling that realize a new class of topological insulators (TIs) with time-reversal (TR) symmetry. All currently known TIs are \emph{uncorrelated} band-insulators \cite{Hasan2010, Qi2010a, Moore2010}. Quantum wells made from these materials feature electron dynamics that somewhat resembles integer quantum Hall states, most notably by exhibiting protected gapless edge modes \cite{Konig2007}. However, in addition to respecting the TR symmetry the new TIs differ from quantum Hall systems by the character of their spectra and by lacking the conservation of ``charge'' whose role is played by the electron's spin. The latter prevents observing a quantized spin Hall conductivity in two spatial dimensions and reduces the number of stable uncorrelated insulating quantum phases from infinity to only two \cite{Kane2005a}.

The subject of this paper are \emph{strongly correlated} TIs in two spatial dimensions whose excitations carry a fraction of electron's charge and exhibit unconventional exchange statistics \cite{Levin2009, Karch2010, Cho2010, Maciejko2010, Swingle2011, Neupert2011, Santos2011, Nikolic2011, Levin2012}. This research is motivated both by the fundamental quest for unconventional quantum states of matter and by potential future applications in spintronics and quantum computing \cite{Kitaev2000, Kitaev2003, Nayak2008, Bonderson2010}. The central problem we address is the classification of topological orders in the ground states of interacting particles. We loosely define topological orders as distinct manifestations of macroscopic many-body quantum entanglement that cannot be altered by tuning topologically unbiased Hamiltonians without going through a quantum phase transition. Topological orders are expressed in the phenomena such as the fractional exchange statistics of quasiparticles and the ground-state degeneracy without symmetry breaking on non-simply connected spaces. We will argue that novel kinds of topological order are made possible by the Rashba spin-orbit coupling in TIs, whose description requires a generalization of the Chern-Simons (CS) effective theory. Our main goal is then to construct a more general topological field theory that can capture a sufficiently broad spectrum of conventional and topological orders. We will discuss examples of spin entanglement that have no analogue in fractional quantum Hall states (FQHS), but a systematic classification of such states is beyond the scope of this paper.

Fractional TIs can exist in various systems, and likely will be observed in the foreseeable future. There are at least three prominent approaches to obtaining fractional TIs in solid state materials. The earliest one relies on Coulomb interactions to facilitate spin-charge separation in materials with geometrically frustrated local magnetic moments \cite{Pesin2010, Rachel2010, Krempa2010, Young2008}. Electrons can be fractionalized into neutral spinons and spinless charge-modes without a spin-orbit coupling, but gapped spinons can additionally exhibit the TI dynamics in the presence of a spin-orbit coupling. A more recent approach explores lattice models with fractional excitations, the so called Chern insulators \cite{Sun2011, Sheng2011, Wang2011, Venderbos2011, Murthy2011, Goerbig2011, Tang2011, Neupert2011a, Neupert2011}. Such models can be TR-invariant and generally rely on narrow bands in the electron spectrum to create favorable conditions for fractional ground states. There are a few proposals of materials that could realize fractionalization using artificially created narrow bands \cite{Bernevig2006a, Xiao2011, Ghaemi2011, Papic2011a, Abanin2012, Papic2012}.

The third approach is to use the currently available band-insulating TI materials and artificially induce electron correlations by a proximity effect in a heterostructure device. For example, a conventional superconductor placed in contact with a TI quantum well can induce superconductivity or leave behind an insulating state inside the TI. A superconductor-insulator quantum phase transition inside the TI can be tuned by a gate voltage, and it turns out that it would belong to the bosonic mean-field or XY universality class in the absence of the spin-orbit coupling \cite{Nikolic2010b}. This quantum critical point is sensitive to perturbations, and correlated ``pseudogap'' topological states can be born out of its quantum critical fan as a result of the spin-orbit coupling \cite{Nikolic2011a}. Candidate states are fractional TIs of spinful $p$-wave Cooper pairs whose existence is allowed by the TIs orbital degrees of freedom and low-energy dynamics enhanced by the spin-orbit coupling. A similar correlated TI of excitons could be envisioned in the device proposed by Seradjeh, \etal \cite{Seradjeh2009}.

Another promising system are ultra-cold gases of bosonic atoms trapped in quasi 2D optical lattices. Superfluid to Mott insulator transitions can be easily arranged to remove any energy scales that could compete with the spin-orbit coupling \cite{Greiner2002}, and thus create similar conditions as in the proximity effect heterostructures. At the same time, the recent development of artificial gauge fields for neutral atoms, created by stimulated Raman transitions between internal atomic states, has not only introduced the effective spin-orbit couplings \cite{Lin2011}, but also looks very promising for generating locally enhanced flux densities needed for fractional states \cite{Campbell2011, Cooper2011}.

The current theoretical studies of two-dimensional strongly correlated TIs are based entirely on adapting the well-known descriptions of FQHS to the TR-symmetry \cite{Qi2008b, Levin2009, Neupert2011, Santos2011, Murthy2011, Lu2012, Levin2012}. This approach is certainly well motivated, but the fact is that no experimental observations of fractional TR-invariant TIs have been made to date. Quantum Hall systems are sufficiently different from the spin-orbit-coupled materials that we must question their validity as an experimental basis for the complete theory of fractional TIs. Specifically, we will argue in this paper that the Dirac spectra of surface electrons in TIs pave the way to topological orders that cannot be fully captured by the standard CS effective theory, which is better suited to systems with Landau levels.

We will instead view the two-dimensional TIs as manifestations of the SU(2) ``quantum Hall physics'', created by an SU(2) ``magnetic field'' that implements the spin-orbit coupling \cite{Frohlich1992}. Some idealized SU(2) incompressible quantum liquids are quantum Hall states because they exhibit a quantized Hall conductivity of spin currents. However, the Hall response quantization is a symmetry-protected feature, lost due to spin non-conserving perturbations that unavoidably exist in materials. Interestingly, the non-commutative character of the SU(2) gauge fields enables incompressible quantum liquids without a quantum Hall effect even in the absence of unwanted perturbations. All presently available two-dimensional TIs can be viewed as the non-quantum-Hall analogues of ``integer'' quantum spin-Hall states, where the Rashba spin-orbit SU(2) ``magnetic flux'' creates a Dirac rather than a Landau-like electron spectrum.

The main purpose of this paper is then to construct and begin exploring a new topological quantum field theory that can naturally describe both the fractional quantum-Hall and non-quantum-Hall states. Our ambition here is to systematically capture the \emph{topological} properties of a broad class of states in a relatively simple manner. This theory will help us predict the topological orders which may be specific to the Rashba spin-orbit coupling. We will take guidance from the experimentally established facts about the available TI materials and construct a theory that can address all of the above candidate systems for TR-invariant fractional incompressible quantum liquids.

This topological field theory will have a general form applicable to interacting elementary particles with arbitrary charge and spin, whose dynamics is restricted to two spatial dimensions and affected by any type of spin-orbit coupling or magnetic field or both. We will use it to show that correlated TIs can feature excitations with fractional charge, spin and exchange statistics, despite the spin non-conservation. We will make predictions about the fractional excitation quantum numbers in relation to symmetries, possible symmetry breaking, as well as Cooper or exciton pairing in topologically-enhanced ground states. We will demonstrate the relationship of this topological field theory to the standard CS gauge theories, and point to limitations of the latter to adequately model all possible TR-invariant TIs. For the purpose of focusing on bulk topological orders, we will view all states of interest here as fractional TIs regardless of whether they have gapless edge states or not, and thus depart from the terminology introduced in Ref.\cite{Levin2009}. We will set up the formalism for analyzing both the Abelian and non-Abelian topological orders, including hierarchical states and incompressible quantum liquids specific to SU($N$) fluxes and the Rashba spin-orbit coupling, which may have no analogue in the quantum Hall states. The field theory we propose can also describe conventional quantum phases, and possibly the universal aspects of phase transitions to topological states. It may be able to provide a broad classification scheme for topological states of quantum matter, analogous to that provided by Landau-Ginzburg theories of symmetry-broken states.

\subsection{Preliminaries}

This introductory section describes the effective field theory of correlated two-dimensional TIs that we propose, explains the principles of its construction, and relates it to other works. The structure of the paper's technical parts is outlined near the section end.

Understanding complex emergent phenomena directly from microscopic models can be extremely difficult. It is often much more practical to study emergent and universal phenomena using effective theories that specialize to the low energy parts of spectra. An effective Lagrangian can be constructed in the continuum limit by introducing field operators to quantize the classical equations of motion, and by collecting all combinations of fields that respect the required symmetries. This method, pioneered in high-energy physics and the theory of critical phenomena, is the basis of the present analysis.

The proposed theory will be written in several different forms throughout the paper, but the initial discussion will be based on the following imaginary-time Lagrangian density $\mathcal{L} = \mathcal{L}_\textrm{LG} + \mathcal{L}_{\textrm{t}}$:
\begin{eqnarray}\label{TopLG}
&& \!\!\!\! \mathcal{L}_\textrm{LG} = \frac{K}{2}\Bigl\vert(\partial_\mu-i\mathcal{B}_\mu)\psi\Bigr\vert^2
       -t|\psi|^2-t'\psi^{\dagger}\Phi_{0}\psi \\
&& ~~~ +u|\psi|^4+v|\psi^{\dagger}\gamma^{a}\psi|^2+v'|\psi^{\dagger}\Phi_{0}\psi|^2
       +\mathcal{L}_{\textrm{M}} \nonumber \\[0.1in]
&& \!\!\!\! \mathcal{L}_{\textrm{t}} = -\frac{i}{8}\, \psi^{\dagger}\epsilon^{\mu\nu\lambda}\Bigl\lbrack
     (\partial_{\mu}-i\mathcal{A}_{\mu})
     \Bigl\lbrace\partial_{\nu}-i\mathcal{A}_{\nu},\Phi_{0}\Bigr\rbrace
     (\partial_{\lambda}-i\mathcal{A}_{\lambda}) \nonumber \\
&& ~~~ +\Bigl\lbrace(\partial_{\mu}-i\mathcal{A}_{\mu})(\partial_{\nu}-i\mathcal{A}_{\nu})
     (\partial_{\lambda}-i\mathcal{A}_{\lambda}),\Phi_{0}\Bigr\rbrace\Bigr\rbrack\psi \ . \nonumber
\end{eqnarray}
We will use Greek indices $\mu,\nu,\lambda\in\lbrace 0,x,y \rbrace$ for space-time directions, Latin indices $i,j,k\in\lbrace x,y \rbrace$ for only spatial directions, and Einstein's notation for sums over repeated indices. The fields $\psi$ are complex spinors with $2S+1$ components whose relationship to physical spin-$S$ particles is established by a duality mapping. Therefore, the ``matter fields'' $\psi$ in this Lagrangian represent vortices of the physical particle currents. The Landau-Ginzburg part $\mathcal{L}_\textrm{LG}$ is the continuum limit of a standard dual theory of lattice bosonic particles \cite{Dasgupta1981, Fisher1989, Sachdev1990, Sachdev2004}, adapted to the presence of internal (spin) degrees of freedom. Densities and currents of particles are represented by the temporal and spatial components respectively of the flux $\Phi_{\mathcal{B}\mu}$ associated with the gauge field matrices $\mathcal{B}_\mu$. There are $2S+1$ independent modes of particle fluctuations that correspond to different states of spin projection on some axis and define the basis vectors for the matrix representation of $\mathcal{B}_\mu$. The Lagrangian is, however, written in the representation-independent form. The dynamics of particles is governed by the Maxwell term $\mathcal{L}_{\textrm{M}}$ in this theory. If the particle spin were conserved, the Maxwell term would have the standard non-compact form:
\begin{equation}\label{Maxwell1}
\mathcal{L}_{\textrm{M}} = \frac{1}{8 \pi^2} \textrm{tr} \left\lbrack
   Q^{-2} \bigl(\Phi_{\mathcal{B}\mu} - \Theta_\mu\bigr)^2 \right\rbrack \ ,
\end{equation}
where $Q$ is a coupling matrix. However, the realistic spin non-conservation in materials requires that certain combinations of $\mathcal{B}_\mu$ modes have compact dynamics. In either case, particle charge and spin densities are allowed to fluctuate near the average values specified by the matrix $\Theta_0$, while the average current densities are zero ($\Theta_i=0$).

The topological term $\mathcal{L}_{\textrm{t}}$, allowed by symmetries, shapes the quantum kinematics of dual topological defects in the $\psi$ field configurations \cite{Nikolic2011}. It is inconsequential in conventional phases such as superconductors and Mott-insulators, but affects the quasiparticle statistics in incompressible quantum liquids. The static U(1)$\times$SU(2) gauge field $\mathcal{A}_\mu$ implements any combination of external electromagnetic fields and spin-orbit couplings. Its components are SU(2) matrices, $\mathcal{A}_\mu^{\phantom{a}} = a_\mu^{\phantom{a}} + A_\mu^a\gamma^a$, where $a_\mu^{\phantom{a}}$ and $A_\mu^a$ are scalars and $\gamma^a$ are three SU(2) generators in the spin-$S$ representation (angular momentum matrices; $a\in\lbrace x,y,z \rbrace$). The flux components of non-Abelian gauge fields are
\begin{equation}\label{Flux}
\Phi^\mu = \epsilon^{\mu\nu\lambda} ( \partial_\nu \mathcal{A}_\lambda - i \mathcal{A}_\nu \mathcal{A}_\lambda )
\end{equation}
in the matrix representation. The temporal ``magnetic'' component $\Phi_0$ of the external flux density is inserted in the topological term $\mathcal{L}_{\textrm{t}}$ to ensure its adequate transformation under TR ($\epsilon^{\mu\nu\lambda}$ is the Levi-Civita tensor in (2+1)D space-time). There are four possible insertion points, and $\mathcal{L}_{\textrm{t}}$ is symmetrized with respect to them using anticommutators (braces). If all components of $\mathcal{A}_\mu$ commute with each other, then the topological term can be reduced to the CS form when the phase fluctuations in the spinor $\psi$ components drive the dynamics. It turns out, however, that the $\mathcal{A}_\mu$ appropriate for solid state TIs have non-commuting components.

The topological term $\mathcal{L}_{\textrm{t}}$ is the main new ingredient in a field theory of this kind and we will devote most of the discussion in this paper to its derivation and consequences. We will derive it using the same field-theoretical principles that yield the CS theories of FQHS, but applied in the context of spinor rather than gauge fields. The standard effective field theory of FQHS is a pure gauge theory in which the CS coupling acts as a topological term that implements a fractional exchange statistics. The CS theory is constructed from the requirement that the action be stationary when the Hall conductivity and incompressible electron density are quantized as observed in FQHS experiments \cite{WenQFT2004}. This requirement can be stated in an alternative form. Electrons in mutually perpendicular electric $\bf E$ and magnetic $\bf B$ fields generally have classical cyclotron trajectories whose orbit centers move at the constant velocity ${\bf v} = {\bf E} \times {\bf B} / |{\bf B}|^2$. The resulting drift current is precisely reproduced by the kinematic equations of motion that make the CS action stationary.

We will seek the analogous drift currents of spin-orbit-coupled particles in the section \ref{secQM}. Our starting point will be the minimal model Hamiltonian of two-dimensional topological band-insulators, which is by now well established experimentally. This Hamiltonian can be written in the form that couples electrons to an external static SU(2) gauge field with a finite ``magnetic'' flux density. We will derive the time-evolution of various current operators in the Heisenberg picture from a generic Hamiltonian of this type. The obtained equations of motion have a direct classical interpretation according to the Ehrenfest's theorem. We will extract from them the topologically protected features of dynamics in the combined U(1)$\times$SU(2) ``electric'' and ``magnetic'' fields. We will discover that topologically quantized constant drift currents and Hall effects are possible only when the appropriate gauge charges (charge and spin) are conserved. Focusing first on this special case, we will show in the section \ref{secTFT} that the drift component of motion agrees with the stationary action condition applied to the topological term of (\ref{TopLG}). This will justify $\mathcal{L}_{\textrm{t}}$ as an effective field theory of quantum Hall and spin-Hall effects that can replace the CS theory (later in the paper we will separately show how $\mathcal{L}_{\textrm{t}}$ can describe fractionalization, hierarchical quantum Hall states, etc.).

If the above had been our only goal, we would have been able to construct the effective topological Lagrangian $\mathcal{L}_{\textrm{t}}$ by directly considering the spin-Hall conductivity. However, we wish to also describe topological states that feature no quantum Hall effect, so only the SU(2) symmetry can guide us. The drift current analysis helps us to transparently construct $\mathcal{L}_{\textrm{t}}$ for any representation of any gauge symmetry group that allows a quantum Hall effect. Having spinors, we can easily describe particles of arbitrary charge and spin moving in any combination of U(1) electromagnetic, SU(2) spin-orbit and other fields. Now that $\mathcal{L}_{\textrm{t}}$ formally has the SU(2) gauge symmetry in any desired representation, we can directly apply it to the fractional TIs with the non-commutative gauge fields of the Rashba spin-orbit coupling. The ensuing spin non-conservation ruins the quantum spin-Hall effect, but any incompressible quantum liquid in which individual particles become microscopic cyclotron vortices will have excitations whose fractional statistics is topologically protected and correctly captured by $\mathcal{L}_{\textrm{t}}$. We will, therefore, have a tool that is more general than the CS theory and capable of handling the Rashba spin-orbit coupling.

The construction of $\mathcal{L}_{\textrm{t}}$ will make it clear that the fields $\psi$ are not the ordinary field operators of particles, but rather the dual field operators that represent vortices. Their dynamics is provided by the non-topological part $\mathcal{L}_\textrm{LG}$ of (\ref{TopLG}) in the most general Landau-Ginzburg form allowed by symmetries. The gauge field $\mathcal{B}$ implements the Magnus force on vortices in this language. Note that $\mathcal{L}_\textrm{LG}$ governs the dynamics of smooth configurations of $\psi$, while $\mathcal{L}_{\textrm{t}}$ is sensitive only to singular configurations. For this reason, the stationary action conditions for $\mathcal{L}_\textrm{LG}$ and $\mathcal{L}_{\textrm{t}}$ are essentially independent. The topological theory dual to (\ref{TopLG}) and expressed in terms of the particle field operators is given by the Lagrangian density (\ref{TopLGp}).

The quantization of classical equations of motion, which we apply to obtain the quantum field theory of TIs, rests upon knowing the exchange statistics of elementary objects. The Lagrangians (\ref{TopLG}) with and without the topological term $\mathcal{L}_{\textrm{t}}$ quantize the same classical system using different state-dependent exchange statistics. Therefore, the role of $\mathcal{L}_{\textrm{t}}$ is to specify quantum statistics and other fundamentally non-classical aspects of dynamics. This is done in a manner that depends on the presence and density of topological defects in the field $\psi$ configuration, which is a desirable property of a general theory that should describe fractional ground states. In contrast, the CS theory has a rigid implementation of statistics, specific to only one particular fractional ground state.

When employed in the context of FQHS, the theory (\ref{TopLG}) is closely related to the CS Landau-Ginzburg Lagrangian of Wen and Niu from the Ref.\cite{Wen1990a}. Going beyond this formal similarity, (\ref{TopLG}) is an effective theory in exactly the same sense as the CS theory in Wen's treatment of FQHS \cite{WenQFT2004}. We will demonstrate that (\ref{TopLG}) naturally generalizes the CS theories of Abelian FQHS to non-Abelian incompressible quantum liquids of particles with arbitrary internal degrees of freedom. Likely descriptions of non-Abelian FQHS in the present formalism \cite{Nikolic2011} seem to be somewhat different than other proposed field theories involving non-Abelian gauge fields \cite{Balatsky1991, Frohlich1992, Lopez1995, Fradkin1998, Fradkin1999, Fradkin2001}. No explicit assumptions about microscopic dynamics, such as the existence of composite bosons or fermions, are made in the construction of (\ref{TopLG}). This marks a contrast to several other approaches to FQHS that use CS gauge fields, including Landau-Ginzburg-CS \cite{Zhang1989, Zhang1989a, Zhang1992}, and ``Hamiltonian'' \cite{Murthy2003, Murthy2012} theories. The present effective field theory is also complementary to microscopic wavefunction constructions \cite{Laughlin1983, Jain1989, Moore1991}. It is better suited for the systematic prediction, classification and qualitative characterization of new possible topological orders, than for describing microscopic realizations of topological states with quantitative accuracy. Being not restricted to topological states in flat bands, this theory is a valuable tool for exploring the uncharted territory of TR-invariant fractional TIs.

The technical part of this paper begins with an introduction to the SU(2) gauge-field description of spin-orbit couplings in the section \ref{secQM}. A simple single-particle quantum mechanics is used there to establish the equation of motion for electrons in external electromagnetic and spin-orbit fields. The following section \ref{secTFT} explains how symmetries and the equations of motion can be used to construct the field theory (\ref{TopLG}), and especially its topological term.

The next major subject of the paper are certain essential properties of the theory and its initial predictions in the context of the simplest Laughlin-type topological orders. We will first apply fundamental principles in the section \ref{secFract} to show that fractionalization is mandated in a class of correlated quantum states that bridge between the phases of maximally localized and maximally delocalized particles. These include the fractional TIs without spin-conservation that could arise in solid-state materials. Then, we will discuss in the section \ref{secCS} how and in what special circumstances the CS theories arise from (\ref{TopLG}) as effective descriptions of fractionalized states. Such circumstances are not met in the currently available TIs, and we will identify in the section \ref{secTopOrd} a special dynamical symmetry of the Rashba spin-orbit coupling that can lead to new but utterly fragile topological quantum phases beyond the pure CS description. The following section \ref{secDeg} verifies the existence of topological order in all these phases by calculating their ground-state degeneracy on a torus and other non-simply connected surfaces. Finally, the section \ref{secDual} formally derives the topological field theory of physical particles dual to (\ref{TopLG}), and takes a bigger perspective on the relationship between quantum Hall states and conventional phases of matter. We will touch upon the possibility of revealing the origins of fractionalization in dynamics. The stability of topological orders against perturbations that violate the SU(2) gauge structure is briefly discussed in the section \ref{secStab} from the duality point of view.

The following segment of the paper goes beyond the Laughlin-type topological orders and explores the ways in which the proposed effective theory (\ref{TopLG}) can be generalized to describe arbitrary Abelian hierarchical quantum Hall states (section \ref{secHier}) and many non-Abelian ones (sections \ref{secNonAbel} and \ref{TInonAbel}). We will demonstrate how the generalized topological term of (\ref{TopLG}) can shape unconventional quantum statistics once the dynamics governed by the Landau-Ginzburg part selects appropriate low-energy fluctuations of the spinor fields. We will discuss in greater detail a class of novel and robust non-Abelian topological orders that can arise specifically due to the Rashba spin-orbit coupling. All conclusions and an outlook of the many remaining issues are summarized in the section \ref{secConcl}.

\section{Effective theory of topological insulators}

\subsection{Classical and quantum mechanics}\label{secQM}

The simplest model of a quantum well made from a topological band-insulator material such as Bi$_2$Se$_3$ or Bi$_2$Te$_3$ is given by the Hamiltonian:
\begin{equation}\label{Bernevig}
H = v\,\hat{{\bf z}} ({\bf S}\times{\bf p}) \tau^{z} + \Delta\tau^{x} - \mu \ .
\end{equation}
The four-component spinor wavefunction $\psi({\bf r})$ captures electron's internal states labeled by the spin projection $\sigma^z$, and the orbital index $\tau^z$ that can be interpreted as the top or bottom surface of the quantum well. The vector spin operator is ${\bf S} = \frac{1}{2} \sigma^a \hat{\bf r}^a$, $a\in\lbrace x,y,z \rbrace$, where $\sigma^a$ and $\tau^a$ are Pauli matrices that act on spin and orbital degrees of freedom respectively (we set $\hbar=1$). In a bulk crystal, the two surfaces would be far apart and decoupled ($\Delta=0$), so their energy spectrum $E(p) = \pm \frac{1}{2} vp - \mu$ would contain massless Dirac states with the helical correlation ${\bf S}=\pm\frac{1}{2} \hat{\bf p} \times \hat{\bf z}$ between momentum and spin. Assuming that the chemical potential $\mu$ were placed well within the bulk bandgap, the above Hamiltonian would then consistently describe the low-energy part of the full spectrum that contains only the surface states. However, electrons in a quantum well can tunnel between the two surfaces ($\Delta \neq 0$), which opens up a gap in the Dirac spectrum of surface states. A two-dimensional band-insulator can be obtained by pushing $\mu$ into this tunneling bandgap in a gated heterostructure. The given Hamiltonian is the minimal model of 2D electrons that both experience a spin-orbit coupling and have a finite bandgap without violating the TR symmetry. It has identical spectrum to the model of HgTe quantum wells introduced by Bernevig, \etal \cite{Bernevig2006}, and may be considered different from it only by the choice of representation. Experimental evidence for the validity of this model comes both from bulk systems and quantum wells \cite{Hsieh2009, Hsieh2009a, Xia2009, Konig2007, Zhang2010}.

The Hamiltonian (\ref{Bernevig}) is related to a gauge theory for electrons in a static external SU(2) gauge field $\boldsymbol{\mathcal{A}}$. Consider:
\begin{equation}\label{GaugeTh}
H' = \frac{({\bf p}-g\boldsymbol{\mathcal{A}})^{2}}{2m} + \Delta\tau^{x} - \mu'
\end{equation}
where
\begin{equation}\label{Asolid}
\boldsymbol{\mathcal{A}}=-mv(\hat{{\bf z}}\times{\bf S}) \quad , \quad g=\tau^{z} \ .
\end{equation}
The SU(2) charge $g$ operates in the orbital subspace, and $\mathcal{A}_\mu$ are SU(2) matrices derived from spin operators. Gauge transformations are specified by three angles $\theta^a({\bf r},t)$ combined into an SU(2) transformation matrix $W({\bf r},t)$:
\begin{equation}\label{GaugeTransf}
W = e^{i \theta^a \gamma^a} \quad,\quad
\mathcal{A}_\mu \to W \mathcal{A}_\mu W^\dagger + \frac{i}{g}W\partial_{\mu}W^{\dagger} \ .
\end{equation}
For spin $S=\frac{1}{2}$ particles, the SU(2) generators are $\gamma^a = \frac{1}{2}\sigma^a$. The Hamiltonians (\ref{Bernevig}) and (\ref{GaugeTh}) produce the same operator equation of motion $\dd {\bf j} / \dd t = i \lbrack H, {\bf j} \rbrack$ in the Heisenberg picture for the current (velocity) operator ${\bf j} = i \lbrack H, {\bf r} \rbrack$:
\begin{equation}\label{EqMot0}
\frac{\dd j_i}{\dd t} = i \lbrack H,j_i \rbrack =
\frac{1}{2}v^{2}\epsilon_{ij}p_{j}\sigma^{z}-\Delta v\epsilon_{ij}\sigma^{j}\tau^{y} \ .
\end{equation}
The symbol $\epsilon_{ij} \equiv \epsilon_{0ij}$ is the 2D antisymmetric tensor that implements vector cross products in the Einstein notation. This equation of motion (written at $t=0$) illustrates the quantum cyclotron dynamics of electrons in its dependence on the spin and orbital index $\tau^z$, which is the fundamental origin of all topological properties. In particular, we can immediately see the tendency of acceleration ${\dd \bf j}/{\dd t}$ to be perpendicular to the particle's momentum $\bf p$, and its dependence on the spin $\sigma^z$ that embodies the TR-invariance. Note, however, that spin precession is not properly taken into account here (will be in the subsequent analysis). From the gauge theory perspective, the cyclotron dynamics is caused by the presence of a finite SU(2) ``magnetic'' flux density:
\begin{equation}\label{U1SU2Flux0}
\Phi^\mu = \epsilon^{\mu\nu\lambda} ( \partial_\nu \mathcal{A}_\lambda - ig \mathcal{A}_\nu \mathcal{A}_\lambda )
         = \frac{1}{2}(mv)^{2}\delta_{\mu 0}\,\tau^z\sigma^z \ .
\end{equation}
The non-Abelian nature of SU(2) gauge fields allows a finite flux even when the gauge field is uniform. Gauge transformations (\ref{GaugeTransf}) merely rotate the flux in a spatially dependent way, $\Phi^\mu \to W \Phi^\mu W^\dagger$.

The model (\ref{GaugeTh}) is different from (\ref{Bernevig}) by the extra $p^2/2m$ term and a constant. The mass parameter $m$ determines the curvature of electron band-dispersions $E(p)$ at larger momenta, which is indeed seen in ARPES experiments \cite{Hsieh2009, Hsieh2009a}. Therefore, we can regard (\ref{GaugeTh}) as a more accurate description of realistic systems than (\ref{Bernevig}), and take advantage of having the parameter $m$ to define the cyclotron frequency scale and flux density. This will prove extremely useful in building the topological field theory of correlated TIs. The gauge model (\ref{GaugeTh}) should be considered valid only below a cutoff momentum scale $\Lambda = \sqrt{(mv)^2-(\Delta/v)^2}$ in order to ensure a true bandgap $2\Delta$ and a natural shape of the valence band. Such a cutoff is indeed produced by the crystal lattice of a realistic system. The presence of a bandgap is essential for the existence of topologically non-trivial insulating states.

We must note that realistic systems do not have the SU(2) gauge symmetry. Still, their topological properties can be protected as long as the perturbations to $H'$ that violate the gauge symmetry do not remove the SU(2) flux. We shall postpone the discussion of gauge-symmetry violations to the section \ref{secStab} and focus first on the pure charge and spin Hall effects. We will explore the combined effects of spin-orbit couplings and external electromagnetic fields on any particles by generalizing the Hamiltonian to the U(1)$\times$SU(2) symmetry group with arbitrary spin $S$ representation.

Band-insulating solid state TIs exhibit a particular realization of an SU(2) ``magnetic'' field. We will consider more general situations in the following, described by the Hamiltonian of particles that have both an electromagnetic U(1) charge $e$ and spin-orbit SU(2) charge $g$:
\begin{equation}\label{U1SU2}
H_0 = \frac{({\bf p} - e {\bf a} - g {\bf A}^a \gamma^a)^2}{2m} - ea_0 - gA_0^a\gamma^a \ .
\end{equation}
This is sufficient for analyzing the cyclotron motion that stands behind all topological phenomena. We will implicitly assume the existence of internal degrees of freedom and microscopic features that are necessary to open a topological gap and stabilize a TI ground state. The general U(1)$\times$SU(2) gauge field $\mathcal{A}_\mu = a_\mu + A_\mu^a \gamma^a$ carries flux:
\begin{equation}\label{U1SU2Flux}
\Phi^\mu = \epsilon^{\mu\nu\lambda} ( \partial_\nu \mathcal{A}_\lambda - ig \mathcal{A}_\nu \mathcal{A}_\lambda )
  = \phi_\mu^{\phantom{a}} + \Phi_\mu^a \gamma^a \ ,
\end{equation}
Its U(1) and SU(2) parts will be labeled by lowercase and uppercase symbols respectively, and $\gamma^a$ for $a\in\lbrace x,y,z \rbrace$ are the three SU(2) generators (angular momentum operators) in any spin-$S$ representation. The temporal $\Phi^0$ and spatial $\Phi^i$ flux matrices correspond to ``magnetic'' $B$ and $90^o$-rotated ``electric'' $E_i$ fields respectively, which together form the field tensor $F_{\mu\nu}$:
\begin{eqnarray}
&& ~~~ B = F_{xy} = -F_{yx} \quad , \quad E_i = F_{0i} = -F_{i0} \nonumber \\
&& F_{\mu\nu}=\partial_{\mu}\mathcal{A}_{\nu}-\partial_{\nu}\mathcal{A}_{\mu}-ig\lbrack\mathcal{A}_{\mu},\mathcal{A}_{\nu}\rbrack
  = \epsilon_{\mu\nu\lambda}\Phi^{\lambda} \nonumber \\
&& ~~~~~~~~~~~~~~~~ \Phi^{\mu}=\frac{1}{2}\epsilon^{\mu\nu\lambda}F_{\nu\lambda} \ .
\end{eqnarray}
The traces of $\Phi^\mu$ contain the U(1) electromagnetic fields $\phi^\mu$, while their traceless parts contain the analogous spin-dependent SU(2) fields. Only the eigenvalues of $\Phi_\mu$ (and $F_{\mu\nu}$) are gauge-invariant. Defining the charge $j_\mu$ and spin $J_\mu^a$ current operators,
\begin{eqnarray}\label{CurrOp}
  j_{0}=1                 & \quad,\quad &    j_{i}=\frac{1}{m}(p_{i}-ea_{i}-gA_{i}^a\gamma^a) \nonumber \\
  J_{0}^{a}=\gamma^{a}    & \quad,\quad &    J_{i}^{a}=\frac{1}{2}\lbrace\gamma^{a},j_{i}\rbrace
\end{eqnarray}
we obtain the following Heisenberg equation of motion for the spatial current components from (\ref{U1SU2}):
\begin{equation}\label{Heisenberg}
\frac{\dd j_{i}}{\dd t} = i \lbrack H_0,j_i \rbrack = \frac{e}{m}\epsilon_{i\nu\lambda}^{\phantom{a}}
  j_{\nu}^{\phantom{a}} \phi_{\lambda}^{\phantom{a}}
  + \frac{g}{2m}\epsilon_{i\nu\lambda}^{\phantom{a}} \lbrace j_{\nu}^{\phantom{a}} ,\Phi_{\lambda}^a\gamma^a\rbrace \ .
\end{equation}
One should keep in mind that all operators in this equation are expressed at time $t$ in the Heisenberg picture, including the flux operators $\Phi_\mu(t) \to e^{i H_0 t} \Phi_\mu e^{-i H_0 t}$ which will precess if $H_0$ and $\Phi_\mu$ do not commute. We will solve this differential equation for $j_i(t)$ treated as a matrix function of time. The expectation value $\langle\psi|j_i(t) |\psi\rangle$ calculated from the solution $j_i(t)$ in any state $|\psi\rangle$ will properly reflect the quantum time-evolution of currents, as well as the behavior of an equivalent classical system according to the Ehrenfest's theorem.

Let us first consider the special case of spin-conserving gauge fields $\mathcal{A}_\mu^{\phantom{z}} = a_\mu^{\phantom{z}} + A_\mu^z \gamma^z$ whose components commute with each other. The resulting flux operators commute with the Hamiltonian, so that $\Phi_\mu(t) = \Phi_\mu = \textrm{const}$ in the Heisenberg picture. Writing
\begin{equation}\label{CurrentExpand}
j_i^{\phantom{z}}(t) = \lambda_i^0(t) + \lambda_i^a(t) \gamma^a
\end{equation}
and organizing the scalars $\lambda_i^0$ and $\lambda_i^a$ into an eight-component vector $\lambda(t)$ reduces (\ref{Heisenberg}) to linear differential equations with constant coefficients whose matrix form and solution are:
\begin{eqnarray}\label{LDE}
&& ~~~~~~ \frac{\dd \lambda}{\dd t} + A \lambda = b \\
&& \lambda(t) = e^{-At}\lambda(0)+A^{-1}b \nonumber \ .
\end{eqnarray}
All eigenvalues of the matrix $A$ are purely imaginary and thus generate cyclotron oscillations. The resulting Heisenberg current operator is
\begin{equation}\label{EqMotSol}
j_{i}(t) = c_{i}e^{i\omega t}e^{i\gamma^{z}\omega^{z}t} + \delta j_i \ ,
\end{equation}
where the first term describes cyclotron motion with frequencies $\omega = e\phi_0/m$, $\omega^z = g\Phi_0^z/m$ and amplitudes $c_y = i c_x$ appropriate for circular classical trajectories. The second term $\delta j_i$ is a constant drift current perpendicular to both ``electric'' and ``magnetic'' fields. Note that $\delta j_i$ is state-independent and thus topologically protected, unlike the cyclotron orbit amplitudes $c_i$.

We will now concentrate on the drift current kinematics. Setting $\dd j_i / \dd t = 0$ in (\ref{Heisenberg}) and $c_i=0$ in (\ref{EqMotSol}), we easily find:
\begin{equation}\label{dJ}
\delta j_{i}^{\phantom{z}}=\Bigl(e\phi_{0}^{\phantom{z}}+g\Phi_{0}^z\gamma^{z}\Bigr)^{-1}
       \Bigl(e\phi_{i}^{\phantom{z}}+g\Phi_{i}^z\gamma^{z}\Bigr)
    =\sum_{k=0}^{2S}u_{i,k}\,(\gamma^{z})^{k} \ .
\end{equation}
It is not hard to recognize that this equation indirectly describes the quantum Hall effect. The amount of drift current is completely determined by the ``magnetic field'' (perpendicular to the sample's plane) and the in-plane ``electric field'' perpendicular to the current flow. The coefficients $u_{i,k}$ can be calculated by expanding both sides of this equation in the powers of $\gamma^z$, and noting that there are only $2S+1$ independent matrices among $(\gamma^z)^n$ in the spin $S$ representation. We will not pursue this expansion. Instead, we will need a slightly different formula
\begin{eqnarray}\label{EqMot}
\delta j_{i}^{\phantom{z}} &\!\!\!=\!\!\!&
    \Bigl\lbrack e^{2}\phi_{0}^{2}+2eg\phi_{0}\Phi_{0}^{z}\gamma^{z}+g^{2}(\Phi_{0}^{z})^2(\gamma^{z})^{2}
    \Bigl\rbrack^{-1} \\
 && \Bigl\lbrack e^{2}\phi_{0}\phi_{i}
    +eg(\phi_{0}\Phi_{i}^{z}+\phi_{i}\Phi_{0}^{z})\gamma^{z}
    +g^{2}\Phi_{0}^{z}\Phi_{i}^{z}(\gamma^{z})^{2}\Bigr\rbrack \ , \nonumber
\end{eqnarray}
which is obtained by inserting $e\phi_{0}^{\phantom{z}} + g\Phi_{0}^z\gamma^{z}$ and its inverse into (\ref{dJ}).

Now let us briefly consider the analogous dynamics of Rashba spin-orbit-coupled electrons. The gauge field (\ref{Asolid}) produces the Hamiltonian (\ref{U1SU2}) that does not commute with the flux operators (\ref{U1SU2Flux0}). Consequently, the Heisenberg-picture operator $\Phi_\mu(t)$ in (\ref{Heisenberg}) has a non-trivial time dependence. The proper way to evolve the gauge field operators is to treat the time evolution in the Heisenberg picture as a generalized gauge transformation that leaves all equations of motion invariant and ensures that all measurable (gauge-invariant) observables evolve according to $\mathcal{O}(t)\to e^{iH_0t}\mathcal{O}e^{-iH_0t}$:
\begin{equation}
\mathcal{A}_{\mu}(t) \to e^{iH_0t}\mathcal{A}_{\mu}e^{-iH_0t}+\frac{i}{g}e^{iH_0t}\partial_{\mu}e^{-iH_0t}
\end{equation}
We can handle the flux precession by formally seeking the time-dependent operator solutions in the Schrodinger picture, where the flux operators are static:
\begin{eqnarray}\label{Schrodinger}
&& ~~~~~~~~~~~~~~~~~ j_i(t) = e^{iH_0t} j_i'(t) e^{-iH_0t} \\
&& \frac{\dd j_i'}{\dd t} + i \lbrack H_0,j_i' \rbrack = \frac{e}{m}\epsilon_{i\nu\lambda}^{\phantom{a}}
  j_\nu' \phi_{\lambda}^{\phantom{a}}
  + \frac{g}{2m}\epsilon_{i\nu\lambda}^{\phantom{a}} \lbrace j_\nu' ,\Phi_{\lambda}^a\gamma^a\rbrace \ . \nonumber
\end{eqnarray}
Like before, we can expand $\delta j_i'$ as in (\ref{CurrentExpand}) to reduce the above equation to the form (\ref{LDE}). Its solution for $\delta j_i'(t)$ must then be used to obtain the Heisenberg-picture operator $j_i(t)$ that properly captures the full dynamics.

It is not useful for the purpose of this paper to calculate the detailed and complicated expression for $j_i(t)$. We will, however, benefit from revealing some qualitative properties of the dynamics shaped by the Rashba spin-orbit coupling. First, the residual commutator $\lbrack H_0,j_i' \rbrack$ in (\ref{Schrodinger}) introduces the momentum operator into the general solutions for $j_i(t)$, because the spin-orbit coupling is proportional to momentum but also contains the spin operators that do not commute with the flux (\ref{U1SU2Flux0}). This means that all aspects of the current dynamics explicitly depend on the electrons' momenta. Second, even the constant drift component of $j_i'$ is turned into an oscillating current in $j_i$. The only way to obtain a constant current that satisfies (\ref{Schrodinger}) is to insist on $\lbrack H, j_i \rbrack = 0$. It can be easily seen that such solutions are possible when $\phi_\mu = 0$, but they are not topologically protected because they can have any amplitude independent of the fluxes. Therefore, this dynamics does not feature a quantum Hall effect. There are certain topologically-protected aspects of the dynamics, but they are buried in the oscillatory and momentum-dependent motion of electrons such as spin precession.

The equivalent expressions (\ref{dJ}) and (\ref{EqMot}) are the most general operators that extract the topologically protected drift charge currents from any quantum Hall state of particles in the external U(1)$\times$SU(2) ``electromagnetic'' fields. By symmetry, these expressions can be generalized to any SU($N$) group. The actual measurable currents of SU($N$) charges are state-dependent.

The equation of motion for a hypothetical classical TI can be obtained from (\ref{Heisenberg}) or (\ref{Schrodinger}) by equating the quantum expectation values of its left and right-hand sides in any wave-packet state. A wave-packet here has a spinor structure that should be interpreted as a representation of the classical spin orientation in some direction. The spin direction can precess, and the equation for that can be similarly derived from the time evolution of spin current operators in the Heisenberg picture. The classical trajectories generally involve spin precession coupled to orbital motion.

\subsection{Quantum field theory construction}\label{secTFT}

We now turn to interacting systems and construct a topological field theory that describes spin $S$ particles and produces the equations of motion (\ref{EqMot}) from its kinematics. We will set $e=g=1$ for simplicity and continue to rely on the full U(1)$\times$SU(2) gauge symmetry in order to emphasize the essential TI physics. No microscopic information is available for a derivation of this field theory, so we must quantize (\ref{EqMot}) the same way it is done in high energy physics.

The Lagrangian we seek is required to respect the U(1)$\times$SU(2) gauge symmetry in the continuum limit of current interest, as well as the translational, rotational (point-group) and TR symmetries unless the external gauge fields violate them explicitly. We wish to express this Lagrangian in terms of a spinor field $\psi$ whose internal degrees of freedom naturally correspond to spin-$S$ particles. The usual approach would then be to associate $\psi^\dagger$ and $\psi$ with the particle's creation and annihilation operators respectively, and construct a second-quantized Lagrangian from the single-particle Hamiltonian such as (\ref{GaugeTh}). However, this is the path to a microscopic formulation of the many-body Lagrangian in which the elementary excitations are not fractionalized and have a pre-determined statistics. Extracting any emergent non-trivial statistics from the quantum vorticity of strongly interacting particles would be extremely difficult.

Instead, our goal is to construct an effective theory that can capture fractionalization in quantum Hall states more directly. This theory must still be consistent with symmetries and classical equations of motion. Being deprived of the usual Lagrangian constructs, we need to consider topological terms that evaluate to zero when the field configuration is smooth (in a simply-connected space). The simplest one allowed by symmetries is given by $\mathcal{L}_{\textrm{t}}$ in (\ref{TopLG}):
\begin{eqnarray}\label{TopTerm}
&& \!\!\!\!\!\!\!\!\! \mathcal{L}_{\textrm{t}} = -\frac{i\eta}{2} \psi^{\dagger}\epsilon^{\mu\nu\lambda}\Bigl\lbrack
     (\partial_{\mu}-i\mathcal{A}_{\mu})
     \Bigl\lbrace\partial_{\nu}-i\mathcal{A}_{\nu},\Phi_{0}\Bigr\rbrace
     (\partial_{\lambda}-i\mathcal{A}_{\lambda}) \nonumber \\
&& \!\!\!\!\! + \Bigl\lbrace(\partial_{\mu}-i\mathcal{A}_{\mu})(\partial_{\nu}-i\mathcal{A}_{\nu})
     (\partial_{\lambda}-i\mathcal{A}_{\lambda}),\Phi_{0}\Bigr\rbrace\Bigr\rbrack\psi \ .
\end{eqnarray}
We will label the components $\psi_s$ of the spinor $\psi$ by the spin projection $s\in\lbrace -S,\dots,S \rbrace$ on the $z$-axis, or the axis selected by the external spin-orbit flux $\Phi_0$. Variants of this expression turn out to be inadequate for our purposes. For example, omitting the gauge fields $\mathcal{A}_\mu$ would fail to produce the desired gauge-invariance and equations of motion, while omitting the SU(2) flux matrix $\Phi_0$ would yield undesired transformation under TR. Note that $\mathcal{L}_{\textrm{t}}$ changes sign under TR ($\psi^{\phantom{*}}_s\to\psi^*_{-s}$) if $\Phi_0 = \phi_0$ contains only the U(1) magnetic field, while it remains invariant if $\Phi_0^{\phantom{z}} = \Phi_0^z \gamma^z$ contains only the spin-orbit coupling. We use anti-commutators to symmetrize $\mathcal{L}_{\textrm{t}}$ with respect to the location of $\Phi_0$, and introduce an unknown coupling constant $\eta$ which cannot be determined from classical considerations. We will treat $\mathcal{L}_{\textrm{t}}$ alone as the Lagrangian that replaces the CS theory in its role to effectively describe topological orders. However, one should keep in mind that it cannot be a complete theory by itself. It acts like a Berry's phase in the full Lagrangian (\ref{TopLG}), being imaginary in imaginary time.

The action is stationary when the field configuration obeys:
\begin{eqnarray}\label{EqMot2}
\psi^{\dagger}\frac{\partial\mathcal{L}_{\textrm{t}}}{\partial\psi^{\dagger}} &=& -\frac{i\eta}{2} \psi^{\dagger}\Bigl\lbrack
  (\partial_{\mu}-i\mathcal{A}_{\mu})\Phi_{0}+\Phi_{0}(\partial_{\mu}-i\mathcal{A}_{\mu})\Bigr\rbrack
  \nonumber \\ && \times \Bigl(\epsilon^{\mu\nu\lambda}\partial_{\nu}\partial_{\lambda}-i\Phi^{\mu}\Bigr)\psi+h.c.=0 \ .
\end{eqnarray}
We used $\epsilon^{\mu\nu\lambda}(\partial_{\nu}-i\mathcal{A}_{\nu})(\partial_{\lambda}-i\mathcal{A}_{\lambda}) = \epsilon^{\mu\nu\lambda}\partial_{\nu}\partial_{\lambda}-i\Phi^{\mu}$ to derive this form. Clearly, the field configurations that satisfy
\begin{equation}\label{EqMot2b}
\Bigl(\epsilon^{\mu\nu\lambda}\partial_{\nu}\partial_{\lambda}-i\Phi^{\mu}\Bigr)\psi=0
\end{equation}
also satisfy (\ref{EqMot2}). Note that the path-integral allows singularities in $\psi$ for which the order of the above two derivatives matters. Only such singularities produce a finite contribution. For example, if $\psi({\bf r}) = e^{i\theta}$ in cylindrical coordinates, then $-i\epsilon^{0\nu\lambda}\psi^\dagger \partial_\nu \partial_\lambda \psi = \epsilon^{0\nu\lambda} \partial_\nu b_\lambda = 2\pi\delta({\bf r})$, where $b_\mu = \partial_\mu \theta$ is the gauge field of a flux tube at the origin. Therefore, the condition (\ref{EqMot2b}) applies to the $\psi$'s topological defects. If we are to interpret it as an equation of motion for particle charge $j_{\pp\mu}$ and spin $J_{\pp\mu}^{a}$ currents, we have no option but to express them as curls of certain $\psi$ field currents:
\begin{eqnarray}\label{pCurrents}
j_{\pp\mu} &=& \epsilon^{\mu\nu\lambda} \partial_{\nu}^{\phantom{a}} \widetilde{j}_{\vv\lambda}^{\phantom{a}} \\ J_{\pp\mu}^{a} &=& \epsilon^{\mu\nu\lambda} \partial_{\nu}^{\phantom{a}} \widetilde{J}_{\vv\lambda}^{a} \ . \nonumber
\end{eqnarray}
Symmetries require that we choose:
\begin{eqnarray}\label{Currents}
\widetilde{j}_{\vv\mu} &=& -\frac{i}{2}\Bigl\lbrack\psi^{\dagger}\Phi_0(\partial_{\mu}\psi)
    -(\partial_{\mu}\psi^{\dagger})\Phi_0\psi\Bigr\rbrack \\
\widetilde{J}_{\vv\mu}^{a} &=& -\frac{i}{2}\Bigl\lbrack\psi^{\dagger}\gamma^{a}\Phi_0(\partial_{\mu}\psi)
    -(\partial_{\mu}\psi^{\dagger})\Phi_0\gamma^{a}\psi\Bigr\rbrack \nonumber \ .
\end{eqnarray}
Inserting the $\Phi_0$ factors is necessary for proper transformations under TR: $j_{\textrm{p}0}\to j_{\textrm{p}0}$, $j_{\textrm{p}i}\to -j_{\textrm{p}i}$, $J_{\textrm{p}0}^a\to -J_{\textrm{p}0}^a$, $J_{\textrm{p}i}^a\to J_{\textrm{p}i}^a$. This is a duality relationship. If (\ref{pCurrents}) are to represent particle currents, (\ref{Currents}) must correspond to vortex currents. Even though the formulas (\ref{Currents}) do not transform properly under gauge transformations, we only care that the particle currents (\ref{pCurrents}) do. When the external flux $\Phi_0$ is uniform and constant in time, we can rewrite the vortex charge current from (\ref{Currents}) as:
\begin{equation}
\widetilde{j}_{\vv\mu} = \frac{i}{2}\partial_{\mu}(\psi^{\dagger}\Phi_0\psi)-i\psi^{\dagger}\Phi_0(\partial_{\mu}\psi) \ ,
\end{equation}
and substitute it in (\ref{pCurrents}) to simplify the particle charge current:
\begin{eqnarray}
&& \!\!\!\!\!\!\!\!\!\!\!
    j_{\pp\mu} = \frac{i}{2}\epsilon^{\mu\nu\lambda}\partial_{\nu}\partial_{\lambda}(\psi^{\dagger}\Phi_0\psi)
      -i\epsilon^{\mu\nu\lambda}(\partial_{\nu}\psi^{\dagger})\Phi_0(\partial_{\lambda}\psi) \nonumber \\
&&    -i\epsilon^{\mu\nu\lambda}\psi^{\dagger}\Phi_0\partial_{\nu}\partial_{\lambda}\psi
    = -i\epsilon^{\mu\nu\lambda}\psi^{\dagger}\Phi_0\partial_{\nu}\partial_{\lambda}\psi \ .
\end{eqnarray}
This simplification comes from the fact that $\psi^\dagger\Phi_0\psi$ is real and cannot expose any vortex singularities of $\psi$ to the double derivative curl $\epsilon^{\mu\nu\lambda}\partial_\nu\partial_\lambda$. Similarly, $\epsilon^{\mu\nu\lambda} (\partial_{\nu} \psi^{\dagger}) \Phi_0 (\partial_{\lambda} \psi)$ vanishes because its singular part reduces to the sum of terms like $\epsilon^{\mu\nu\lambda} b_{s\nu} b_{s\lambda}$, where $b_{s\mu} = \partial_\mu \theta_s$ are obtained from the phases $\theta_s$ of the individual spinor $\psi$ components (expressed in the representation that diagonalizes $\Phi_0$). In summary:
\begin{eqnarray}\label{pCurrents2}
j_{\pp\mu} &=& -i\epsilon^{\mu\nu\lambda}\psi^{\dagger}\Phi_0\partial_{\nu}\partial_{\lambda}\psi \\
J_{\pp\mu}^{a} &=& -\frac{i}{2}\epsilon^{\mu\nu\lambda}\psi^{\dagger} \lbrace \Phi_0,\gamma^a \rbrace
                    \partial_{\nu}\partial_{\lambda}\psi \ . \nonumber
\end{eqnarray}

Knowing the symmetry-restricted form of currents, we can interpret the equation of motion (\ref{EqMot2b}). We emphasized earlier that this is sensible only in quantum spin-Hall states, which require spin conservation and commuting gauge field and flux components, $\Phi_\mu^a = \Phi_\mu^z \delta_{az}^{\phantom{a}}$. If we multiply (\ref{EqMot2b}) from the left by $\psi^\dagger (\gamma^z)^n \Phi_0$ for $n=0,1$ and extract the currents (\ref{pCurrents2}) from the obtained expressions, we find
\begin{eqnarray}\label{EqMot3}
\!\!\!\!\!\! j_{\pp\mu}^{\phantom{z}} &=& \phi_{0}^{\phantom{z}}\phi_{\mu}^{\phantom{z}}\Gamma_{0}^{\phantom{z}}
  + (\Phi_{0}^z\phi_{\mu}^{\phantom{z}}+\phi_{0}^{\phantom{z}}\Phi_{\mu}^z)\Gamma_{1}^{\phantom{z}}
  + \Phi_{0}^z\Phi_{\mu}^z\Gamma_{2}^{\phantom{z}} \\[0.03in]
\!\!\!\!\!\! J_{\pp\mu}^{z} &=& \phi_{0}^{\phantom{z}}\phi_{\mu}^{\phantom{z}}\Gamma_{1}^{\phantom{z}}
  + (\Phi_{0}^z\phi_{\mu}^{\phantom{z}}+\phi_{0}^{\phantom{z}}\Phi_{\mu}^z)\Gamma_{2}^{\phantom{z}}
  + \Phi_{0}^z\Phi_{\mu}^z\Gamma_{3}^{\phantom{z}} \ , \nonumber
\end{eqnarray}
where we introduced the symbols $\Gamma_{n} = \psi^{\dagger} (\gamma^z)^n \psi$. These are the many-body particle currents in the stationary action state, which depend on the external magnetic and spin-orbit fluxes as well as the average vortex densities $\Gamma_n$. We are now ready to show that these second-quantized equations of motion are equivalent to the first-quantized ones obtained in the previous section. We can interpret (\ref{EqMot}) as the renormalized current $\delta j_i = j_i / j_0$ that describes a single particle $\delta j_0 = 1$. The ensuing many-body quantum average $\langle j_i \rangle = \langle j_0 \delta j_i \rangle$ of the single-quantized formalism reproduces (\ref{EqMot3}) after replacements $\langle (\gamma^z)^n \rangle \to \Gamma_n$, $\langle j_\mu \rangle \to j_{\pp\mu}$. Analogous correspondence between equations of motion is found for all topologically protected drift currents, including spin currents and even currents with arbitrary powers of $\gamma^z$ placed in (\ref{pCurrents2}).

Therefore, $\mathcal{L}_{\textrm{t}}$ captures the kinematics of any topologically protected drift currents in the combined U(1) and SU(2) ``electromagnetic'' fields. Conversely, $j_{\pp\mu}$ and $J_{\pp\mu}^a$ given by (\ref{pCurrents}, \ref{pCurrents2}) are only the drift components of the particle charge and spin currents respectively. Recall that the CS theory is related to the classical drift motion in the same manner as $\mathcal{L}_{\textrm{t}}$. The fluctuating currents are described by the gauge field $\mathcal{B}_\mu$ in (\ref{TopLG}). We will show later that $\mathcal{L}_{\textrm{t}}$ also determines the topological order of incompressible quantum liquids. Its ability to do so transcends the quantum Hall states that we used to derive it.

The full many-body equation of motion (\ref{EqMot2}) implements the conservation of the non-drifting component of particle currents. However, its additional solutions beyond (\ref{EqMot3}) exhibit spatial and temporal changes of particle densities or currents. These are suppressed in incompressible quantum liquids by the Landau-Ginzburg part of (\ref{TopLG}). Such dynamics can be prominent only in conventional quantum phases, but then the entire topological term of (\ref{TopLG}) is irrelevant as we will explain shortly.

\section{The essential properties of Laughlin states}

\subsection{Fractionalization in incompressible quantum liquids}\label{secFract}

The conventional quantum phases of bosonic particles that the Lagrangian (\ref{TopLG}) can describe are superconductors and Mott insulators. Superconductors can admit localized vortices at the expense of expelling particles from the nearest vicinity of vortex singularities (cores). This tends to marginalize the topological term $\mathcal{L}_{\textrm{t}}$ of (\ref{TopLG}), because $\mathcal{L}_{\textrm{t}}$ thrives on having a finite vortex density $\psi^\dagger \psi \neq 0$ and particle density (\ref{pCurrents2}) in the same regions of space, according to (\ref{EqMot2}). The analogous conclusion holds in Mott insulators from the dual point of view. A Mott insulator is a superfluid of vortices whose smooth field $\psi$ configurations cannot generate a finite $\mathcal{L}_{\textrm{t}}$, except at regions in space where physical particles (topological defects of $\psi$) are localized. However, vortex currents are expelled from such regions ($\psi^\dagger \psi \to 0$). It takes strong quantum fluctuations to intermix particle and vortex densities and make $\mathcal{L}_{\textrm{t}}$ important.

Incompressible quantum liquids are characterized by having an overlapping uniform particle density and diffused flux density. Both densities are incompressible and this can be formally stated by two conditions: (A) the density $\rho_s$ fluctuations are suppressed in all spinor components $\psi_s = \sqrt{\rho_s} \exp(i\theta_s)$ of the vortex field $\psi$; (B) vortices are locally coherent so that the phase gradients $b_{s\mu} = \partial_\mu\theta_s$ follow the fluctuations of the gauge field $\mathcal{B}_\mu$ (whose spinor component curls represent dynamical particle currents in individual spin channels). In the case of bosonic particles, the first condition is realized in superfluid and superconducting states, while the second condition holds in Mott insulators (vortex condensates). By duality, (A) and (B) tend to be mutually exclusive. However, quantum Hall states allow both conditions to be satisfied at the time and length scales that are probed in the following analysis. We will show that the theory (\ref{TopLG}) unavoidably gives rise to quasiparticle excitations with fractional amounts of electron's charge and spin when the ground state meets both conditions. Note that the condition (A) is more restrictive than necessary, since it leads to Abelian quantum Hall liquids of the Laughlin type. Generalizations to other incompressible quantum liquids are postponed until the section \ref{secGen}.

The condition (B) means that the ``drift'' currents (\ref{pCurrents}) can be considered equivalent to the appropriate dynamical particle currents given by the fluxes $\Phi_{\mathcal{B}\mu}$ of the gauge field $\mathcal{B}_\mu$ in (\ref{TopLG}). It allows us to use $j_{\pp\mu}$ from (\ref{pCurrents}) to express the amount of charge $Q = j_{\pp 0}dA_0$ located within a small sample area $dA_0$ during a very short interval of time. We can also define space-time oriented surface elements $dA_i = dl_i dt$ and express by $\Delta Q_{dt} = j_{\pp i}dA_i$ the amount of charge pushed through the sample's line segment $dl_i$ in the time interval $dt$. Now consider a quantum measurement of $Q$ or $\Delta Q_{dt}$. The outcome is random, but always equal to an integer multiple of the elementary charge $e$. We can similarly extract the amount of spin $S^z = J_{\pp\mu}^zdA_{\mu}^{\phantom{z}}$, which must appear quantized in any measurement.

Let us denote by $dC$ the oriented space-time contour that bounds $dA_\mu$. The phases of the spinor $\psi_s$ can have only integer winding numbers $n_s$ around the loop $dC$ if $\psi$ is to be single-valued:
\begin{equation}
n_{s}=\frac{1}{2\pi}\oint\limits _{dC}\dd l_{\mu}\,\partial_{\mu}\theta_{s} \in \mathbb{Z} \ .
\end{equation}
We can use (\ref{pCurrents}) and (\ref{Currents}) to express the measured charge and spin in terms of $n_s$ and vortex densities $\rho_s$, which are kept constant by the condition (A):
\begin{eqnarray}\label{FracExc}
Q &=& \oint\limits _{dC}\dd l_{\mu}\,\widetilde{j}_{\vv\mu} =
      \sum_{s=-S}^S 2\pi n_{s} (\phi_0^{\phantom{z}}+s\Phi_0^z) \rho_{s} \\
S^{z} &=& \oint\limits _{dC}\dd l_{\mu}\,\widetilde{J}_{\vv\mu}^{z} =
      \sum_{s=-S}^S 2\pi n_{s} (\phi_0^{\phantom{z}}+s\Phi_0^z) s \rho_s \nonumber \ .
\end{eqnarray}
Microscopic excitations are characterized by quantum numbers $(Q,S^z)$, where $Q=1$ and $S^z \in \lbrace -S,\dots,S \rbrace$ in units $e=\hbar=1$. The fixed densities $\rho_s$ surely cannot depend on measurement outcomes $(Q,S^z)$. Hence, we can view (\ref{FracExc}) as a system of equations for $n_s$, which are integers that depend on the measurement outcomes $(Q,S^z)$. We can solve these equations by requiring that only one of the $2S+1$ integers $n_s$ be non-zero for each microscopic excitation:
\begin{equation}
n_s(1,S^z) = m_s \delta_{s,S^z} \quad , \quad m_s = \frac{1}{2\pi \rho_s (\phi_0^{\phantom{z}} + s\Phi_0^z)} \ .
\end{equation}
The linearity of (\ref{FracExc}) then generates the solutions $(n_{-S},\dots,n_S)$ for general $(Q,S^z)$ by adding the solutions for microscopic excitations. We see that the $2S+1$ numbers $m_s$ must be integers. This imposes a restriction on the allowed values for densities:
\begin{equation}\label{DensityQuant}
\rho_s = \frac{1}{2\pi m_s (\phi_0^{\phantom{z}} + s\Phi_0^z)} \ .
\end{equation}
We have indeed obtained $\rho_s$ that are independent of the measurement outcomes, but depend on the external fluxes and a set of integers $m_s$ that must, therefore, characterize the ground state. There are no other physically acceptable solutions of (\ref{FracExc}).

The ground state charge and spin densities extracted from (\ref{EqMot3}) are:
\begin{equation}\label{EqState}
j_{\pp 0} = \sum_{s=-S}^{S}\frac{\phi_0+s\Phi_0^z}{2\pi m_s} \quad , \quad
J_{\pp 0}^z = \sum_{s=-S}^{S}\frac{s\phi_0+s^2\Phi_0^z}{2\pi m_s} \ .
\end{equation}
Various combinations of $m_s$ can lead to states with broken particle-hole symmetry $j_{\pp 0}\neq 0$ or magnetization $J_{\pp 0}^z\neq 0$, especially in combined magnetic $\phi_0$ and spin-orbit $\Phi_0^z$ fluxes. While these symmetries may be easily explicitly broken independently of any quantum Hall physics, the above equations of state are purely a result of orbital motion and can describe spontaneous symmetry breaking when particles are relativistic and/or not Zeeman-coupled to external fields. The symmetry properties of ground states are related to the quantum numbers and statistics of quasiparticle excitations via $m_s$.

Since both charge and spin are delocalized in quantum Hall states, a finite area of the sample can contain any amounts of them on average. However, (\ref{FracExc}) still relates the amounts of charge and spin to quantized winding numbers $n_s$ of dual-vortices. A single dual-vortex $n_s = \delta_{s,\sigma}$ in the spin channel $\sigma$ can be isolated in principle in some quantum state, for example the state of being localized inside a small area of the sample. If an experimentalist managed to suppress the fluctuations of $n_s$ in this localized state, he or she would measure on average a fractionally quantized amount of charge $\delta Q$ and spin $\delta S^z$. More generally, a bundle of dual-vortices with arbitrary $n_s$ would look like a quasiparticle with charge and spin (in units $e=\hbar=1$):
\begin{equation}\label{QuantNum}
\delta Q = \sum_{s=-S}^S \frac{n_s}{m_s} \quad , \quad \delta S^z = \sum_{s=-S}^S \frac{n_s}{m_s}s \ .
\end{equation}
The fractionalization formulas (\ref{FracExc}) and (\ref{QuantNum}) are independent of any fluctuations of the ``quantum numbers'' $n_s$. A spin-orbit coupling such as (\ref{Asolid}) can favor quasiparticles that exist in superpositions of the above states with different $\delta S^z$. Further degradation of these ``quantum numbers'' can occur in the presence of perturbations beyond the spin-orbit coupling that do not conserve spin. However, vortex excitations, which are dual to the above quasiparticles, can survive as protected fractional degrees of freedom because their ``charges'' are always conserved (see sections \ref{secStab} and \ref{secHier}).

We will show in the section \ref{secDual} that fractionalization is dynamically related to vortex ``charge''. If low-energy vortices carry an integer number $m$ of flux quanta $hc/e$, then charge fluctuations exhibit the fractionalized quantum $e/m$, observable for example in shot-noise transport measurements. The analysis in this section actually exploits this fact. The fractional quasiparticles do not a priori have an unconventional exchange statistics in generic systems. However, quantum Hall states effectively bind a fractionalized amount of charge to a singly-quantized vortex, and the resulting quasiparticle is an anyon. The topological term in (\ref{TopLG}) regulates the statistics of these quasiparticles in the present formalism (and simultaneously gives rise to the ground-state degeneracy on a torus). Specifically, the fractional statistics is generated by the $\psi^\dagger \epsilon^{\mu\nu\lambda} \partial_\mu \partial_\nu \partial_\lambda \psi$ part of the topological term. The following section will reveal that this is in fact the sum of CS self-couplings
\begin{equation}
-\frac{i \epsilon^{\mu\nu\lambda}}{8}\, \psi^{\dagger} \Bigl(
     \Phi_0 \partial_\mu \partial_\nu \partial_\lambda + \cdots \Bigr) \psi \to
  \sum_s \frac{i \epsilon^{\mu\nu\lambda}}{4\pi m_s} b_{s\mu} \partial_\nu b_{s\lambda} \nonumber
\end{equation}
for each ``gauge field'' $b_{s\mu} = \partial_\mu \theta_s$ in ground states with incompressible vortices. Assuming that $b_{s\mu}$ follow the physical particle currents per condition (B), each quasiparticle in the spin-channel $\sigma$ acts as a source of $2\pi/m_\sigma$ flux in the same spin channel. Consider two fractional quasiparticles with identical quantum numbers $(\delta Q, \delta S^z)$ specified by the integers $n_s = \delta_{s,\sigma}$. Their two-body wavefunction $\xi_\sigma({\bf r}_1,{\bf r}_2)$ acquires the factor $\exp(i\gamma_\sigma)$ when they are exchanged, given by the statistical angle of the spin channel $\sigma$:
\begin{equation}\label{StatAngle}
\gamma_{\sigma} = \frac{\pi}{m_\sigma} \ .
\end{equation}
Note that $m_\sigma=1$ corresponds to integer quantum Hall states of fermionic particles.

For spin $S = \frac{1}{2}$ particles, having no spin-orbit coupling $\Phi_0^z=0$ and choosing $m_{\pm 1/2} \in \mathbb{Z}$ such that $\nu=2m_{+1/2}^{-1}=2m_{-1/2}^{-1}$ yields the Laughlin sequence of fractional quantum Hall states in the external magnetic field $\phi_0$. The ground state particle density is $j_{\pp 0} = \nu \phi_0 / 2\pi$, there is no magnetization $J_{\pp 0}^z = 0$, and the fundamental quasiparticle excitations carry fractional charge $\delta Q = \nu/2$ and spin $\delta S^z = \pm\nu/4$. We see that spin must be fractionalized just like charge, effectively reducing $\hbar$ by an integer. A correlated TR-invariant TI ($\Phi_0^z \neq 0$) in zero magnetic field $\phi_0=0$ exhibits the same combined spin and charge fractionalization when $m_{+1/2} = -m_{-1/2}$. Generally, $m_s = -m_{-s}$ is required if the ground state is to be invariant under TR, which ties together the charge and spin fractions. Independent fractionalization of charge and spin generally requires TR symmetry breaking, even in the zero magnetic field.

\subsection{Chern-Simons theory of quantum Hall states}\label{secCS}

In this section we derive from (\ref{TopLG}) a simplified effective theory in which the fluctuations of spinor $\psi$ amplitudes are neglected. The topological term turns into a CS coupling when all U(1)$\times$SU(2) gauge fields and their flux matrices can be simultaneously diagonalized. This physically corresponds to having a conserved spin projection in addition to the conserved charge. We will later show that pure CS gauge theories are not equally well suited for systems without this symmetry, and eventually argue that new topological orders could arise from the Rashba spin-orbit coupling in solid-state TIs.

The spinor field $\psi = (\psi_{-S},\dots,\psi_S)$ has $2S+1$ complex components $\psi_s = \sqrt{\rho_s}\exp(i\theta_s)$ in the theory of spin $S$ particles. We choose to express it in the representation that diagonalizes the flux matrix $\Phi_0^{\phantom{z}} = \phi_0^{\phantom{z}} + \Phi_0^z \gamma^z$, and assume for now that all gauge field matrices $\mathcal{A}_\mu$ and $\mathcal{B}_\mu$ are also diagonal in this representation. This will ensure that all $\mathcal{A}_\mu$ commute with each other in the given gauge. A diagonal $\mathcal{B}_\mu = \textrm{diag}(\beta_{-S,\mu},\dots,\beta_{S,\mu})$ is sensible only if the background particle ``density'' $\Theta_\mu = \textrm{diag}(\Theta_{-S,0},\dots,\Theta_{S,0}) \delta_{\mu,0}$ is also diagonal in this representation; then the gauge fields $\beta_{s\mu}$ independently represent the conserved $2S+1$ spin states of physical particles. When density fluctuations are small and mostly confined to microscopic scales, we may approximate $\rho_s = \psi_s^\dagger\psi_s^{\phantom{\dagger}}$ in all spin channels $s$ by their averages, and capture dynamics solely via the fluctuations of phases $\theta_s$. In the continuum limit with translational symmetry, the average $\rho_s$ are uniform, so the Landau-Ginzburg part of (\ref{TopLG}) becomes:
\begin{eqnarray}\label{XY}
\mathcal{L}_\textrm{LG} &\approx& \sum_{s=-S}^{S} \biggl\lbrack
    \frac{K\rho_s}{2}\bigl(\partial_\mu\theta_s-\beta_{s\mu}\bigr)^2 \\
&& +\frac{1}{8\pi^2 q_s^2}\bigl(\epsilon^{\mu\nu\lambda}\partial_\nu\beta_{s\lambda}-\Theta_{s0}\delta_{\mu,0}\bigr)^2
   +\cdots \biggr\rbrack \ . \nonumber
\end{eqnarray}
Here, $q_s$ are the diagonal matrix elements of $Q$ in (\ref{Maxwell1}), which is the appropriate Maxwell term in the presence of spin conservation. The dots denote terms that depend only on $\rho_s$, and can be used to determine $\rho_s$ in a saddle-point approximation. We will relate this to an XY model on a lattice in the section \ref{secDual}, which couples $2S+1$ independent XY fields $\theta_s$ to the non-compact gauge fields $\beta_{s\mu}$. The present continuum limit is best suited for describing smooth fluctuations of $\theta_s$, but one should keep in mind that the more accurate formulation of this model requires a lattice.

Now we treat the topological term $\mathcal{L}_{\textrm{t}}$ from (\ref{TopTerm}) to the same approximation. $\mathcal{L}_{\textrm{t}}$ is sensitive only to the topological defects of the spinor field $\psi$. Let us integrate by parts the left-most derivative of $\mathcal{L}_{\textrm{t}}$ in (\ref{TopTerm}) and write $\mathcal{L}_{\textrm{t}} = \mathcal{L}_{\textrm{t}}' + \delta\mathcal{L}_{\textrm{t}}$, where $\delta\mathcal{L}_{\textrm{t}}$ is the total derivative of a field bilinear. By Gauss' theorem, $\delta\mathcal{L}_{\textrm{t}}$ picks monopoles $\partial_\mu (\epsilon^{\mu\nu\lambda} \partial_\nu b_{s\lambda}) \neq 0$ of the ``gauge fields'' $b_{s\mu} = \partial_\mu \theta_s$. However, the bulk monopole-charge density is zero, so $\delta\mathcal{L}_{\textrm{t}}$ can contribute to the path-integral only at the system boundaries. This can be seen from $\delta\mathcal{L}_{\textrm{t}} \propto \partial_\mu j_{\pp\mu}$, where we interpret the (2+1)D divergence $\partial_\mu j_{\pp\mu} \to 0$ of the conserved charge current as monopole-charge density according to (\ref{pCurrents2}). The only bulk contribution to the topological term comes from $\mathcal{L}_{\textrm{t}}'$, which is sensitive to the vortex singularities $\phi_{s\mu} = \epsilon^{\mu\nu\lambda} \partial_\nu b_{s\lambda} \neq 0$ and yields CS effective theories in incompressible states.

Let us organize the phase gauge fields $b_{s\mu}$ into a diagonal matrix $B_\mu = \textrm{diag}(b_{s\mu})$ whose flux is $\Phi_B^\mu = \textrm{diag}(\phi_s^\mu)$. The bulk topological $\theta_s$ kinematics is captured by:
\begin{eqnarray}\label{EffTh}
\mathcal{L}_{\textrm{t}}' &\!\!\!=\!\!\!&
  \frac{i\eta}{2}\Bigl\lbrack(\partial_{\mu}-i\mathcal{A}_{\mu})\psi\Bigr\rbrack^\dagger \Phi_0
  \Bigl\lbrack(\epsilon^{\mu\nu\lambda}\partial_{\nu}\partial_{\lambda}-i\Phi^{\mu})\psi\Bigr\rbrack + \cdots \nonumber \\
&\!\!\!=\!\!\!& \frac{i\eta}{2}\textrm{tr}\Bigl\lbrack(B_\mu-\mathcal{A}_\mu)\Phi_0
  (\Phi_B^\mu-\Phi_{\phantom{B}}^\mu)(\psi\psi^\dagger)\Bigr\rbrack + \cdots \ .
\end{eqnarray}
We emphasized only one of the four symmetrization terms in (\ref{TopTerm}), but the other three denoted by dots generate the same expression.

In the $S=0$ representation, all spinors and matrices have a single component, and the external flux $\Phi_0 = \phi_0$ can describe only the U(1) magnetic field. This immediately yields the CS theory of a Laughlin quantum Hall state \cite{WenQFT2004}, although not normalized in the standard fashion. We have seen in the section \ref{secFract} that the density $\rho \equiv \rho_0 = |\psi|^2$ must be quantized in quantum Hall states as $\rho = (2\pi m \phi_0)^{-1}$, where $m$ is a positive integer that specifies the winding number of the phase $\theta$ associated to a single localized electron excitation. With this in mind, we may redefine the ``gauge field'' $b_\mu = m c_\mu$ to associate a single electron to one flux quantum of $c_\mu$. The charge current (\ref{pCurrents2}) becomes
\begin{equation}
j_{\pp\mu} = \frac{1}{2\pi} \epsilon^{\mu\nu\lambda}\partial_\nu c_\lambda \ ,
\end{equation}
and
\begin{equation}\label{CSTh}
\mathcal{L}_{\textrm{t}}' \to \mathcal{L}_\textrm{CS} = -4i\eta
  \left\lbrack-\frac{m}{4\pi}\epsilon^{\mu\nu\lambda}c_\mu\partial_\nu c_\lambda+j_{\pp\mu}a_\mu\right\rbrack \ .
\end{equation}
We have assumed that the external gauge field satisfies $\epsilon^{\mu\nu\lambda}a_\mu\partial_\nu a_\lambda=0$, which typically is the case. This is still different from the standard form \cite{WenQFT2004}. First, the factor of $-i$ is present because the Lagrangian is expressed in the imaginary time; converting it to real time removes $-i$. Second, there is an overall coupling constant $\eta$ that could naively have any value. However, the CS self-coupling $-\frac{m}{2}j_{\pp\mu}c_\mu$ defines the quantum exchange statistics of particles because $j_{\pp\mu}$ is the flux of $c_\mu$. The value of $4\eta m$ must be an even integer to reproduce the bosonic statistics, and an odd integer to yield the fermionic statistics of elementary particles. Since $m$ is already an integer, we conclude that $\eta=\frac{1}{4}$. This value has been already applied in (\ref{TopLG}). It should be emphasized again that $j_{\pp\mu}$ are local drift currents of particles, which however are identified with the actual fluctuating particle currents $\epsilon^{\mu\nu\lambda} \partial_\nu \beta_\lambda$ in the quantum Hall states where (\ref{XY}) is still capable of locking $b_\mu = \partial_\mu \theta$ to $\beta_\mu$ ($K$ is large enough).

If particles have spin $S=\frac{1}{2}$, then $\psi\psi^\dagger$ in (\ref{EffTh}) is a $2\times 2$ matrix with elements $\sqrt{\rho_s\rho_{s'}} \exp\lbrack i(\theta_s-\theta_{s'})\rbrack$, where $s, s' = \pm \frac{1}{2}$. We will here restrict ourselves to the cases where all components of the SU(2) gauge fields and their fluxes can be simultaneously diagonalized. This implies $\lbrack \Phi_0 , \Phi_i \rbrack = 0$ and allows choosing the gauge $\mathcal{A}_{0} = -\epsilon_{ij} x_{i} \Phi_{j}$, $\mathcal{A}_{i} = -\frac{1}{2} \epsilon_{ij} x_{j} \Phi_{0}$. Now the only non-diagonal matrix in (\ref{EffTh}) is $\psi\psi^\dagger$, so its off-diagonal components do not matter under the trace. Its diagonal elements are equal in a quantum Hall state that respects the TR symmetry ($\Phi_0^{\phantom{z}} = \Phi_0^z\gamma^z$) and we have discovered their quantization $\rho_s = (\pi m \Phi_0^z)^{-1}$ with positive integer $m$ in the section \ref{secFract}. The CS theory will have two decoupled sectors corresponding to two spin states. In order to obtain the ``background field'' (BF) representation of the CS theory \cite{Cho2010, Neupert2011}, we can decompose the ``gauge field'' matrix $B_\mu = \frac{m}{2} (c_\mu^{\textrm{s}} + c_\mu^{\textrm{c}} \sigma^{z})$ into the charge-like $c_\mu^{\textrm{c}}$ and spin-like $c_\mu^{\textrm{s}}$ scalar components:
\begin{equation}
\mathcal{L}_{t}' \to  \mathcal{L}_\textrm{BF} = -i \left\lbrack
  -\frac{m}{4\pi}\epsilon_{\phantom{\mu}}^{\mu\nu\lambda}c_{\mu}^{\textrm{c}}\partial_{\nu}^{\phantom{a}}c_{\lambda}^{\textrm{s}}
  +j_{\pp\mu}^{\phantom{z}}a_{\mu}^{\phantom{a}}+J_{\pp\mu}^{z}A_{\mu}^{z} \right\rbrack
\end{equation}
The gauge fields have been defined again to represent one unit of charge and spin by a single flux quantum:
\begin{equation}
j_{\pp\mu} = \frac{1}{2\pi} \epsilon_{\phantom{\mu}}^{\mu\nu\lambda} \partial_{\nu}^{\phantom{a}} c_{\lambda}^{\textrm{c}}
 \quad,\quad
J_{\pp\mu}^{z} = \frac{1}{4\pi} \epsilon_{\phantom{\mu}}^{\mu\nu\lambda} \partial_{\nu}^{\phantom{a}} c_{\lambda}^{\textrm{s}} \ .
\end{equation}
This theory predicts a quantized spin Hall conductivity $\sigma_{xy}^{\textrm{s}} = J_{\pp i}^z/\phi_i = (2\pi m)^{-1}$, pertaining to the Laughlin spin-Hall liquids with the conserved $S^z$ spin projection.

It should be pointed out that the CS theories obtained here are slightly different than the standard ones. The present CS ``gauge fields'' $b_{s\mu}$ are gradients of spinor phases $\theta_s$, so their configurations admit only quantized flux loops by the requirement that $\exp(i\theta_s)$ be single-valued. A flux quantum corresponds to the smallest fractional amount of charge or spin. In order to obtain a standard CS theory, one must integrate out short length-scale fluctuations in the path-integral. The resulting coarse-grained CS theory can allow flux to diffuse and become consistent with unconstrained gauge fields.

\subsection{Fragile topological phases from the Rashba spin-orbit coupling}\label{secTopOrd}

The precise form of the external gauge field may determine certain features of the topological ground state that are not as robust as topological order, but depend on symmetries. For example, the SU($N$) Hall conductivity is ``topologically'' quantized only if the appropriate SU($N$) charge is conserved, while the bulk topological order can exist without charge conservation. Here we wish to shed some more light on these issues.

CS theories describe naturally only the charge-conserving situations. An attempt to derive a CS theory appropriate for the gauge field (\ref{Asolid}) of solid-state TIs quickly runs into a difficulty. First note that (\ref{EffTh}) with fixed vortex densities $\rho_s$ expresses the topological term $\mathcal{L}_{\textrm{t}}$ for any external SU(2) gauge field $\mathcal{A}_\mu$ in the representation that diagonalizes its ``magnetic'' flux $\Phi_0$. However, the $\mathcal{A}_\mu$ of (\ref{Asolid}) is not diagonal in that representation. The effective model for small density fluctuations contains not only the CS gauge fields $c_{s\mu}$, but also explicit functions of $\theta_s$ which come from the off-diagonal elements of $\mathcal{A}_\mu$. In other words, the resulting effective theory is not a pure gauge theory in this language. The phases $\theta_s$ generally fluctuate in a correlated state-dependent manner, so it is not permissible to simply neglect them or average them out.

The CS theory depends on the external gauge field $\mathcal{A}_\mu$ only through its flux (\ref{U1SU2Flux}), which reduces to $\Phi^\mu = \epsilon^{\mu\nu\lambda}\partial_\nu \mathcal{A}_\lambda$ under the previously imposed restriction that all gauge field and flux components can be simultaneously diagonalized (and thus commute with each other). However, the same SU(2) flux can be obtained from different gauge fields that cannot be related by a gauge transformation. Compare for example:
\begin{eqnarray}\label{A1A2}
&& \mathcal{A}_{1\mu} = g A_\mu^z \gamma^z
   \quad,\quad \epsilon^{\mu\nu\lambda} \partial_\nu A_\lambda^z = \Phi\delta_{\mu,0} \\
&& \mathcal{A}_{2\mu} = \sqrt{\Phi} (\delta_{\mu,x}\gamma^y - \delta_{\mu,y}\gamma^x)
   \nonumber \ .
\end{eqnarray}
These two gauge fields have the same flux
\begin{equation}\label{A1A2Flux}
\Phi^\mu = \epsilon^{\mu\nu\lambda} ( \partial_\nu \mathcal{A}_\lambda - ig \mathcal{A}_\nu \mathcal{A}_\lambda )
  = g\Phi\gamma^z \delta_{\mu,0} \ ,
\end{equation}
but only the first one has commuting components. The second gauge field is actually applicable to solid-state TIs. If we used $\mathcal{A}_{1\mu}$ in the Hamiltonian
\begin{equation}
H_g = \frac{({\bf p} - g\boldsymbol{\mathcal{A}})^2}{2m} \ ,
\end{equation}
the spectrum would consist of macroscopically degenerate Landau levels, while $\mathcal{A}_{2\mu}$ would produce a fundamentally different Dirac particle spectrum. This indicates that there is no SU(2) gauge transformation that converts $\mathcal{A}_{1\mu}$ to $\mathcal{A}_{2\mu}$, despite the fact that their fluxes are the same. Clearly, the eigenvalues of the flux matrices are not the only gauge-invariant quantities that characterize the SU(2) gauge fields.

The qualitative distinction between $\mathcal{A}_{1\mu}$ and $\mathcal{A}_{2\mu}$ can be related to global symmetries. The ideal SU(2) gauge Hamiltonian $H_1$ with $\mathcal{A}_{1\mu}$ conserves the $z$-projection of spin $C_1 = \gamma^z$, while the same Hamiltonian $H_2$ with $\mathcal{A}_{2\mu}$ conserves ``helical spin'' $C_2 = p_x \gamma^y - p_y \gamma^x$, where $p_{x,y}$ are momentum operator components. The commutators $\lbrack H_i, C_i \rbrack=0$ and $\lbrack C_1, C_2 \rbrack \neq 0$ indicate that each Hamiltonian (gauge field configuration) has a global symmetry, but the two symmetries are incompatible and cannot be established simultaneously. The two global symmetries can be regarded as gauge-dependent: gauge transformations alter the Hamiltonians as $H_i \to W H_i W^\dagger$, so that the conserved operators must be transformed as $C_i \to W C_i W^\dagger$ in order to keep all commutators intact. Note, however, that there is no gauge transformation that could convert $C_1$ into $C_2$, that is $C_1 \neq W C_2 W^\dagger$ for all $W=\exp(i\theta^a\gamma^a)$.

Topologically ordered but otherwise featureless many-body ground states $|0_i\rangle$ of the ideal second-quantized Hamiltonians $\mathcal{H}(\mathcal{A}_{i\mu})$ are bound to have the respective incompatible symmetries. The symmetry of $\mathcal{A}_{1\mu}$ is ``geometrical'', while the symmetry of $\mathcal{A}_{2\mu}$ is ``dynamical''. $|0_1\rangle$ and $|0_2\rangle$ could also be the ground states of a globally SU(2) symmetric Hamiltonian in different parameter regimes, in which case they would be separated by at least one symmetry-breaking phase transition. The qualitative difference between $|0_1\rangle$ and $|0_2\rangle$ could be very deep, depending on the dynamics. For example, one would expect that the most stable topological orders in $\mathcal{A}_{1\mu}$ typically feature Abelian quasiparticle statistics, while $\mathcal{A}_{2\mu}$ could dynamically prefer quantum liquids with non-Abelian statistics that we will discuss in the section \ref{TInonAbel}. However, here we will consider the minimal possible difference between $|0_1\rangle$ and $|0_2\rangle$. We will assume that both ground states can be characterized by the same quantized vortex densities $\rho_s$ in the $s=\pm\frac{1}{2}$ spin channels, according to (\ref{DensityQuant}). We will justify the validity of this assumption only in the section \ref{secStab}.

The simplest minimally different ground states $|0_i\rangle$ are both uncorrelated or Laughlin quantum liquids whose excitations can exhibit charge and spin fractionalization given by (\ref{QuantNum}). However, their specific symmetries affect the statistics of measurement outcomes in an observable way. Suppose that one could prepare the system by exciting a particular fractional quasiparticle $|q_i\rangle$ with desired quantum numbers above the ground state. The excitation $|q_1\rangle$ of the ideal Hamiltonian $\mathcal{H}(\mathcal{A}_{1\mu})$ will have a good quantum number $S^z$ and additional quantum number(s) derived from orbital motion. The eigenstates with opposite $S^z$ have the same energy but different orbital quantum numbers in spin-orbit-coupled TR-invariant TIs. Measuring $S^z$ of $|q_1\rangle$ many times would produce a fractionally quantized average value $\langle S^z \rangle$ because $S^z$ is conserved. Analogous quantum measurements of any other spin projection in $|q_1\rangle$ with fixed orbital quantum numbers would yield non-quantized averages that smoothly depend on the orientation of the spin-projection axis. The macroscopically degenerate Landau levels of $\mathcal{H}(\mathcal{A}_{1\mu})$ offer many choices of orbital states in which this measurement statistics could be observed. The quasiparticle excitations $|q_2\rangle$ of the ideal Hamiltonian $\mathcal{H}(\mathcal{A}_{2\mu})$ behave differently. There, one must prepare $|q_2\rangle$ in a momentum $\bf p$ eigenstate, and measure the spin projection along the $\hat{\bf z}\times\hat{\bf p}$ axis in order to observe the fractional quantization of average values. All other projection axes or orbital states would spoil the observation of spin fractionalization.

\begin{figure}
\subfigure[{}]{\includegraphics[width=1.5in]{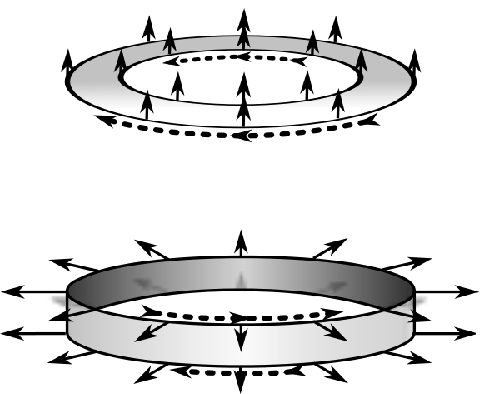}}
\subfigure[{}]{\includegraphics[width=1.5in]{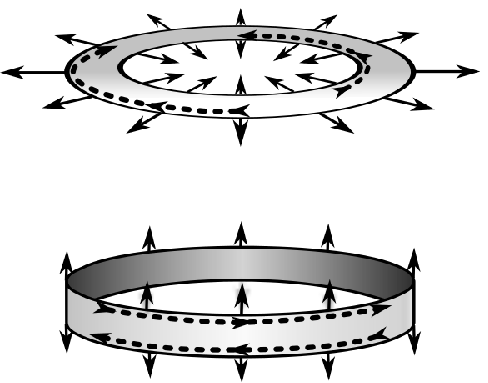}}
\caption{\label{Rings}Symmetry-protected edge spin-current states in the two SU(2) gauge fields given by (\ref{A1A2}): (a) $\mathcal{A}_{1\mu}$ that conserves $S^z$, and (b) $\mathcal{A}_{2\mu}$ that conserves $(\hat{\bf z}\times{\bf p}){\bf S}$. The direction of spin-current flow at each edge is shown by the dashed arrow, and the local projection of spin that flows is indicated by short arrows. The shown flat and cylindrical geometries are different not by topology, but by whether the edge is curved in the plane of the system or not ($\hat{\bf z}$ is always perpendicular to the sample plane). The two topological states (a) and (b) differ in each geometry by which spin projection flows and by whether the edge exerts a torque on the particle spin. Note that torque is perpendicular to spin in all cases, so it does not cause dissipation that would jeopardize the existence of gapless edge states.}
\end{figure}

A sharp distinction and a phase transition between the minimally different ground states $|0_1\rangle$ and $|0_2\rangle$ exists only when the symmetries of both are not jeopardized by the fundamental dynamics (e.g. when both states arise from spontaneous symmetry breaking in the same SU(2) symmetric many-body Hamiltonian). This is reflected in the nature of their topological symmetry-protected edge modes that we compare in the Figure \ref{Rings}. An entire class of smooth local perturbations has an effect on the spin density and current flows along edges that is equivalent to their local spin rotations, or an SU(2) gauge transformation. Since no gauge transformation can connect the edge modes and bulks of $|0_1\rangle$ and $|0_2\rangle$, these idealized ground states are different quantum phases. In fact, there are infinitely many different quantum phases of both kinds, characterized for example by the number of gapless edge modes (a spin Chern number in the uncorrelated lattice case of $|0_1\rangle$).

The spin-Hall conductivity is quantized only when the spin projection ($S^z$) parallel to the external SU(2) flux ($\Phi_0 \propto S^z$) is conserved. Therefore, the ideal $|0_1\rangle$ is a quantum spin-Hall state while $|0_2\rangle$ is not. However, the spin projection of a single particle is not conserved in realistic systems at least due to interactions. The ``dynamical'' spin symmetry of $|0_2\rangle$ is even more fragile. It depends on the single-particle momentum conservation, which is jeopardized by interactions, disorder, and even bending of the system's edges. Spin-fractionalization is only approximate in realistic systems. Its manifestations could be visible at such short time or length scales that allow neglecting all scattering events of a single quasiparticle excitation that alter its spin projection. The ground state incompressibility is helpful in this regard since it endows the low-energy quasiparticles with an infinite lifetime.

Apart from these imperfections of realistic systems, the fractionalization of the quasiparticle exchange statistics and the spectrum of symmetry-protected quantum numbers are the same in the two minimally different ground states $|0_i\rangle$. We will also show in the next section that the ground-state degeneracy on a torus is the same in these two Laughlin quantum liquids. Therefore, they have the same topological order because they are distinguished only by properties that are not topologically protected. We have seen that one of such properties is even the quantized spin-Hall conductivity of the ideal $|0_1\rangle$. Only the U(1) symmetry allows its respective Hall conductivity to be topologically protected. Any perturbation that removes the defining symmetries of $|0_1\rangle$ and $|0_2\rangle$ can open a gap in the edge state spectrum and thus ruin the fragile phase transition between them. The TR-symmetry alone protects only a Z$_2$ edge-state distinction between topological states with the same topological order \cite{Levin2009, Neupert2011, Levin2012}. The lack of symmetry also spoils the bulk measurements of fractional spin discussed above, but does not jeopardize the topological order expressed via the topologically protected numbers $m_s$ in (\ref{DensityQuant}) which determine the details of many-body quantum entanglement.

\subsection{Topological ground state degeneracy}\label{secDeg}

Here we calculate the ground state degeneracy of fractional TIs on non-simply connected surfaces. We will focus on the Laughlin sequence of TR-invariant fractional states of spin $S=\frac{1}{2}$ particles. Our main goal is to establish the existence of spin-related topological orders despite the fact that the Rashba spin-orbit coupling of solid-state materials spoils spin conservation. This is motivated by our interest in topological orders that are created by the Rashba spin-orbit coupling, rather than the ones merely perturbed by it.

\begin{figure}
\subfigure[]{\includegraphics[width=1.3in]{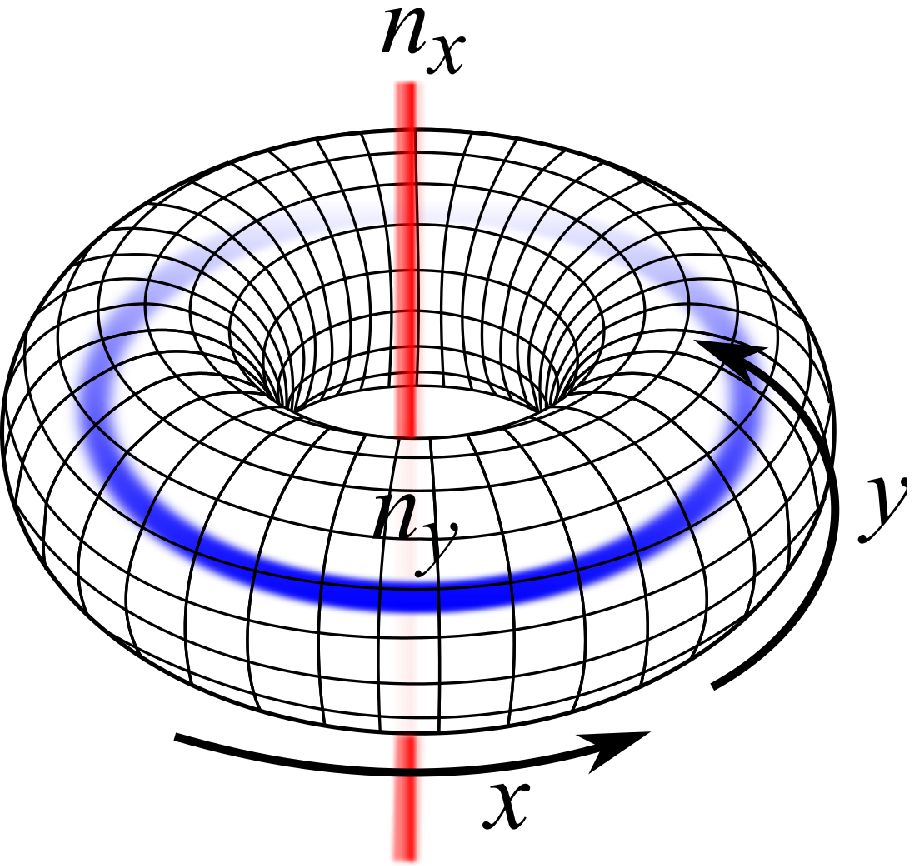}}
\hspace{0.2in}
\subfigure[]{\includegraphics[width=1.5in]{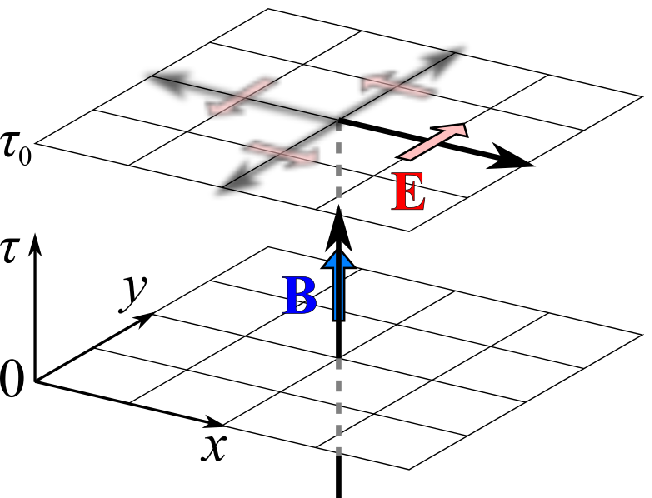}}
\caption{\label{FluxBend}(a) Magnetic flux loops with $n_{x}$ and $n_{y}$ quanta threaded through the torus openings. This is a depiction of the field configurations (\ref{GSconf}) that have a vanishing energy cost in the limit $l\to\infty$ of infinite torus size. (b) The emergence of the Faraday's law in the (2+1)D path-integral. A flux line stretching along the imaginary time direction $\tau$ represents the scalar magnetic field $B = \epsilon^{0ij} \partial_i b_j$ (thick blue arrow). The magnetic field is constant in this drawing until $\tau = \tau_0$ when it vanishes. Flux must be conserved, so the flux line (thin black arrow) must bend into a spatial plane in order for $B$ to vanish. A spatially-stretching flux line represents the vector electric field $E_i = - \partial_i b_0 - \partial_0 b_i$ (thick red arrow), which is related to the spatial flux component by the right-hand-rule, $\epsilon^{i\mu\nu} \partial_\mu b_\nu = \epsilon_{ij} E_j$. The flux line may extend in any direction in the $xy$ plane with equal probability. Whenever the path-integral dynamics leads to flux diffusion, these direction choices are averaged so that the coarse-grained electric field lines form closed loops and have magnitudes fixed by the flux conservation to the values predicted by the Faraday's law $\epsilon_{ij} \partial_i E_j = - \partial_0 B$. This drawing illustrates a segment of the torus space-time, but the analogous Faraday's law is generated by instanton events in which the quantized flux tubes threaded through the torus openings in the figure (a) bend to change the magnetic field in time and penetrate the torus space.}
\end{figure}

Let us first review the procedure for extracting the topological degeneracy in the well-understood situations when spin and charge are conserved \cite{Wen1990a}. Consider the TR-invariant SU(2) Hall effect shaped by the external gauge field $\mathcal{A}_{1\mu}$ in (\ref{A1A2}). The two spin projections are decoupled and experience opposite magnetic fields. This admits a pure CS gauge theory description of low energy dynamics in quantum Hall states. The TR-invariant Laughlin state discussed in the section \ref{secFract} is characterized by vortex densities $\rho_s = \psi_s^\dagger \psi_s^{\phantom{\dagger}} = 1/(\pi m \Phi_0^z)$, where $s=\pm\frac{1}{2}$ for spin $S=\frac{1}{2}$ particles, and $m$ is a positive integer. We assume that vortex fields $\psi_s = \sqrt{\rho_s} \exp(i\theta_s)$ are locally coherent in quantum Hall states, so their phase gradients $b_{s\mu} = \partial_\mu \theta_s$ are locked to the gauge field $\mathcal{B}_\mu$ that represents the physical particle currents. We may, therefore, integrate out $\mathcal{B}_\mu$ in (\ref{TopLG}) and use (\ref{EffTh}) to arrive at the following effective CS theory:
\begin{eqnarray}\label{LagDen1}
\mathcal{L}_{CS} &=& \sum_s \biggl\lbrack \frac{i\sigma_s}{4\pi m} \epsilon^{\mu\nu\lambda} b_{s\mu} \partial_{\nu} b_{s\lambda}
   -\frac{i}{4\pi m} \epsilon^{\mu\nu\lambda} b_{s\mu}\partial_{\nu} A_{\lambda}^{z} \nonumber \\
&& +\frac{1}{8 \pi^2 q^2}(\epsilon^{\mu\nu\lambda} \partial_{\nu} b_{s\lambda} - \Theta_{s0}\delta_{\mu 0})^{2} \biggr\rbrack \ .
\end{eqnarray}
We have defined $\sigma_s = 2s = \pm 1$, and omitted the generated vortex Coulomb and current-current interactions. The latter is justified by our exclusive interest in the ground states and the fact that vortices are always gapped. Recall that the ``gauge fields'' $b_{s\mu} = \partial_\mu \theta_s$ admit only quantized flux loops by the requirement that $\exp(i\theta_s)$ be single-valued. This constraint enables a sharp distinction between gauge field configurations that correspond to the ground-state manifold and excitations. Any quantized flux tube is bound to cost a finite amount of energy due to its narrow core. The exception are flux tubes that reside outside of the space in which the Lagrangian density is defined. If the space is shaped as a torus, a flux tube can be threaded through one of its openings without paying any core energy, as illustrated in the Figure \ref{FluxBend}(a).

The situation is temporarily complicated by the presence of $\Theta_{s0} \neq 0$ in (\ref{LagDen1}), which is related to the background density of particles. In normal circumstances, this nucleates flux lines that stretch along the path-integral's time direction at an average spatial density determined by $\Theta_{s0}$. The presence of a net bulk magnetic flux is reflected  by the corresponding circulating currents along the system's boundary. However, we will consider a torus geometry of space, which has no boundaries. The torus geometry frustrates the system by allowing only the configurations of temporal flux lines whose net flux is zero. A finite $\Theta_{s0}$ is, therefore, fairly innocuous on a torus. It is better to not have any temporal flux lines at all then to compensate every flux line by an anti-line. The exception are, again, flux tubes threaded through the torus openings. We conclude that ground-states on a torus are shaped by phase $\theta_s$ configurations that wind an integer number of times around any torus opening.

If the torus sizes in both $x$ and $y$ spatial directions are $l$, then the relevant ground-state configurations are:
\begin{equation}\label{GSconf}
\theta_{s}({\bf r})=\frac{2\pi}{l}(n_{sx}x+n_{sy}y) ~~ \Rightarrow ~~ b_{si}=\frac{2\pi n_{si}}{l} \ ,
\end{equation}
where $n_{si}$ are the integer winding numbers. Since $b_{s\mu} = \partial_\mu \theta_s$, we also find
\begin{equation}
b_{s0} = \frac{2\pi}{l} \left( x\partial_0 n_{sx} + y\partial_0 n_{sy} \right) \ ,
\end{equation}
and $\epsilon^{\mu\nu\lambda} \partial_\nu b_{s\lambda} = 0$ indicates that, indeed, no flux penetrates the torus surface. The configurations (\ref{GSconf}) are, however, not the only relevant ones because they prohibit electric fields (spatial flux) on the torus. If the magnetic flux threaded through a torus opening changes in time, $\partial_0 n_{si} \neq 0$, then electric field loops should appear on the torus surface according to the Faraday's law. This is physically required because a threaded magnetic flux is merely a circulating supercurrent flowing around the torus, so changing it requires an electric field pulse.

The Faraday's law is a consequence of flux conservation in the path-integral. A flux-tube stretching in the time direction is a constant magnetic field, while its bending toward a spatial direction turns it into an electric field flux line that unavoidably coincides with the change of magnetic field. Electric field is actually perpendicular to the spatial flux vector, so the flux-tube bending into various spatial directions generates circulating electric fields around the place where the magnetic field changes. It is necessary to integrate out short length- and time-scale fluctuations in the path-integral in order to recover the continuum Faraday's law from the diffusion of quantized flux loops. This is visualized in the Figure \ref{FluxBend}(b).

Now we can look for the missing important field configurations in (\ref{GSconf}). If one viewed the (2+1)D torus space-time as the boundary of a four-dimensional hyperspace, then a quantized magnetic flux tube threaded through a torus opening could bend in the ``fourth dimension'' to accommodate $\partial_0 n_{si} \neq 0$. This would cost energy whenever the tube touched or punctured the torus, but the cost can be arbitrarily small if the entry and exit points are sufficiently close together. We may regard such tube intrusions into the torus space-time as instanton configurations of $\theta_s$. They must be included in the low energy dynamics, but there is no need to construct them explicitly. We may simply coarse-grain the action on the torus by integrating out the short-scale fluctuations, and require that the Faraday's law be dynamically obtained from instanton events. The coarse-graining will allow flux to diffuse on the torus surface, but the threaded magnetic flux through the torus opening remains quantized. Now, note that keeping $b_{si}$ from (\ref{GSconf}) while setting $b_{s0}$ to zero would yield the configurations
\begin{eqnarray}\label{Faraday}
&& ~~~~~~~~~~~~~~~~~ b_{s0}=0 \quad,\quad b_{si}=\frac{2\pi n_{si}}{l} \\
&& \Rightarrow \quad\epsilon^{0ij}\partial_{i}b_{sj}=0 \quad,\quad
\epsilon^{i\mu\nu}\partial_{\mu}b_{s\nu}=-\frac{2\pi}{l}\epsilon_{ij}\partial_{0}n_{sj} \nonumber
\end{eqnarray}
that precisely implement the correct Faraday's law. Therefore, these are the full coarse-grained gauge field configurations that adequately represent the dynamics of the ground states.

The coarse-grained Lagrangian density has the same form as (\ref{LagDen1}), but with a renormalized coupling of the Maxwell term (the CS couplings are topological and never renormalized). If we substitute (\ref{Faraday}) in the coarse-grained Lagrangian density, convert it to real time and integrate over the spatial coordinates of the torus, we obtain the Lagrangian of the spatially-uniform low-energy modes:
\begin{equation}\label{Lag1}
L = \sum_s \Bigl\lbrack \frac{\pi\sigma_s}{m} \epsilon_{ij} n_{s i} \partial_{0} n_{s j}
   +\frac{1}{2q^2}(\partial_{0} n_{s i})^{2} \Bigr\rbrack \ .
\end{equation}
Note that $\Theta_{s0}$ drops out completely due to the torus geometry, and hence affects only the dynamics of excitations. The canonical coordinates are $q_{si} \equiv n_{si}$, so the corresponding canonical momenta are:
\begin{equation}
p_{si}=\frac{\partial L}{\partial(\partial_{0}q_{si})}=
  \frac{1}{q^2} \partial_{0}q_{si} - \frac{\pi\sigma_s}{m} \epsilon_{ij} q_{sj} \ ,
\end{equation}
and the Hamiltonian is:
\begin{equation}
H = \sum_{s} p_{si}\partial_{0}q_{si} - L = \frac{q^2}{2}\sum_{s}
  \left(p_{si}+\frac{\pi \sigma_s}{m}\epsilon_{ij}q_{sj}\right)^{2} \ .
\end{equation}
This is equivalent to the quantum mechanics of a single particle with an internal spin-state, in a spin-dependent external magnetic field. The particle is allowed to move only on a discrete square lattice whose lattice constant is 1. The equivalent gauge fields $\widetilde{a}_{si}$ and their magnetic fields $\widetilde{b}_{s}$ are:
\begin{equation}
\widetilde{a}_{si} = -\frac{\pi \sigma_s}{m} \epsilon_{ij} q_{sj} \quad\Rightarrow\quad
\widetilde{b}_{s}=\epsilon_{ij} \partial_{i} \widetilde{a}_{j} = \frac{2\pi \sigma_s}{m}
\end{equation}
The amount of flux per plaquette is $\pm 2\pi/m$, so that there are $1/m$ flux quanta per plaquette. The above Hamiltonian, thus, corresponds to a Hofstadter problem, whose spectrum is $m$-fold degenerate in each spin sector \cite{Hofstadter1976}. The total ground-state degeneracy is $m^{2}$.

We are now ready to analyze the main problem of interest. Consider the gauge field $\mathcal{A}_{2\mu}$ from (\ref{A1A2}), which captures the Rashba spin-orbit coupling in solid-state TIs. An attempt to derive the CS theory from (\ref{EffTh}) ends with the following effective theory:
\begin{eqnarray}\label{LagDen2}
\mathcal{L} &=& \sum_s \biggl\lbrack \frac{i\sigma_s}{4\pi m} \epsilon^{\mu\nu\lambda} b_{s\mu} \partial_{\nu} b_{s\lambda}
   -\frac{i\Phi_0^z}{8\pi m} b_{s0} \\
&& ~~~~~ +\frac{1}{8 \pi^2 q^2}(\epsilon^{\mu\nu\lambda} \partial_{\nu} b_{s\lambda} - \Theta_{s0}\delta_{\mu 0})^{2}
   \biggr\rbrack \nonumber \\
&& +\frac{i}{4\pi m}\frac{\sqrt{\Phi_0^z}}{2}\Bigl\lbrack e^{i(\theta_{\uparrow}-\theta_{\downarrow})}
    (\epsilon^{y\nu\lambda}-i\epsilon^{x\nu\lambda})\partial_{\nu}b_{\uparrow\lambda} \nonumber \\
&& ~~~~~~~~~~~~~~ -e^{-i(\theta_{\uparrow}-\theta_{\downarrow})}
    (\epsilon^{y\nu\lambda}+i\epsilon^{x\nu\lambda})\partial_{\nu}b_{\downarrow\lambda}\Bigr\rbrack \nonumber \ .
\end{eqnarray}
The spin non-conservation should introduce compact components in the Maxwell term (\ref{Maxwell1}), but we have expanded them to the quadratic order given that the relevant flux densities (\ref{Faraday}) are extremely small in the thermodynamic limit $l \to \infty$. Even though vortex current-current interactions are omitted again, this is not a pure gauge theory because $S^z$ is not conserved. Nevertheless, the relevant $\theta_s$ configurations in the ground-state manifold are still given by (\ref{GSconf}) and instantons for the same reasons as before. Instantons always introduce flux lines into the torus space-time and hence rotate $\theta_s$ in a manner that averages $\exp(i\theta_s)$ to zero over space-time. We have no means to capture this exactly in the complicated Lagrangian density (\ref{LagDen2}). However, we can consider a hypothetical worst-case scenario in which instantons fail to annihilate the $\exp(i\theta_s)$ factors even after coarse-graining. If instantons and coarse-graining only produced (\ref{Faraday}) as the relevant gauge field configurations $b_{s\mu}$ (which are now liberated from $\theta_s$ due to flux diffusion), then integrating out (\ref{LagDen2}) over the spatial coordinates on the torus would yield the real-time Lagrangian:
\begin{eqnarray}\label{Lag2}
L &=& \sum_s \Bigl\lbrack \frac{\pi\sigma_s}{m} \epsilon_{ij} n_{s i} \partial_{0} n_{s j}
   +\frac{1}{2q^2}(\partial_{0} n_{s i})^{2} \Bigr\rbrack \\
&& -\frac{l\sqrt{\Phi_0^z}}{4m}\Bigl\lbrack(\partial_{0}n_{\uparrow x}
   -\partial_{0}n_{\downarrow x})+i(\partial_{0}n_{\uparrow y}+\partial_{0}n_{\downarrow y})\Bigr\rbrack \times \nonumber \\
&& \times \delta_{n_{\uparrow x},n_{\downarrow x}}\delta_{n_{\uparrow y},n_{\downarrow y}} \nonumber \ .
\end{eqnarray}
The factors $e^{i(\theta_\uparrow - \theta_\downarrow)}$ are averaged to zero unless $n_{\uparrow i}=n_{\downarrow i}$ because of the windings (\ref{GSconf}). Consequently, the second part of the Lagrangian always vanishes, so this Lagrangian is actually identical to (\ref{Lag1}) and we are tempted to conclude that the ground-state degeneracy on a torus is again $m^2$. But, before we make any conclusions we must address the gauge-dependence of (\ref{LagDen2}) because choosing a gauge different than (\ref{GSconf}) would invalidate the above averaging of the $e^{i(\theta_\uparrow - \theta_\downarrow)}$ factors. Since we are not dealing with a true gauge theory, different gauges are really different physical states. Let us recall that spatially varying $\theta_s$ configurations generally represent vortex currents (\ref{Currents}) in the original theory (\ref{TopLG}). We focused earlier only on the circulating vortex currents that produce the quantized flux tubes of $b_{s\mu}$. However, any vortex density or current produced by a deviation from (\ref{GSconf}) implies the presence of vortex excitations. Vortices are fully gapped by the assumption that the ground state is not a Mott insulator (and they are gapped even in Mott insulators by the Anderson-Higgs mechanism according to our simplified model that neglects the physical photon fluctuations). Therefore, the ``gauge'' we used in deriving (\ref{Lag2}) is exactly pertinent to the ground-state manifold and sharply separated from other ``gauge'' choices by an energy gap. The only important deviations from (\ref{GSconf}) are instantons, but they are bound to help rather than hinder the averaging of $e^{i(\theta_\uparrow - \theta_\downarrow)}$ to zero.

We conclude that the ground-state degeneracy of a TR-invariant SU(2) Laughlin state on a torus is always $m^2$. The specific spin-non-conserving form of the external gauge field $\mathcal{A}_{2\mu}$ only determines the global symmetry of the ground state, and the details of spatial fluctuations in the Lagrangian density that are relevant for the excitation spectrum. We are also assuming that $\mathcal{A}_{2\mu}$ shapes the energy landscape in a way that gives rise to the specific edge states discussed in the previous section. Quasiparticles are fractionalized, but their conserved spin-like quantum number is the spin-projection perpendicular to momentum according to the right-hand rule.

A more general derivation of the topological ground-state degeneracy relies on the ``topological symmetry'' transformations of the effective Lagrangian in quantum Hall states. This procedure reveals the degeneracy on Riemann surfaces of arbitrary genus $g$. We will not discuss it here because it essentially follows the calculation of the Ref.\cite{Wen1990a} and arrives at the analogous conclusion that the degeneracy of all TR-invariant Laughlin states is $m^{2g}$. One would have to start from the Lagrangian (\ref{TopLGp}), which is dual to (\ref{TopLG}) and describes physical particles directly. Then, one would derive an effective theory of quantum Hall states from this Lagrangian (by fixing the particle densities in all spin channels in the topological term). This effective theory is essentially the same as the Landau-Ginzburg theory with a Chern-Simons coupling from the Ref.\cite{Wen1990a}. The only difference is in the choice of a (singular) gauge and an implicit constraint for the flux quantization of the CS gauge field, but this does not affect any symmetries. Therefore, the same symmetry analysis can then be performed to demonstrate topological orders on Riemann surfaces. The ground-state degeneracy is shaped by the low-energy dynamics of particle or vortex fields, and not directly by the specific spin conserving or non-conserving form of the external SU(2) gauge flux.

\subsection{Duality}\label{secDual}

This section discusses the duality relationship between the vortex theory (\ref{TopLG}) and the theory that directly describes the dynamics of physical particles (\ref{TopLGp}). We will assume that the particles are bosons and formally derive duality using the XY and CS descriptions of dynamics \cite{Dasgupta1981, Fisher1989, Sachdev1990, Sachdev2004}, which are not always available as we have seen in the previous sections. This and the symmetries will guide us in making a conjecture about an approximate duality that holds generally at the level of the spinor Lagrangians (\ref{TopLG}) and (\ref{TopLGp}) for any type of particles.

When the phase fluctuations of the spinor $\psi$ components dominate the dynamics, the Landau-Ginzburg part of (\ref{TopLG}) can be viewed as the continuum limit of the gauged quantum XY model on the square lattice. Its action can be formulated in the discretized imaginary time:
\begin{eqnarray}\label{Sxy}
S_{\vv\textrm{XY}} &=& \sum_s \biggl\lbrack -\frac{1}{g_s} \sum_{{\bf r},\mu} \cos \left( \Delta_\mu \theta_{s} - \beta_{s\mu} \right) \\
&& ~~~~~ + \frac{1}{2e_s^2} \sum_\square \left( \epsilon^{\mu\nu\lambda} \Delta_\nu \beta_{s\lambda} \right)^2 \biggr\rbrack \ . \nonumber
\end{eqnarray}
The angles $\theta_s$ are the phases of the complex spinor components $\psi_s$ in the representation that diagonalizes the gauge field matrix $\mathcal{B}_\mu = \textrm{diag}(\beta_{-S,\mu},\dots,\beta_{S,\mu})$. In this microscopic model, however, $\theta_s$ live on the (2+1)D space-time lattice sites ${\bf r}$, and $\beta_{s\mu}$ live on the lattice bonds, where $\mu \in \lbrace 0,x,y \rbrace$ labels the three bond orientations and $\beta_{s,\mu} \equiv \beta_{s,\langle {\bf r},{\bf r}+\hat{\boldsymbol{\mu}} \rangle} = -\beta_{s,-\mu} \equiv -\beta_{s,\langle {\bf r}+\hat{\boldsymbol{\mu}},{\bf r} \rangle}$. The discrete lattice derivative is $\Delta_\mu \theta_{s,{\bf r}} = \theta_{s,{\bf r} + \hat{\boldsymbol{\mu}}} - \theta_{s,{\bf r}}$ and the discrete curl $\epsilon^{\mu\nu\lambda} \Delta_\nu \beta_{s\lambda}$ is given by the oriented sum of $\beta_{s,\langle {\bf rr}' \rangle}$ on the bonds $\langle {\bf rr}' \rangle$ of a square plaquette.

The sum of all gauge fields $\beta_{s\mu}$ is non-compact as implied by (\ref{TopLG}), because the physical particles must be able to exist in fully delocalized states such as superfluids \cite{Sachdev2004}. Certain linear combinations of $\beta_{s\mu}$ are allowed to be compact since spin need not be conserved, but we will assume for simplicity that every $\beta_{s\mu}$ is individually non-compact in (\ref{Sxy}). We will also neglect the electromagnetic field fluctuations to which electrons are coupled; otherwise, the sum of $\beta_{s\mu}$ associated with charge fluctuations would be gapped out by the Anderson-Higgs mechanism. The formal duality transformation maps (\ref{Sxy}) into the XY model of particles \cite{Dasgupta1981} whose continuum limit is given by the Landau-Ginzburg part of (\ref{TopLGp}).

The gauged XY model (\ref{Sxy}) represents the dynamics of vortices, while its dual XY model represents the dynamics of particles. According to duality, the transition from the superfluid state of particles to a Mott insulator can be viewed as the condensation of quantized vortices \cite{Fisher1989, Fisher1989a}. Particles are mobile and coherent in the superfluid state, while vortices are gapped and localized into a vortex lattice if their density is finite. A Mott insulator is a dual reflection of the superfluid where particles and vortices exchange their behavior.

We can qualitatively view incompressible quantum liquids of bosons, including quantum Hall states, as ``arrested'' Mott transitions in which both particles and vortices are abundant and mobile, yet uncondensed and controlled by the cyclotron scales. Duality allows simultaneous mobility of both particles $j_{\pp\mu}$ and vortices $j_{\vv\mu}$ only if vortices are ``attached'' to particles and both have incompressible densities. This prevents a vortex from passing through a particle, and thereby prohibits strong phase fluctuations that would localize the particles (and vice versa by duality). We must imagine that the superfluid-like correlations are not locally lost in an incompressible quantum liquid, but particles have nevertheless begun localizing with respect to other nearby particles. In the presence of external fluxes, the wavefunction of a localized particle must acquire vorticity, and this constitutes the microscopic mechanism for ``flux attachment''. The quantum Hall regime is achieved when every particle becomes a microscopic ``cyclotron'' vortex. The motion of particles is then governed by Lorentz and Magnus forces, whose effect is captured by the topological (Chern-Simons) terms in the Lagrangian.

An example of conditions that could produce an incompressible quantum liquid is found beyond a first-order transition from a vortex lattice superconductor to a correlated insulator. Vortex lattice melting is generally first-order \cite{herbut95}, but this implies that the superconducting system enters an insulating state while its order parameter magnitude is still finite. In other words, the superconductor's phase coherence survives at short length-scales given by the separation between vortices in their liquid state. This is enough to define isolated vortices in the first place and give rise to Magnus forces on them, provided that vortices are small and compact as is the case when the number of particles per vortex is small.

We will now include the topological term $\mathcal{L}_{\textrm{t}}$ of (\ref{TopLG}) in the duality analysis, in order to discover its form in the direct theory of physical particles. Unfortunately, the form of $\mathcal{L}_{\textrm{t}}$ is precisely defined only in the continuum limit. This can be appreciated by considering an equivalent CS coupling $\epsilon^{\mu\nu\lambda} b_\mu \partial_\nu b_\lambda$, which couples the particle current $j_{\pp\mu} \propto \epsilon^{\mu\nu\lambda} \partial_\nu b_\lambda$ to the vortex current $j_{\vv\mu} \sim b_\mu$. If a CS term were to be defined microscopically, then the particle current would have to live on the bonds of the direct (physical) lattice, while the vortex currents would live on the links of the dual lattice. These two lattices do not coincide, so the lattice CS coupling $j_{\pp\mu} j_{\vv\mu}$ is ambiguous. Instead of trying to deal with this issue, we will carry out the duality mapping in the continuum limit. Hence, we will give up keeping track of the detailed lattice effects on the particle and vortex dynamics, but they are anyway known from the exact duality mapping in the absence of topological terms. Our interest is mainly to discover the continuum limit of the topological term in the particle theory, and we will obtain the adaptation of the result from the Ref.\cite{Wen1990a} to the specific features of the present formalism.

Our starting point is the effective theory given by (\ref{XY}) and (\ref{EffTh}). We require the presence of the U(1)$^{2S+1}$ symmetry that decouples different spin channels in the diagonal representation of $\Phi_0$. In that case, all XY fields $\theta_s$ fluctuate independently, and we need to only focus on the dynamics of one of them at a time. A single copy of the CS term looks like (\ref{CSTh}) with the topologically fixed coupling $\eta=\frac{1}{4}$, where the normalization $\partial_\mu\theta_s = b_{s\mu} = m_s c_{s\mu}$ is used for the CS gauge field. Therefore, the effective theory of vortices in a single spin channel $s \in \lbrace -S,\dots,S \rbrace$ takes the form:
\begin{eqnarray}\label{L1}
\mathcal{L}_\vv^s \!\! &=& \!\! \frac{K\rho_s}{2}\bigl(\partial_\mu\theta_s-\beta_{s\mu}\bigr)^2
   \! + \! \frac{1}{8\pi^2 q_s^2}\bigl(\epsilon^{\mu\nu\lambda}\partial_\nu\beta_{s\lambda}-\Theta_{s0}\delta_{\mu,0}\bigr)^2
    \nonumber \\
&& +\frac{i}{4\pi m_s}\epsilon^{\mu\nu\lambda}b_{s\mu}\partial_\nu b_{s\lambda}
   -\frac{i}{2\pi m_s}\epsilon^{\mu\nu\lambda}A_{s\mu} \partial_\nu b_{s\lambda} \ .
\end{eqnarray}
Even though $b_{s\mu}$ reflects all local fluctuations of $\theta_s$, only its singular (gauge-invariant) part contributes to the CS coupling. It is therefore convenient to carry out a singular gauge transformation \cite{Franz2000, Vafek2001} and separate two kinds of fluctuations $\theta_s = \theta_s'+\theta_s''$, where $\theta_s'$ are smooth and $\theta_s''$ are singular phase configurations:
\begin{eqnarray}
&& ~~~~~~~ \epsilon^{\mu\nu\lambda}\partial_\nu^{\phantom{i}} (\partial_\lambda^{\phantom{i}} \theta_s') = 0 \ , \\
&& b_{s\mu}^{\phantom{i}} = \partial_\mu^{\phantom{i}} \theta_s''
   \quad , \quad \epsilon^{\mu\nu\lambda} \partial_\nu^{\phantom{i}} b_{s\lambda}^{\phantom{i}} \neq 0 \ . \nonumber
\end{eqnarray}
We have redefined the ``gauge field'' $b_{s\mu}$ by fixing its gauge. In order for vortex spinors to be single-valued, $\exp(i\theta_s)$ must be single-valued and hence the space-time configurations of $b_{s\mu}$ can allow flux to be concentrated only in quantized loops as if the dynamics of $b_{s\mu}$ were extremely compact. Monopole configurations of $b_{s\mu}$ would have been allowed if only they didn't have to be compensated by $\beta_{s\mu}$, which is non-compact. We can now promote $b_{s\mu}$ to arbitrary gauge-field configurations and impose the constraint of ``compactness'' via the Poisson summation formula, by introducing an integer-valued fluctuating current field $\mathcal{J}_{s\mu}$ in the path-integral $z$:
\begin{eqnarray}\label{Poisson}
z \!\! &=& \!\! \int \dd\theta_s' \dd b_{s\mu}^{\phantom{a}} \dd\beta_{s\mu}^{\phantom{a}} \, e^{-S} \prod_{s\mu} \sum_{n_{s\mu}=-\infty}^\infty \!\!\!\!
  \delta \left( \epsilon^{\mu\nu\lambda} \partial_\nu b_{s\lambda} - 2\pi n_{s\mu} \right) \nonumber \\
\!\! &\propto& \!\! \int \dd\theta_s' \dd b_{s\mu}^{\phantom{a}} \dd\beta_{s\mu}^{\phantom{a}} \sum_{\mathcal{J}_{s\mu}=-\infty}^\infty \!\!\!\!
  e^{-S + i \mathcal{J}_{s\mu} \epsilon^{\mu\nu\lambda} \partial_\nu b_{s\lambda}} \ .
\end{eqnarray}
An integer $\mathcal{J}_{s\mu}$ is associated with every lattice bond in the microscopic lattice formulation of the path-integral, but here we must treat $\mathcal{J}_{s\mu}$ as an integer-valued vector function of continuous space-time coordinates. The transformed Lagrangian (\ref{L1}) reads:
\begin{equation}
\mathcal{L}_\vv^s = \frac{K\rho_s}{2}\bigl(\partial_\mu^{\phantom{i}}\theta_s' + b_{s\mu}^{\phantom{i}}
  - \beta_{s\mu}^{\phantom{i}} \bigr)^2 - i \mathcal{J}_{s\mu} \epsilon^{\mu\nu\lambda} \partial_\nu b_{s\lambda} + \cdots
\end{equation}
Let us decouple the two XY-model-like terms in (\ref{L1}) by applying the Hubbard-Stratonovich transformation in the path-integral:
\begin{equation}
\exp\left(\frac{a}{2}x^2\right) \propto \int \dd y \, \exp\left(-\frac{y^2}{2a} + xy\right)
\end{equation}
We need to introduce two real-valued Hubbard-Stratonovich current fields that appear in the transformed Lagrangian, $j_{s\mu}$ for particles and $\widetilde{j}_{s\mu}$ for vortices:
\begin{eqnarray}\label{L2}
&& \!\!\!\!\!\!\!\!\!\!\!\!\! \mathcal{L}^s = \frac{1}{2K\rho_s} \widetilde{j}_{s\mu}^2
   -i\,\widetilde{j}_{s\mu}^{\phantom{i}} \bigl(\partial_\mu^{\phantom{i}}\theta_s' + b_{s\mu}^{\phantom{i}} -
   \beta_{s\mu}^{\phantom{i}} \bigr) +\frac{(2\pi q_s)^2}{2} j_{s\mu}^2 \nonumber \\
&& -i j_{s\mu}^{\phantom{i}} \bigl(\epsilon^{\mu\nu\lambda}\partial_\nu^{\phantom{i}}\beta_{s\lambda}^{\phantom{i}}
     -\Theta_{s0}^{\phantom{i}}\delta_{\mu,0}^{\phantom{i}}\bigr) \nonumber
   -i\mathcal{J}_{s\mu} \epsilon^{\mu\nu\lambda} \partial_\nu b_{s\lambda}  \\
&& +\frac{i}{4\pi m_s}\epsilon^{\mu\nu\lambda}b_{s\mu}\partial_\nu b_{s\lambda}
   -\frac{i}{2\pi m_s}\epsilon^{\mu\nu\lambda}A_{s\mu} \partial_\nu b_{s\lambda} \ .
\end{eqnarray}
Then, integrating out $\theta_s'$ and $\beta_{s\mu}$ multiplies the measure of the path-integral by the Dirac delta functions that implement the following constraints:
\begin{equation}\label{Constr}
\partial_\mu \widetilde{j}_{s\mu} = 0 \quad , \quad
\epsilon^{\mu\nu\lambda}\partial_\nu j_{s\lambda} = \widetilde{j}_{s\mu} \ .
\end{equation}
Strictly speaking, only the transverse (flux-changing) modes of $\beta_{s\mu}$ are physical and fluctuating, but integrating them out imposes the constraint $\widetilde{j}_{s\mu}^{\perp} = \epsilon^{\mu\nu\lambda}\partial_\nu j_{s\lambda}$ only on the transverse part $\widetilde{j}_{s\mu}^{\perp}$ of the full vortex current $\widetilde{j}_{s\mu}^{\phantom{i}} = \widetilde{j}_{s\mu}^{\perp} + \widetilde{j}_{s\mu}^{\parallel}$. The transverse and longitudinal currents are defined by restrictions $\partial_\mu^{\phantom{i}} \widetilde{j}_{s\mu}^{\perp} = 0$ and $\epsilon^{\mu\nu\lambda} \partial_\nu^{\phantom{i}} \widetilde{j}_{s\lambda}^{\parallel} = 0$ respectively, so that integrating out $\theta_s'$ quenches the longitudinal current by imposing $\partial_\mu \widetilde{j}_{s\mu}^{\parallel} \to 0$. Note that $\theta_s'$ does not include the singular configurations that produce a non-zero flux of $b_{s\mu} = \partial_\mu \theta_s$, but this does not affect its coupling to the longitudinal current. Once the longitudinal current is suppressed, we can identify the total vortex current with its transverse part and solve the resulting constraints (\ref{Constr}) by introducing a scalar and a gauge field:
\begin{equation}
\widetilde{j}_{s\mu}^{\phantom{i}} = -\frac{1}{2\pi m_s}
  \epsilon^{\mu\nu\lambda} \partial_\nu^{\phantom{i}} \alpha_{s\lambda}^{\phantom{i}} \quad,\quad
j_{s\mu} = \frac{1}{2\pi m_s} \bigl( \partial_\mu^{\phantom{i}} \varphi_s' - \alpha_{s\mu}^{\phantom{i}} \bigr) \ .
\end{equation}
The scalar field $\varphi_s'$ is not allowed to have singular configurations because the curl of $j_{s\mu}$ must be generated entirely by $\alpha_{s\mu}$. Substituting the above expressions in (\ref{L2}) gives:
\begin{eqnarray}\label{L3}
&& \!\!\!\!\!\!\!\!\! \mathcal{L}_{\textrm{p}}^s =
       \frac{1}{2e_s^2} \bigl( \epsilon^{\mu\nu\lambda} \partial_\nu \alpha_{s\lambda} \bigr)^2
   + \frac{\widetilde{K}_s n_s}{2} \bigl(\partial_\mu^{\phantom{i}} \varphi_s' - \alpha_{s\mu}^{\phantom{i}}\bigr)^2 \nonumber \\
&& +i n_s \bigl( \partial_0 \varphi_s' - \alpha_{s0}^{\phantom{i}} \bigr)
   -i\mathcal{J}_{s\mu} \epsilon^{\mu\nu\lambda} \partial_\nu b_{s\lambda} \\
&& +\frac{i\epsilon^{\mu\nu\lambda}}{4\pi m_s} \Bigl\lbrack b_{s\mu}\partial_\nu b_{s\lambda}
   -2 A_{s\mu} \partial_\nu b_{s\lambda}
   +2 b_{s\mu}^{\phantom{i}}\partial_\nu^{\phantom{i}}\alpha_{s\lambda}^{\phantom{i}}
   \Bigr\rbrack \ , \nonumber
\end{eqnarray}
where we have introduced symbols
\begin{equation}\label{DualSymb}
n_s = \frac{\Theta_{s0}}{2\pi m_s} \quad , \quad \widetilde{K}_s = \frac{q_s^2}{m_s^2 n_s^{\phantom{2}}} \quad , \quad
  e_s^2 = (2\pi m_s)^2 K \rho_s ~
\end{equation}
to simplify notation (note that $n_s>0$ restricts the allowed values of $\Theta_{s0}$ and $m_s$). Now we are ready to integrate out $b_{s\mu}$, the last field that originated in the vortex Lagrangian (\ref{L1}). This can be done by completing the square:
\begin{eqnarray}
&& \!\!\!\!\!\!\! \epsilon^{\mu\nu\lambda} b_{s\mu} \partial_{\nu} b_{s\lambda}
   - 2 \epsilon^{\mu\nu\lambda} b_{s\mu}\partial_{\nu} \bar{a}_{s\lambda} \\
&& \!\!\!\!\!\! = \epsilon^{\mu\nu\lambda} \left( b_{s\mu} - \bar{a}_{s\mu} \right) \partial_{\nu}
                              \left( b_{s\lambda} - \bar{a}_{s\lambda} \right)
   - \epsilon^{\mu\nu\lambda} \bar{a}_{s\mu} \partial_{\nu} \bar{a}_{s\lambda} \nonumber
\end{eqnarray}
where
\begin{equation}
\bar{a}_{s\mu} = a_{s\mu} + 2\pi m_s \mathcal{J}_{s\mu} \quad,\quad a_{s\mu} = A_{s\mu} - \alpha_{s\mu}
\end{equation}
Integrating out $\bar{b}_{s\mu} \equiv b_{s\mu} - \bar{a}_{s\mu}$ merely multiplies the path-integral by a constant and leaves behind the effective Lagrangian of physical particles:
\begin{eqnarray}\label{L4}
&& \!\!\!\!\!\!\!\!\!\!\!\!\!\! \mathcal{L}_{\textrm{p}}^s =
       \frac{1}{2e_s^2} \bigl( \epsilon^{\mu\nu\lambda} \partial_\nu \alpha_{s\lambda} \bigr)^2
   + \frac{\widetilde{K}_s n_s}{2} \bigl(\partial_\mu^{\phantom{i}} \varphi_s' - \alpha_{s\mu}^{\phantom{i}}\bigr)^2 \nonumber \\
&& + i n_s \bigl( \partial_0 \varphi_s' - \alpha_{s0}^{\phantom{i}} \bigr)
   - \frac{i}{4\pi m_s} \epsilon^{\mu\nu\lambda} \bar{a}_{s\mu} \partial_{\nu} \bar{a}_{s\lambda} \ .
\end{eqnarray}
Expanding the CS coupling for $\bar{a}$ yields:
\begin{eqnarray}\label{MakeCompact}
&& \!\!\!\!\!\!\!\!\!\!
   -\frac{i}{4\pi m_s} \epsilon^{\mu\nu\lambda} \bar{a}_{s\mu} \partial_{\nu} \bar{a}_{s\lambda} =
   -\frac{i}{4\pi m_s} \epsilon^{\mu\nu\lambda} a_{s\mu} \partial_{\nu} a_{s\lambda} \\
&& ~~~  -i\mathcal{J}_{s\mu} \epsilon^{\mu\nu\lambda} \partial_\nu a_{s\lambda}
   -i\pi m_s \epsilon^{\mu\nu\lambda} \mathcal{J}_{s\mu} \partial_\nu \mathcal{J}_{s\lambda} \ . \nonumber
\end{eqnarray}
Since $\mathcal{J}_{s\mu} \in \mathbb{Z}$, the last term in this expression is an integer multiple of $2\pi i$ for bosonic particles that require $m_s$ to be an even integer. Then, this term does not affect the path-integral and can be ignored (because $e^{2\pi i n}=1$). In contrast, fermionic particles have an odd $m_s$, so for them the last term introduces interference of the Z$_2$ kind in the path-integral. We will not attempt to study this interference here; that can be done only in the lattice limit, and may prove to be a formidable challenge since duality mappings for fermions are not known. Instead, we will simply focus on bosons and drop the last term in (\ref{MakeCompact}). The summation over $\mathcal{J}_{s\mu}$ has no quadratic weight any more and thus forces the gauge field $a_{s\mu}$ to collect flux only into quantized vortex loops according to (\ref{Poisson}). From this point on, we can view $a_{s\mu}$ as the gradient of a singular phase. Let us define an XY field $\varphi_s^{\phantom{i}}$ via $\partial_\mu^{\phantom{i}} \varphi_s^{\phantom{i}} = \partial_\mu \varphi_s' + a_{s\mu}^{\phantom{i}}$ which is allowed to contain singularities, and fully eliminate $\alpha_{s\mu}$ from (\ref{L4}) in favor of $a_{s\mu}$:
\begin{eqnarray}\label{dualXY}
&& \!\!\!\!\!\!\!\! \mathcal{L}_{\textrm{p}}^s =
     \frac{1}{2e_s^2} \left( \epsilon^{\mu\nu\lambda} \partial_\nu^{\phantom{i}} a_{s\lambda}^{\phantom{i}} - \Phi_s^\mu \right)^2
   + \frac{\widetilde{K}_s n_s}{2} \left( \partial_\mu \varphi_s - A_{s\mu} \right)^2 \nonumber \\
&& + i n_s \left( \partial_0 \varphi_s - A_{s0} \right)
   - \frac{i}{4\pi m_s} \epsilon^{\mu\nu\lambda} a_{s\mu} \partial_\nu a_{s\lambda} \ .
\end{eqnarray}
$\Phi_s^\mu = \epsilon^{\mu\nu\lambda} \partial_\nu^{\phantom{i}} A_{s\lambda}^{\phantom{i}}$ is the externally imposed static flux in the chosen spin channel. If we pick the gauge in which $a_{s\mu} = \partial_\mu \varphi_s$, this Lagrangian becomes entirely analogous to the starting combination (\ref{L1}) of a continuum ``XY'' model and a CS term. Note that the first term ($1/2e_s^2$) is redundant because it is obtained merely by taking the curl of the second term. Particles are here coupled only to a non-fluctuating external gauge field $A_{s\mu}$.

The continuum limit (\ref{dualXY}) by itself does not define a charge quantum, since the field $\varphi_s$ can be arbitrarily renormalized. A reference for the normalization of $\varphi_s$ is provided only via the topological CS term, which specifies the statistics of (vortex) excitations. However, the microscopic lattice rendition of (\ref{dualXY}) is compact in such a manner that the charge quantum is the fundamental electron's charge. The natural (compact) XY fields are the angles $\varphi_s$ that we wrote, judging by how they couple to the external gauge field $A_{s\mu}$. If the particle density were an integer number per lattice site in each spin channel, we would have $n_s=0$ and the lattice XY action of particles without the topological term would be:
\begin{equation}
S_{\pp\textrm{XY}} = - \sum_s g_s \sum_{{\bf r},\mu} \cos \left( \Delta_\mu \varphi_{s} - A_{s\mu} \right) \ .
\end{equation}

We are now ready to construct the continuum-limit Lagrangian of the physical particle spinor field $\xi$, whose limit of small density fluctuations is given by (\ref{dualXY}). We will immediately write it for particles of arbitrary spin $S$. Defining the particle spinor $\xi = (\xi_{-S},\dots,\xi_S)$, where $\xi_s = \sqrt{n_s} \exp(i \varphi_s)$, we obtain:
\begin{eqnarray}\label{TopLGp}
&& \!\!\!\!\!\!\! \mathcal{L}_\textrm{\pp LG} = \xi^\dagger (\partial_0 - i\mathcal{A}_0) \xi
   + \Bigl\lbrack(\partial_{i}-i\mathcal{A}_{i})\xi\Bigr\rbrack^{\dagger} \frac{\widetilde{K}}{2}
     \Bigl\lbrack(\partial_{i}-i\mathcal{A}_{i})\xi\Bigr\rbrack \nonumber \\
&& ~~~~~  -\xi^{\dagger}\widetilde{t}\,\xi+\widetilde{u}|\xi|^4+\widetilde{v}|\xi^{\dagger}\gamma^{a}\xi|^2 \\[0.1in]
&& \!\!\!\!\!\!\! \mathcal{L}_{\textrm{\pp t}} = \frac{i}{8}\, \xi^{\dagger}\epsilon^{\mu\nu\lambda}\Bigl\lbrack
   \partial_\mu^{\phantom{i}} \Bigl\lbrace\partial_\nu^{\phantom{i}},\Theta_{0}^{-1}\Bigr\rbrace \partial_\lambda^{\phantom{i}}
  +\Bigl\lbrace\partial_\mu^{\phantom{i}}\partial_\nu^{\phantom{i}}\partial_\lambda^{\phantom{i}},\Theta_{0}^{-1}
     \Bigr\rbrace\Bigr\rbrack\xi \ . \nonumber
\end{eqnarray}
This Lagrangian exhibits the non-relativistic dynamics that we found in the duality mapping, but relativistic dynamics is also possible in special circumstances (integer particle filling of lattice sites). The chemical potential matrix $\widetilde{t}$ determines the average particle densities $n_s$ in all spin channels, while the kinetic energy of particles in different internal states is determined by the matrix $\widetilde{K} = \textrm{diag} (\widetilde{K}_{-S},\dots,\widetilde{K}_S)$ that equals a constant multiple of the unit-matrix in normal circumstances. The embedded factor of $\Theta_0^{-1}$ in the topological term assures its proper transformation under TR. The dynamical particle charge and spin currents take the usual non-relativistic form in terms of the particle spinors:
\begin{eqnarray}\label{pCurrents3}
j_0 &=& \xi^\dagger \xi \\
j_{i} &=& -\frac{i}{2}\Bigl\lbrack\xi^{\dagger}(\partial_{i}\xi)
    -(\partial_{i}\xi^{\dagger})\xi\Bigr\rbrack-\xi^\dagger\mathcal{A}_{i}\xi \nonumber \\
J_0^a &=& \xi^\dagger \gamma^a \xi \nonumber \\
J_{i}^{a} &=& -\frac{i}{2}\Bigl\lbrack\xi^{\dagger}\gamma^{a}(\partial_{i}\xi)
    -(\partial_{i}\xi^{\dagger})\gamma^{a}\xi\Bigr\rbrack
    -\frac{1}{2}\xi^\dagger\lbrace\mathcal{A}_{i},\gamma^a\rbrace\xi \nonumber \ .
\end{eqnarray}

It should be pointed out that the topological term $\mathcal{L}_{\textrm{\pp t}}$ is gauge-invariant, but written in the gauge fixed by the non-dynamical spin-orbit coupling embedded in $\mathcal{A}_\mu$. Strictly speaking, the derivatives $\partial_\mu \to \partial_\mu - i\mathcal{V}_\mu$ in $\mathcal{L}_{\textrm{\pp t}}$ are covariant, where $\mathcal{V}_\mu$ is the static background gauge field dual to $\mathcal{A}_\mu$ and minimally coupled to the vortex spinor $\psi$ in (\ref{TopLG}). We never wrote $\mathcal{V}_\mu$ before because it equals zero in the chosen natural gauge, but it is formally added to the purely transverse fluctuations of the dynamical gauge field $\mathcal{B}_\mu$ in order to carry the gauge transformations of $\mathcal{B}_\mu$ when the spinor $\psi$ is transformed. While present for general SU($N$) symmetry groups, the gauge-invariance of $\mathcal{L}_{\textrm{\pp t}}$ can be explicitly revealed in the U(1) duality by merely shifting $\beta_{s\mu} \to \beta_{s\mu} + v_{s\mu}$ by a constant background gauge field $v_{s\mu}$, where $\mathcal{V}_\mu = \textrm{diag}(v_{-S,\mu},\dots,v_{S,\mu})$.

We conjecture that the topological field theory (\ref{TopLGp}) of physical particles is generally dual to the corresponding theory of vortices (\ref{TopLG}), despite the fact that we could derive it only in the case of bosonic particles whose dynamics respects the global spin U(1)$^{2S+1}$ symmetry at low energies. The duality relationship between (\ref{TopLG}) and (\ref{TopLGp}) is established in the following qualitative sense. The spin $S$ particles are directly represented by the spinor field $\xi$ in (\ref{TopLGp}), which is minimally coupled to the external U(1)$\times$SU(2) gauge field $\mathcal{A}_\mu$ that embodies magnetic fields and spin-orbit interactions. The topological term regulates the quantum statistics of vortices, which are line-like topological defects in the (2+1)D space-time configurations of $\xi$. The dual theory (\ref{TopLG}) directly describes the dynamics of vortices, represented there by the spinor field $\psi$. Each spin channel of particles corresponds to a spin channel of vortices. By duality, vortices must be minimally coupled to a gauge field $\mathcal{B}_\mu$ that represents the physical particles via its fluxes. The topological term of (\ref{TopLG}) regulates the statistics of physical particles.

There are several interesting features worth pointing out. First of all, the lattice version of the theory (\ref{TopLGp}) applied to bosonic particles can describe conventional superconducting and Mott-insulating phases. The written continuum limit represents by $n_s^{\phantom{\dagger}} = \xi_s^\dagger \xi_s^{\phantom{\dagger}}$ the excess density of particles in the spin state $s$ relative to an integer number per lattice site \cite{tesanovic04, balents05}. Assuming the U(1)$^{2S+1}$ symmetry at low energies, the dual ``flux'' $\Theta_\mu = \Theta_0 \delta_{\mu,0}$ depends on these densities according to (\ref{DualSymb}) and thus acts as a source of Magnus forces on vortices in the dual theory (\ref{TopLG}). By duality, $\Theta_0$ must exhibit the same kind of symmetry transformations as the SU(2) ``magnetic'' flux $\Phi_0$. A non-zero $\Theta_0$ in the presence of TR-symmetry must have a form such as $\Theta_0 \propto \gamma^z$, which implies a reduction of the full spin SU(2) symmetry at least down to U(1) in the theory (\ref{TopLGp}). Condensates of $\xi$, or insulating phases with independent fluctuations of individual spinor $\xi$ components in some particular representation, are thus consistent with $\Theta_0 \neq 0$. On the other hand, SU(2) symmetric insulators with fixed $\xi^\dagger \xi$ are consistent only with $\Theta_0=0$. This includes the special case of Mott insulators with an integer number of particles in each spin-projection state. It also includes Mott insulators with arbitrary density whose SU(2) symmetry is restored by fluctuations (in which case our detailed duality derivation is not applicable).

Magnus forces on vortices are proportional to particle densities rather than the dual ``flux'' $\Theta_0$ as naively portrayed by (\ref{TopLG}) and (\ref{L1}). This can be revealed by renormalizing the gauge fields $\beta_{s\mu} \to m_s \bar{\beta}_{s\mu}$ that capture ordinary matter fluctuations in the dual theory. Using (\ref{DualSymb}), the first two terms of (\ref{L1}) become:
\begin{equation}
\frac{K\rho_s}{2}\bigl(\partial_\mu\theta_s-m_s\bar{\beta}_{s\mu}\bigr)^2
   \! + \! \frac{m_s^2}{8\pi^2 q_s^2}\bigl(\epsilon^{\mu\nu\lambda}\partial_\nu\bar{\beta}_{s\lambda}
   - 2\pi n_s\delta_{\mu,0}\bigr)^2 \ . \nonumber
\end{equation}
We see that the numbers $m_s$ are vortex charges. In conventional phases $m_s = \pm 1$ and there is no fractionalization. However, the interplay between various coupling constants in the Landau-Ginzburg theory (\ref{TopLGp}) can in principle set such densities and dispersions of particle modes that the charges $m_s$ become non-trivial via (\ref{DualSymb}). This is not a complete picture of how fractionalization arises dynamically, but it is more general than a rigid kinematic CS description of fractionalized states. We will not make any further attempts to study the dynamical origin of fractionalized states in this paper.

We found in the section \ref{secFract} that vortex densities $\rho_s$ are locked into the values of $m_s$ in quantum Hall states, and similar locking between the particle densities $n_s$ and $m_s$ can be extracted using the duality analysis from this section. Namely, $m_s^{-1}$ are filling factors, so that $n_s = \Phi_{0s}/2\pi m_s$ and $\Theta_0 = \Phi_0$  according to (\ref{DualSymb}) in quantum Hall liquids. Therefore, $\Theta_0$ is rigidly determined at least in all Laughlin quantum Hall liquids, irrespective of which fractional charges are selected by dynamics. The appearance of $\Theta_0^{-1} \to \Phi_0^{-1}$ in the topological term of (\ref{TopLGp}) is not alarming because this relationship between $\Theta_0$ and $\Phi_0$ does not hold in conventional states of matter that always become stable in the $\Phi_0 \to 0$ limit. In such conventional states, $\Theta_0$ is determined by the particle density as discussed above.

The theory (\ref{TopLGp}) might apply to a much more general context than the one we derived it in. For example, we could formulate it for fermionic particles even though duality mapping is then not known. Furthermore, we will see in the sections \ref{secHier} and \ref{secNonAbel} what generalizations are necessary for hierarchical Abelian and non-Abelian quantum Hall states.

\subsection{The stability of topological orders}\label{secStab}

We have analyzed simple topological orders of spinful particles using an SU(2) gauge theory. In several occasions we first considered Laughlin states with conserved spin, and then naively generalized the analysis to the Rashba spin-orbit coupling that does not conserve spin. However, real materials do not feature the SU(2) gauge structure and contain perturbations that violate all symmetries of the ideal spin-orbit couplings. We would like to explain here using duality why none of these violations of spin-conservation or the SU(2) symmetry jeopardize the conclusions so far, or why the spin-related topological orders are stable.

We found in the section \ref{secFract} that generic U(1)$\times$SU(2) Laughlin incompressible quantum liquids feature quasiparticle excitations with fractional quantum numbers given by (\ref{QuantNum}), regardless of whether the spin-orbit coupling conserves spin or not. What matters in the derivation of that result are only the conditions (A) and (B) stated at the beginning of the section \ref{secFract}. Specifically, the vortex densities $\rho_s$ in all spin channels must be effectively frozen. Note that the conditions (A) and (B) make no reference to the possible conservation of the $S^z$ spin projection. If $S^z$ fluctuates, then the winding numbers $n_s$ from (\ref{FracExc}) and (\ref{QuantNum}) fluctuate accordingly in a state where all $\rho_s$ are fixed. The crucial point is that the dual-spin of vortices $\psi$ in (\ref{TopLG}) is conserved, even when the spin of particles $\xi$ in (\ref{TopLGp}) is not conserved. This is what allows the vortex densities $\rho_s$ in various dual-spin channels to have quantized incompressible values.

The conservation of dual-spin can be made explicit by emphasizing its nature in the context of the particle Lagrangian (\ref{TopLGp}). A vortex with a definite dual-spin projection is a topological defect of a particular particle spinor component $\xi_s = |\xi_s| \exp(i\varphi_s)$ in which the phase $\varphi_s$ winds by $2\pi\times$integer about the singularity. The conservation of dual-spin corresponds to the conservation of vorticity in the particle theory (\ref{TopLGp}). If the particle Lagrangian is allowed to have only local terms, then only vortex-antivortex pairs and other topologically neutral deformations of the fields in any given spin channel can cost finite action (energy). Furthermore, no local operator written in terms of $\xi_s^{\phantom{\dagger}}$ and $\xi_s^\dagger$ can create or annihilate a single vortex, because it would have to qualitatively alter the field configuration arbitrarily far away from the vortex core.

A local theory can allow the non-conservation of vorticity only if it is a gauge theory with deconfined monopoles (typically a compact gauge theory). A gauge field coupled to the matter field is necessary in order to compensate the macroscopic energy cost of a vortex configuration in the matter field alone. Then, creating or annihilating a vortex at a particular point in space and time requires compensation by a monopole gauge-field configuration. In our case there are $2S+1$ flavors of vortices, and we could introduce that many compact gauge fields $\mathcal{X}_{s\mu}$ in the model (\ref{TopLGp}) to spoil the conservation of vorticity. Alternatively, we could make the gauge fields $\mathcal{X}_{s\mu}$ non-compact and combine them in the following kind of the Maxwell term:
\begin{equation}\label{XX}
\left( \sum_s \epsilon^{\mu\nu\lambda} \partial_\nu \mathcal{X}_{s\lambda} \right)^2 \ .
\end{equation}
This would prohibit individual monopoles, but allow monopole-antimonopole pairs in arbitrary two spin channels that correspond to dual-spin flips (at the expense of the SU(2) symmetry). Both scenarios require dynamical gauge fields, which are simply absent from the theory (\ref{TopLGp}) of particles. At best, the particles may be charged and coupled to the conventional U(1) gauge field of electrodynamics (which has non-compact dynamics), but the spin-orbit SU(2) gauge field is fundamentally static so that spin-flavored vorticity must be conserved. Note that the choice of the spin projection axis is arbitrary in the above argument, so the full SU(2) dual-spin symmetry can be partially lowered in (\ref{TopLG}) only through the external fluxes $\Phi_0$ and $\Theta_0$. This is at the heart of the robustness of topological orders against perturbations such as disorder and spin non-conserving collisions. Perturbations do not introduce dynamical gauge fields. Topological order can be destroyed only via the Landau-Ginzburg part of the Lagrangian by perturbations that can overcome the TI's gap and make a conventional state more energetically favorable.

Contrast this to the situation of the dual Lagrangian of vortices (\ref{TopLG}). It does feature a dynamical gauge field $\mathcal{B}_\mu$, which describes particle fluctuations in all spin states and provides the means to allow the non-conservation of dual vorticity, that is the physical particle spin. One weighted trace of $\mathcal{B}_\mu$ is related to charge and must be non-compact due to charge conservation, but certain combinations of $\mathcal{B}_\mu$ eigenmodes can be compact and implement spin-changing events via appropriate modifications of (\ref{Maxwell1}). An example of such a Maxwell term is:
\begin{eqnarray}\label{Maxwell2}
\mathcal{L}_{\textrm{M}} &=& \frac{1}{8\pi^2 q^2} \left\lbrack \sum_s
    \Bigl( \epsilon^{\mu\nu\lambda} \partial_\nu \bar{\beta}_{s\lambda} - 2\pi n_s \delta_{\mu,0} \Bigr) \right\rbrack^2 \\
&&  - \sum_s t_s \cos
    \Bigl( \epsilon^{\mu\nu\lambda} \partial_\nu \bar{\beta}_{s\lambda} - 2\pi n_s \delta_{\mu,0} \Bigr) \ , \nonumber
\end{eqnarray}
which is written using the continuum notation but ultimately defined on a lattice, using the representation $\mathcal{B}_\mu = \textrm{diag} (m_{-S}\bar{\beta}_{-S,\mu}, \dots, m_S\bar{\beta}_{S,\mu})$ that was introduced in the previous section. Generally, the compact cosine terms have to reflect the specific ways in which the spin is not conserved, and therefore depend on the spin-orbit coupling and various spin non-conserving perturbations. We have no means to derive by duality the precise form of the Maxwell couplings in these general circumstances. Nevertheless, we can understand qualitatively the effect of fluctuations that are made possible by the compact gauge fields in the vortex Lagrangian (\ref{TopLG}).

A compact gauge field can admit $2\pi$-quantized flux lines of arbitrary length without any energy cost, implying that monopoles (ends of semi-infinite flux tubes) are allowed. The ensuing fluctuations of quantized flux tubes and monopoles lead to the quantization of local density fluctuations in all charge and spin channels that couple to the compact gauge fields (in the spirit of the Dirac's charge quantization by the existence of monopoles). The meaning of density quantization is that an isolated volume of space can contain only integer multiples of charge/spin quanta. The quanta can be mobile, but the suppression of smooth density fluctuations produces an incompressible ground state. The specific compact components of the dual Maxwell term $\mathcal{L}_{\textrm{M}}$ dictate which particular vortex density fluctuations become quantized in the theory (\ref{TopLG}). For example, (\ref{Maxwell2}) would quantize all vortex density fluctuations that carry a definite dual-spin projection on the $z$-axis. A different kind of spin non-conserving dynamics, translated by duality to a different $\mathcal{L}_{\textrm{M}}$, could quantize other dual-spin projections, possibly dependent on position or momentum.

The Landau-Ginzburg part of the Lagrangian (\ref{TopLG}) has to determine the incompressible vortex densities that shape the topological order while respecting the imposed density quantization rules. We cannot predict the outcome of this without knowing the precise form of $\mathcal{L}_{\textrm{M}}$, but we can rest assured that only the dual-spin densities of vortices can be affected. Consequently, the quantization (\ref{DensityQuant}) of $\rho_s$ holds against all forms of spin non-conservation at least in the TR-invariant Laughlin states, because the resulting average dual-spin density is zero and properly quantized. However, spin non-conservation affects the quantum numbers of excitations. The fractional quantum numbers of quasiparticles discussed in the section \ref{secFract} can be jeopardized, but fractional vortex excitations (typically discussed in the FQHS literature and explained in the section \ref{secHier}) are protected by the dual-spin conservation.

\section{Generalizations of topological orders}\label{secGen}

This section demonstrates how the theory (\ref{TopLG}) and its dual (\ref{TopLGp}) can describe a broad range of topological insulating states. We will first develop the formalism for describing hierarchical Abelian FQHS and immediately extend it to the general Abelian SU(2) states of fractional TIs. This formalism will also provide a natural description of uncorrelated ``integer'' quantum (spin) Hall states. Then we will discuss non-Abelian states and generalizations to arbitrary symmetry groups and representations. We will demonstrate how the topological term of (\ref{TopLGp}) can produce non-Abelian statistics of excitations depending on the character of low-energy fluctuations, and devote a special attention to new topological orders that could be obtained from the Rashba spin-orbit coupling.

Many important properties of the ground state are decided by the dynamical (Landau-Ginzburg) part of the Lagrangian. They include particle densities, excitation spectra, vortex winding numbers and the corresponding fractions of elementary charge or spin that are expressed in quantum fluctuations. A class of ground states in external gauge fields features frustration which is resolved by nucleating vortices and ``binding'' them to particles in an incompressible quantum liquid. Instead of trying to microscopically understand this ``binding'' process, we qualitatively capture its most important outcomes via the topological term. The unique feature of (\ref{TopLG}) is that its topological term is very general and does not by itself specialize to any concrete topological state of matter. Instead, the topological term describes the quantum entanglement due to particle-vortex ``binding'' once the conventional dynamical properties of the ground state are known (or chosen for classification purposes). The deducible manifestations of the many-body quantum entanglement are the fractional statistics of quasiparticles and the ground-state degeneracy on Riemann surfaces such as torus. Therefore, the topological term alone is responsible for describing topological orders, and can be used to classify them. We will here discuss only topological orders without asking what dynamical conditions are necessary for stabilizing them.

The section \ref{secFract} characterizes some entanglement effects in a particular set of incompressible topological states whose vortex densities are given by (\ref{DensityQuant}). We can now imagine more complicated incompressible states that feature multiple low energy excitations labeled by some ``flavor'' quantum number $i=1,\dots,n$ that corresponds to emergent symmetries. The appearance of emergent symmetries in the low-energy dynamics is the only kind of ground state reorganization that overcomes the very restrictive condition (\ref{DensityQuant}) without leading to a conventional state such as superconductor or Mott insulator. The resulting possible topological states of matter form a hierarchy based on the emergent symmetries, and are generally obtained form (\ref{TopLG}) when the background flux $\Phi_0$ and density $\Theta_0$ matrices do not commute.

More generally, the low-energy dynamics can lead to non-Abelian topological orders. Only the topological term in (\ref{TopLG}) can capture them properly, while the Landau-Ginzburg part written there is specialized for the dynamics that favors Abelian statistics. We will not attempt in this paper to generalize this Landau-Ginzburg part because its form is known only to the extent allowed by the duality transformation of the physical particle theory (\ref{TopLGp}), and the duality mapping is currently available only for the cases with Abelian statistics. Instead, we will explore the non-Abelian topological states starting from the particle Lagrangian (\ref{TopLGp}) whose Landau-Ginzburg part can be readily constructed by understanding the microscopic system of interest. The goal we pursue in this paper is very modest and limited to the construction of effective theories of a few highly-entangled topological liquids, should they be stable ground states. Many important issues will be left untouched, most notably the characterization of any observable properties. A systematic study of fractional non-Abelian states and their properties is left for future work.

The main message will be that the topological spinor Lagrangian provides a formalism of much greater flexibility than the standard CS theory. Its topological term can be reduced to an effective CS form by coarse-graining in an incompressible quantum liquid, but the CS gauge fields may be arbitrary functions of any number of independent parameters that are restricted by the system's fundamental symmetry group and its representation. These parameters need not be sufficient to generate fully unconstrained fluctuations of the CS gauge fields. They instead represent the physical low-energy degrees of freedom, while the CS gauge fields are merely the mathematical tool that endows the physical excitations with non-trivial statistics via the topological term. We will show that such a flexibility is necessary to describe new topological orders that could arise in the Rashba spin-orbit-coupled TIs.

\subsection{Hierarchical states}\label{secHier}

First, we will consider Abelian fractional quantum Hall states of electrons in external magnetic fields (without a spin-orbit coupling), and obtain their most general CS descriptions from (\ref{TopLG}). The crucial assumption we need to make is that the low energy dynamics exhibits an emergent U(1)$^n$ symmetry. The number of low energy modes, and the level of the effective CS theory, is equal to the dimension of the U(1)$^n$ symmetry representation. We will consider only the minimal $n$-dimensional representations (conserved quantum numbers), so the vector $\psi = (\psi^1,\dots,\psi^n)$ in (\ref{TopLG}) will group together $n$ vortex flavors. The corresponding flavors of particle modes are captured by the $n$-dimensional gauge field matrix $\mathcal{B}_\mu$ that depends on $n$ fluctuating eigenmodes. These can be either microscopic particles in different internal states or any emergent low-energy modes that carry conventional quantum numbers (charge, spin, band/orbital index, etc.). The currents $j_{\pp\mu}^i$ of particle modes can be related to vortex currents $\widetilde{j}_{\vv\mu}^i$ via a linear transformation:
\begin{eqnarray}\label{C1}
j_{\pp\mu}^i &=& \sum_{j} Y^{ij} \epsilon^{\mu\nu\lambda} \partial_{\nu}^{\phantom{i}} \widetilde{j}_{\vv\lambda}^j \\
  &=& -\frac{i}{2} \sum_{j} Y^{ij} \phi_j \epsilon^{\mu\nu\lambda} \partial_{\nu} \Bigl(\psi^{j\dagger}(\partial_{\lambda}\psi^j)-(\partial_{\lambda}\psi^{j\dagger})\psi^j\Bigr) \ , \nonumber
\end{eqnarray}
where the coefficients $Y^{ij}$ have a dynamical origin in $\mathcal{L}_{\textrm{LG}}$ and are related to vortex charges along the lines hinted in the section \ref{secDual}. We will also consider a generic static U(1)$^n$ gauge field $\mathcal{A}_\mu$ coupled to vortex flavors in the topological term. This extended gauge field generally depends on the physical U(1) gauge field $a_\mu$. All space-time components of $\mathcal{A}_\mu$ commute with one another, but need not commute with $\mathcal{B}_\mu$, in which case we have a non-trivial linear relationship (\ref{C1}).

By the assumed symmetry, we can choose to work in the representation that simultaneously diagonalizes all components of $\mathcal{A}_\mu$ and $\Phi_\mu$. Let us define $\psi^i = \sqrt{\rho^i} \exp(i\theta^i)$ in this representation, and $\mathcal{A}_\mu = \textrm{diag}(A_\mu^1, \dots, A_\mu^n)$, $\Phi_0 = \textrm{diag}(\phi^1, \dots, \phi^n)$. If the fluctuations of all densities $\rho^i$ are suppressed, then we can repeat the analysis from the section \ref{secFract} and relate the local measurements of microscopic particle quantum numbers (charges) to the quantized windings of $\theta_i$ in the low-energy vortex field configurations. From (\ref{C1}) we find:
\begin{equation}\label{C2}
Q^i = \sum_{j} Y^{ij} \oint\limits _{dC}\dd l_{\mu}\,\widetilde{j}_{\vv\mu}^j = 2\pi \sum_{j} Y^{ij} m^j \phi^j \rho^j \ .
\end{equation}
It is convenient to define $Z^{ij} = 2\pi Y^{ij} \phi^j \rho^j$ and switch to matrix notation $Q=Zm$, where the vectors $Q = (Q^1,\dots,Q^n)$ and $m = (m^1,\dots,m^n)$ have integer components. The detection of a physical particle excitation in the flavor state $k$ corresponds to the vector $Q(k)$ with components $Q^i = \delta_{i,k}$, and is related to a particular combination of dual-vortex winding numbers contained in the vector $m(k)$. We find $m(k) = Z^{-1} Q(k)$, so that all matrix elements of $Z^{-1}$ must be integers. A particular compliant matrix $Z$ characterizes the topological insulating ground state and determines the fractional quantum numbers $\delta Q = (\delta Q^1,\dots,\delta Q^n)$ of various quasiparticle excitations enumerated by integer-valued vectors $l$:
\begin{equation}\label{Cquant}
\delta Q = Z l \qquad , \qquad (Z^{-1})^{ij} \; , \; l^i \in \mathbb{Z} \ .
\end{equation}

The constraints on $Z$ still leave significant freedom for the matrix $Y$. The main restriction on $Y$ comes from the definition $Z^{ij} = 2\pi Y^{ij} \phi^j \rho^j$, which implies that $Z^{ij}/Y^{ij}$ can depend only on the index $j$, that is $Y^{ij} = Y^{1j}(Z^{ij}/Z^{1j})$. Otherwise, we are free to impose other requirements without any loss of generality. For example, we can require that the matrix $Y$ be orthogonal. This amounts to a choice of normalization for $\psi^i_{\phantom{\mu}}$. The ground state is characterized by incompressible vortex densities:
\begin{equation}\label{Cvorden}
\rho^j = \frac{1}{2\pi\phi^j} \, \frac{Z^{1j}}{Y^{1j}} \ .
\end{equation}

By fixing $\rho^i$, the currents (\ref{C1}) and the topological term in the Lagrangian (\ref{TopLG}) reduce to:
\begin{eqnarray}\label{C3a}
j_{\pp\mu}^i &=& \sum_{j} Y^{ij} \phi^j \rho^j \epsilon^{\mu\nu\lambda} \partial_\nu^{\phantom{j}} b_\lambda^j \\
\mathcal{L}_{\textrm{t}} &=& -i \sum_j \phi_{\phantom{\mu}}^j \rho_{\phantom{\mu}}^j \left(
  - \frac{1}{2} \epsilon^{\mu\nu\lambda}_{\phantom{\mu}} b_\mu^j \partial_\nu^{\phantom{j}} b_\lambda^j
  + \epsilon^{\mu\nu\lambda}_{\phantom{\mu}} A_\mu^j \partial_\nu^{\phantom{j}} b_\lambda^j \right) \ . \nonumber
\end{eqnarray}
This can be expressed in the matrix form by introducing a diagonal matrix $R = \textrm{diag}(\phi^1\rho^1, \dots, \phi^n\rho^n)$ and vectors $b_\mu = (\partial_\mu \theta^1, \dots, \partial_\mu \theta^n)$, $j_{\pp\mu} = (j_{\pp\mu}^1, \dots, j_{\pp\mu}^n)$:
\begin{eqnarray}\label{C3b}
j_{\pp\mu} &=& \frac{1}{2\pi} Z \epsilon^{\mu\nu\lambda} \partial_\nu b_\lambda \\
\mathcal{L}_{\textrm{t}} &=& \frac{i}{2} \epsilon^{\mu\nu\lambda} b_\mu^T R \, \partial_\nu^{\phantom{c}} b_\lambda^{\phantom{c}}
  - i\epsilon^{\mu\nu\lambda} A_\mu^T R \, \partial_\nu^{\phantom{c}} b_\lambda^{\phantom{c}} \ . \nonumber
\end{eqnarray}
The currents of particle modes can be directly represented as fluxes of a different set of CS gauge fields $c_\mu = (c_\mu^1, \dots, c_\mu^n)$:
\begin{eqnarray}\label{C3c}
j_{\pp\mu} &=& \frac{1}{2\pi} \epsilon^{\mu\nu\lambda} \partial_\nu c_\lambda \\
\mathcal{L}_{\textrm{t}} &=& -i \left\lbrack
  -\frac{1}{4\pi} \epsilon^{\mu\nu\lambda} c_\mu^T K \partial_\nu^{\phantom{c}} c_\lambda^{\phantom{c}}
  +\frac{1}{2\pi} \epsilon^{\mu\nu\lambda} a_\mu q^T \partial_\nu c_\lambda \right\rbrack \ , \nonumber
\end{eqnarray}
where $c_\mu = Z b_\mu$, and the matrix $K$ and vector $q$ are defined by:
\begin{equation}\label{C3d}
K = 2\pi (Z^{-1})^T R Z^{-1}  \quad,\quad
a_\mu q^T = 2\pi A_\mu^{T} R Z^{-1} \ .
\end{equation}
The topological Lagrangian in (\ref{C3c}) now has the standard CS form, with the coupling matrix $K$ being symmetric by definition. Further restrictions on $K$ and $q$ follow from the requirement that $j_{\pp\mu}$ be the physical particle modes with conventional quantum numbers, which can be created or annihilated by local combinations of physical electron operators. It can be easily seen from (\ref{C2}) and subsequent definitions that measurable (integer) quantum numbers of physical particle modes correspond to flux quanta of the $c_\mu$ gauge fields. One of the particle modes directly corresponds to electrons and thus carries the U(1) charge that couples to the external gauge field $a_\mu$ and has fermionic statistics. Let this mode be labeled by $i=1$ in our representation, so that the vector $q$ is given by $q = (1,0,0, \dots, 0)$. The other modes are normally neutral (particle-hole) and hence must have bosonic statistics. The CS self-coupling implements the exchange statistics of excitations via the matrix $K$, and since the physical modes are flux quanta of $c_\mu$, the matrix elements of $K$ must be integers. Specifically, $K^{11}$ must be odd and $K^{ii}$, $i>1$ must be even. Other choices for $q$ and $K$ are appropriate for systems with different types of particle modes. Note that $(Z^{-1})^{ij} \in \mathbb{Z}$ is not enough to make $K^{ij} \in \mathbb{Z}$, the latter requirement further constrains the possible vortex densities $\rho^i$ in quantum Hall states. From the equation of motion one obtains the filling factor $\nu = q^T K^{-1} q$.

The literature on fractional quantum Hall effects \cite{WenQFT2004} generally considers a different kind of fractional excitations than the ones captured by (\ref{Cquant}). These excitations are ``fractional vortices'' whose currents $j_\mu = (j_\mu^1, \dots, j_\mu^n)$ minimally couple to the CS gauge fields $c_\mu$. Their U(1) charge $\delta q$ and statistical angle $\delta \theta$ are given by:
\begin{equation}\label{C4}
\delta q = q^T K^{-1} l \qquad,\qquad \delta\theta = \pi l^T K^{-1} l \ ,
\end{equation}
where $l$ is any vector of integers in the flavor space. The quasiparticles of (\ref{Cquant}) minimally couple to the gauge field $\mathcal{A}_\mu$ and thus are dual to vortices.

It should be emphasized that the topological Lagrangian $\mathcal{L}_{\textrm{t}}$ and the CS theory as its special limit do not by themselves determine the character of fluctuations that are detected in any particular experiment. The nature of low-energy modes is (also) shaped by the Landau-Ginzburg part of the Lagrangian. The eigenstates of the appropriate many-body Hamiltonian always carry an integer-quantized total charge, which however may be spatially distributed in fractional lumps. The topological Lagrangian only imposes some constraints on what kinds of lumps are possible and how their relative locations affect the many-body wavefunction's phase. The dynamics of lumps is beyond the topological term's reach, but affects the statistics of measurement outcomes. In that sense, one cannot easily make predictions about what kind of excitations would a particular experiment be sensitive to, fractionalized particles, vortices, or some other. A prediction we can make is that if an experimentalist successfully localizes a single fractional vortex, the observable amount of charge in its vicinity will be given by (\ref{C4}).

The effective Lagrangian in (\ref{C3c}) is the most general CS theory of hierarchical Abelian quantum Hall states. We have seen that such topological states can resolve the frustration of electron's kinetic energy at virtually any fractional particle density of $\nu$ particles per flux quantum, provided that the low-energy dynamics spontaneously develops multiple internal degrees of freedom for quasiparticle and vortex excitations. Analogous hierarchy of fractional states can now be constructed for TR-invariant topological insulators. The hierarchical states of spin $S$ particles feature the SU(2)$^n$ symmetry, with the $(2S+1)$-dimensional representation of the SU(2) subgroup. A $(2S+1)$-component sub-spinor $\psi^i = (\psi^{i,-S}, \dots, \psi^{i,S})$ is needed for each vortex flavor $i$ to determine vortex ``charge'' $\widetilde{j}_{\vv\mu}^i$ and ``spin'' $\widetilde{J}_{\vv\mu}^{ia}$ currents analogous to (\ref{Currents}):
\begin{eqnarray}
\widetilde{j}_{\vv\mu}^i &=& -\frac{i}{2}\Bigl\lbrack\psi^{i\dagger}\Phi_0^i(\partial_{\mu}\psi^i)
    -(\partial_{\mu}\psi^{i\dagger})\Phi_0^i\psi^i\Bigr\rbrack \\
\widetilde{J}_{\vv\mu}^{ia} &=& -\frac{i}{2}\Bigl\lbrack\psi^{i\dagger}\gamma^{a}\Phi_0^i(\partial_{\mu}\psi^i)
    -(\partial_{\mu}\psi^{i\dagger})\Phi_0^i\gamma^{a}\psi^i\Bigr\rbrack \nonumber \ .
\end{eqnarray}
We can allow a flavor-dependent SU(2) flux $\Phi_0^i$ to ensure the proper TR transformations. Then, we can generalize the physical particle charge and spin currents (\ref{C1}) as:
\begin{eqnarray}\label{C5}
j_{\pp\mu}^i &=& \epsilon^{\mu\nu\lambda} \partial_{\nu}^{\phantom{j}}
  \Bigl( Y^{ij}_{00} \, \widetilde{j}_{\vv\lambda}^j + Y^{ij}_{0a} \, \widetilde{J}_{\vv\lambda}^{ja} \Bigr) \\
J_{\pp\mu}^{ia} &=& \epsilon^{\mu\nu\lambda} \partial_{\nu}^{\phantom{j}}
  \Bigl( Y^{ij}_{a0} \, \widetilde{j}_{\vv\lambda}^j + Y^{ij}_{ab} \, \widetilde{J}_{\vv\lambda}^{jb} \Bigr) \ . \nonumber
\end{eqnarray}
We used here the Einstein's notation for all indices. The spin-orbit coefficients $Y^{ij}_{0a}$ and $Y^{ij}_{a0}$ must be either zero, or operators that change sign under TR. If in addition to the emergent SU(2)$^n$ symmetry the low-energy dynamics features mutually commuting emergent gauge fields $\mathcal{A}_\mu^i$, and their fluxes $\Phi_\mu^i$ in all flavors, then the topological term $\mathcal{L}_{\textrm{t}}$ has an even higher U(1)$^{(2S+1)n}$ symmetry and reduces to one of the CS theories discussed in Ref.\cite{Santos2011} when the density fluctuations are small. It is also straight-forward to construct the CS theories of hierarchical Abelian fractional quantum Hall states for electrons that experience both a strong magnetic field and spin-orbit coupling.

The TR-invariant topological gauge theory of spin $S=\frac{1}{2}$ particles obtained in this manner can be written in the BF form:
\begin{eqnarray}\label{BFfract}
j_{\pp\mu}^{\phantom{c}} &=& \frac{1}{2\pi} \epsilon^{\mu\nu\lambda} \partial_\nu^{\phantom{c}} c_\lambda^c \\
J_{\pp\mu}^z &=& \frac{1}{4\pi} \epsilon^{\mu\nu\lambda} \partial_\nu^{\phantom{c}} c_\lambda^s \nonumber \\
\mathcal{L}_{\textrm{t}} &=& -i\left\lbrack
  -\frac{1}{4\pi} \epsilon^{\mu\nu\lambda} c_\mu^{cT} K \partial_\nu^{\phantom{c}} c_\lambda^s
  + a_\mu^{\phantom{c}} q^T j_{\pp\mu}^{\phantom{c}} + A_\mu^z s^T J_{\pp\mu}^z \right\rbrack \nonumber
\end{eqnarray}
The superscripts $c$ and $s$ label the CS gauge field vectors that represent charge and spin currents respectively.

\subsection{Non-Abelian states}\label{secNonAbel}

Low-energy dynamics can support conditions for non-Abelian fractional incompressible quantum liquids. It is not presently clear how to describe such conditions in the vortex Lagrangian (\ref{TopLG}), so we will construct the non-Abelian effective theories in the language of the particle Lagrangian (\ref{TopLGp}). The prominent fluctuations of the $n$-component particle spinor field $\xi$ could be captured in certain topological states by $m$ fluctuating phases $\varphi^a$, where $a\in\lbrace 1,\dots,m \rbrace$ and $m \le n$. A fairly general form of such fluctuations can be written as:
\begin{equation}
\xi = e^{i\varphi^{a}\eta^{a}} \left( \begin{array}{c} f^1 \\ \vdots \\ f^n \end{array} \right)
\end{equation}
in some representation, where $\eta^a$ are a set of linearly independent Hermitian $n \times n$ matrices, and $f^a$ are non-fluctuating complex or Grassmann amplitudes. We will be interested in non-commuting $\eta^a$ (the above low-energy fluctuations generated by mutually commuting $\eta^a$ could produce only Abelian topological orders). If we define the gauge field matrix
\begin{eqnarray}\label{CSB0}
\mathcal{Z}_\mu &=& -i \Bigl( \partial_\mu e^{i \varphi^a \eta^a} \Bigr) e^{-i \varphi^a \eta^a} \\
  &=& \left( \partial_\mu \varphi^a \right) \int\limits_0^1 \dd x \, e^{ix \varphi^b \eta^b} \eta^a e^{-ix \varphi^c \eta^c}
  \nonumber
\end{eqnarray}
we can write:
\begin{equation}\label{CSB}
\partial_\mu \xi = i \mathcal{Z}_\mu \xi \ .
\end{equation}
The charge and spin currents (\ref{pCurrents3}) are:
\begin{eqnarray}\label{pCurrents4}
&& \!\!\!\!\!\!\!\! j_i = \xi^\dagger (\mathcal{Z}_i - \mathcal{A}_i) \xi
  = \textrm{tr} \Bigl\lbrack (\mathcal{Z}_i - \mathcal{A}_i) (\xi\xi^\dagger) \Bigr\rbrack \\
&& \!\!\!\!\!\!\!\! J_i^a = \frac{1}{2} \xi^\dagger \lbrace \mathcal{Z}_i - \mathcal{A}_i , \gamma^a \rbrace \xi
  = \frac{1}{2} \textrm{tr} \Bigl\lbrack \lbrace \mathcal{Z}_i - \mathcal{A}_i , \gamma^a \rbrace (\xi\xi^\dagger)
    \Bigr\rbrack \ . \nonumber
\end{eqnarray}
In the limit of suppressed amplitude $f^a$ fluctuations, the topological term from (\ref{TopLGp}) becomes:
\begin{equation}\label{NACS}
\mathcal{L}_{\textrm{pt}} =
 -\frac{i}{8} \textrm{tr}\Bigl\lbrack
    \bigl\lbrace \mathcal{Z}_\mu^{\phantom{a}},\Theta_0^{-1} \bigr\rbrace
    \bigl\lbrace \Phi_Z^\mu,\xi\xi^{\dagger}\bigr\rbrace
 \Bigr\rbrack
\end{equation}
up to a total derivative, where
\begin{equation}\label{CSBflux}
\Phi_{Z}^{\mu} = \epsilon^{\mu\nu\lambda}(\partial_{\nu}\mathcal{Z}_{\lambda}-i\mathcal{Z}_{\nu}\mathcal{Z}_{\lambda}) \ .
\end{equation}
It was pointed out in the section \ref{secDual} that $\mathcal{L}_{\textrm{pt}}$ is gauge-invariant, but written in the natural gauge where the external gauge field $\mathcal{A}_\mu$ is directly derived from the fixed form of the Rashba (or other) spin-orbit coupling. Making (\ref{NACS}) manifestly gauge-invariant requires replacing $\mathcal{Z}_\mu$ by $\mathcal{Z}_\mu-\mathcal{V}_\mu$, where $\mathcal{V}_\mu$ is a static flux-less background gauge field coupled to vortices. The dependence of
\begin{equation}
\xi\xi^{\dagger} = e^{i\varphi^{a}\eta^{a}}\left(\begin{array}{ccc}
  |f^1|^2 & \cdots & f^1 f^{n*} \\
  \vdots & \ddots & \vdots \\
  f^{1*} f^n & \cdots & |f^n|^2
\end{array}\right)e^{-i\varphi^{a}\eta^{a}}
\end{equation}
on $\varphi^a$ generally seeps into the Lagrangian (\ref{NACS}), which therefore is not a pure gauge theory. Nevertheless, the kinematics shaped by (\ref{NACS}) features excitations with non-Abelian fractional statistics.

The fluctuations of $\varphi^a$ contain both singular and non-singular components, and the topological kinematics of the singular ones is produced entirely via the non-Abelian gauge field $\mathcal{Z}_\mu$. Being interested only in the qualitative aspects of topological kinematics, we are tempted to convert (\ref{NACS}) into an approximate pure gauge theory. This can be done by coarse-graining. We will integrate out the short length-scale fluctuations in the path-integral that average out the non-singular fluctuations of $\varphi^a$ and allow the CS flux to diffuse. We need to identify the independent degrees of freedom $\varphi^a$ that adequately span the prominent low-energy fluctuations, and this amounts to choosing the proper set of matrices $\eta^a$ and amplitudes $f^a$ that match the dynamics (we are not prescribing the method to do this here). Then, we can formally treat $\varphi^a$ as independent random variables. We will also replace $\xi\xi^\dagger$ in (\ref{NACS}) by its average $\langle \xi\xi^\dagger \rangle$ under the assumption that fluctuations are abundant and chaotic at short length and time scales. This amounts to neglecting the subtle microscopic correlations between $\xi\xi^\dagger$ and $\mathcal{Z}_\mu$ (that is $\varphi^a$ and $\partial_\mu \varphi^a$).

We will consider two examples of the above coarse-graining procedure. The first example is a level-2 hierarchical state of electrons in magnetic field:
\begin{equation}\label{Psi1}
\xi=e^{i(\varphi_{x}\sigma^{x}+\varphi_{y}\sigma^{y})} \left(\begin{array}{c} f^1 \\ f^2 \end{array}\right)
    \equiv e^{i\boldsymbol{\varphi\sigma}} \left(\begin{array}{c} f^1 \\ f^2 \end{array}\right) \ ,
\end{equation}
where $\boldsymbol{\varphi} = (\varphi_x, \varphi_y, 0 )$ and $\boldsymbol{\sigma}$ is the vector of Pauli matrices $\sigma^a$. This two-component spinor represents prominent fluctuations of two particle modes in some unspecified quantum liquid. The gauge field (\ref{CSB}) is:
\begin{eqnarray}\label{Psi1B}
\mathcal{Z}_\mu &=& \left\lbrack 1-\frac{\sin(2|\varphi|)}{2|\varphi|}\right\rbrack
  \frac{(\boldsymbol{\varphi}\partial_{\mu}\boldsymbol{\varphi})(\boldsymbol{\varphi\sigma})}{|\varphi|^{2}} \\
&& +\frac{\sin(2|\varphi|)}{2|\varphi|}\boldsymbol{\sigma}\partial_{\mu}\boldsymbol{\varphi}
   -\left(\frac{\sin|\varphi|}{|\varphi|}\right)^{2}
    (\boldsymbol{\varphi}\times\partial_{\mu}\boldsymbol{\varphi})\boldsymbol{\sigma} \nonumber \\[0.1in]
&=& Z_\mu^a(\boldsymbol{\varphi},\partial_\mu\boldsymbol{\varphi}) \sigma^a \ . \nonumber
\end{eqnarray}
All components $Z_\mu^a$ are non-zero for generic values of $\boldsymbol{\varphi}$, so the gauge field is ``microscopically'' non-Abelian. Integrating out the short-wavelength fluctuations of $\boldsymbol{\varphi}$ will coarse-grain the gauge field in a manner that is hard to extract analytically due to the non-linearity of $Z_\mu^a(\boldsymbol{\varphi},\partial_\mu\boldsymbol{\varphi})$. However, we can rest assured that even the coarse-grained gauge field is non-Abelian because it is traceless while being dependent on two angles per point in space-time. One of these two angles would have to control the fluctuations of $\textrm{tr}(\mathcal{Z}_\mu)$ in order for $\mathcal{Z}_\mu$ to be Abelian in its two-dimensional representation.

We can approximately coarse-grain $\xi$ and $\xi\xi^\dagger$ by averaging them over the independent local random variables $\varphi_x({\bf r})$ and $\varphi_y({\bf r})$. Assuming that the probability distribution is symmetric under $\varphi_i \to -\varphi_i$, we find:
\begin{equation}\label{Psi1Av}
\langle\xi\rangle = \langle\cos|\varphi|\rangle\left(\begin{array}{c}f^1\\f^2\end{array}\right)
\end{equation}
and
\begin{eqnarray}\label{Psi1Den}
&& \!\!\!\!\!\!\!\!\!\!\!\!\!\!\!\! \langle\xi\xi^{\dagger}\rangle = \frac{|f^1|^{2}+|f^2|^{2}}{2}
   +\frac{|f^1|^{2}-|f^2|^{2}}{2}\langle\cos(2|\varphi|)\rangle\sigma^{z} \\
&& \!\!\!\!\!\!\!\!\!\!\! +\textrm{Re}\left(f^2 f^{1*}\right)\langle\cos^{2}|\varphi|\rangle\sigma^{x}
   +\textrm{Im}\left(f^2 f^{1*}\right)\langle\cos^{2}|\varphi|\rangle\sigma^{y} \nonumber \ .
\end{eqnarray}
This can be now substituted in (\ref{NACS}) to obtain a pure non-Abelian effective CS theory. The amplitudes $f^i$ can be determined in the spirit of the section \ref{secFract}, pending some information about the microscopic physical excitations that (\ref{Psi1}) represents. For illustration purposes, let us make the simplest assumption that the isolated phase windings in $\varphi_x$ or $\varphi_y$ correspond to microscopic (non-fractional) bosonic vortices. If we generate only one type of vortices, via $\varphi_x$ for example, then the other phase $\varphi_y$ does not have singularities and does not contribute to the topological term. The gauge field (\ref{Psi1B}) created solely by $\varphi_x$ becomes Abelian, $\mathcal{Z}_\mu = \sigma^x \partial_\mu \varphi_x \equiv \zeta_\mu \sigma^x$. In the present representation, this Abelian $\mathcal{Z}_\mu$ and the most general $\Theta_0^{-1} = t_0^{\phantom{a}} + \frac{1}{2} T_0^a \sigma^a$ produce the following effective theory from (\ref{NACS}):
\begin{eqnarray}
&& \!\!\!\!\!\!\!
   \mathcal{L}_{\textrm{pt}}^x = -\frac{1}{2}(\epsilon^{\mu\nu\lambda}\zeta_\mu \partial_\nu \zeta_\lambda) \\
&& \times \left\lbrack t_0\left(|f^1|^2+|f^2|^2\right)
  +T_0^x \; \textrm{Re}\left(f^2f^{1*}\right)\langle\cos^2|\varphi|
  \rangle\right\rbrack \ . \nonumber
\end{eqnarray}
We can now deduce the constraints on $f^i$ by reading out the statistics of excitations from $\mathcal{L}_{\textrm{pt}}^x$ and the analogous $\mathcal{L}_{\textrm{pt}}^y$:
\begin{eqnarray}
&& t_0\left(|f^1|^2+|f^2|^2\right)
  +T_0^x \;\textrm{Re}\left(f^2f^{1*}\right)\langle\cos^2|\varphi|\rangle
  = \frac{1}{2\pi m_x} \nonumber \\
&& t_0\left(|f^1|^2+|f^2|^2\right)
  +T_0^y \;\textrm{Im}\left(f^2f^{1*}\right)\langle\cos^2|\varphi|\rangle
  = \frac{1}{2\pi m_y} \nonumber \ .
\end{eqnarray}
The (even) integers $m_x$ and $m_y$ characterize the topologically ordered ground state whose fractional vortex excitations have mutually non-Abelian statistics.

It should be noted that the non-Abelian CS theory (\ref{NACS}) applied to the present example (\ref{Psi1}) is not a true gauge theory even though it looks like one. The actual fluctuations are generated by two local variables rather than a real SU(2) gauge field that requires three local variables (for three ``gauge boson'' modes) to fully explore its phase-space. The measurable currents associated with the two particle modes are obtained by substituting (\ref{Psi1}) and (\ref{Psi1B}) into (\ref{pCurrents4}).

A variation of this example is the non-Abelian incompressible quantum liquid of spin $S=\frac{1}{2}$ particles whose spinor components in (\ref{Psi1}) are identified with the spin $S^z$ amplitudes, $f^1 = f_\uparrow$ and $f^2 = f_\downarrow$. This state is generally not TR-invariant, but it can be TR-invariant if certain conditions are met. The necessary conditions for the TR symmetry are that the averages $\langle \cos(n|\varphi|) \rangle$ vanish, and one of the amplitudes $f_\uparrow$ or $f_\downarrow$ be zero. The first condition is naturally obtained from the wave-like oscillations $\varphi_i \sim {\bf pr} - \omega t$. Then we would get $\langle \xi \rangle = 0$ and $\langle \xi\xi^\dagger \rangle \propto 1$ consistent with the TR symmetry.

The second example is a TR-invariant topological insulator of $S=1$ particles such as $p$-wave Cooper pairs. We can construct a symmetric non-Abelian pure gauge theory starting from the normalized spinor:
\begin{equation}\label{Psi2}
\xi=e^{i\varphi^{a}\gamma^{a}}\left(\begin{array}{c} 0 \\ 1 \\ 0 \end{array}\right)
\end{equation}
in the representation that diagonalizes the external spin-orbit SU(2) flux $\Phi_0$. The matrices $\gamma^a$ are here the three-dimensional representations of the three SU(2) generators. The fluctuations generated by $\varphi^a$ span the full three-dimensional spin space in a manner that treats differently the $xy$ plane and the $z$-direction, but respects the TR and $xy$ rotation symmetries. Therefore, the symmetry of fluctuations is consistent with the spin U(1) symmetry of the model, which is anyway established by the presence of the external spin-orbit SU(2) flux $\Phi_{0}^{\phantom{z}}=\Phi_{0}^{z}\gamma^{z}$. By following the same procedure as before, the coarse-grained gauge field and density matrices are found to be:
\begin{equation}
\mathcal{Z}_\mu^{\phantom{a}} = Z_\mu^a(\varphi^x,\varphi^y,\varphi^z)\gamma^{a}
  \quad,\quad
\langle\xi\xi^{\dagger}\rangle=\alpha+\beta\Phi_{0}^{2} \quad ,
\end{equation}
where $\alpha$ and $\beta$ are numerical constants (note that $(\gamma^{z})^{2} \neq 1$ in the $S=1$ representation). No symmetry is violated, and one obtains a pure non-Abelian gauge theory by substituting this into (\ref{NACS}):
\begin{eqnarray}\label{Psi2CS}
\mathcal{L}_{\textrm{pt}} &=& -\frac{i\alpha}{4} \textrm{tr} \Bigl\lbrack
  \bigl\lbrace \mathcal{Z}_\mu^{\phantom{a}},\Theta_{0}^{-1} \bigr\rbrace
  \Phi_Z^\mu\Bigr\rbrack \\
&& -\frac{i\beta}{8} \textrm{tr} \Bigl\lbrack
  \bigl\lbrace \mathcal{Z}_\mu^{\phantom{a}},\Theta_{0}^{-1} \bigr\rbrace
  \bigl\lbrace \Phi_{Z}^\mu,\Phi_{0}^{2} \bigr\rbrace \Bigr\rbrack \ . \nonumber
\end{eqnarray}
We will not pursue in this and subsequent examples the quantization of constants such as $\alpha$ and $\beta$ that reproduces the correct statistics of microscopic excitations. An $n$-dimensional gauge field representation can be Abelian only if it depends on up to $n$ parameters and one them controls the $\textrm{tr}(\mathcal{Z}_\mu)$. But, the gauge field here belongs to a three-dimensional representation and depends on three parameters while being strictly traceless, so it is guarantied to remain non-Abelian after coarse-graining. Its scalar components $Z_\mu^a$ can be regarded as independently fluctuating fields after coarse-graining, so (\ref{Psi2CS}) is a true SU(2) non-Abelian CS gauge theory, which respects the TR symmetry.

We would like to remark here that the obtained non-Abelian gauge theories have a self-coupling of the dynamical gauge field that differs from the usually discussed \cite{Balatsky1991, Frohlich1992, Lopez1995, Fradkin1998, Fradkin1999, Fradkin2001} (TR-violating) CS form of the Yang-Mills gauge field:
\begin{equation}\label{nonAbelCS}
\mathcal{L}_{\textrm{cs}} = \frac{ik}{4\pi}\textrm{tr}\left(
  \epsilon^{\mu\nu\lambda}\mathcal{Z}_{\mu}\partial_{\nu}\mathcal{Z}_{\lambda}
  -i\frac{2g}{3}\epsilon^{\mu\nu\lambda}\mathcal{Z}_{\mu}\mathcal{Z}_{\nu}\mathcal{Z}_{\lambda}\right) \ .
\end{equation}
This is gauge-invariant up to the boundary term, and the Wess-Zumino-Witten term \cite{Witten1984, Ardonne2004} whose gauge-invariance requires the quantization of the constant $k$. In contrast, the form (\ref{NACS}) is gauge-invariant because it contains a (hidden) background gauge field $\mathcal{V}_\mu$ bundled into $\mathcal{Z}_\mu-\mathcal{V}_\mu$. Even though we could set the gauge $\mathcal{V}_\mu=0$, the fundamental presence of $\mathcal{V}_\mu$ is very much in the spirit of the duality between (\ref{TopLG}) and (\ref{TopLGp}). The formally analogous topological terms of (\ref{TopLG}) and (\ref{TopLGp}) are essential for reproducing the correct drift currents via equations of motion. The CS coupling (\ref{nonAbelCS}) might require a different kind of a topological term in the spinor Lagrangian, possibly tailored to a different kind of physics than discussed in this paper.

\subsection{Non-Abelian topological orders from the Rashba spin-orbit coupling}\label{TInonAbel}

The Rashba spin-orbit coupling $H_{\textrm{so}} = v \hat{\bf z} ({\bf p} \times {\bf S})$ in quantum wells does not conserve any fixed spin projection, but respects a special dynamical symmetry that we discussed in the section \ref{secTopOrd}. If no perturbations or interactions spoil this symmetry, it will define the quantum numbers of excitations. Our goal here is to rudimentarily explore the highly entangled incompressible quantum liquids consistent with this symmetry, and establish that they host quasiparticles with a non-Abelian fractional statistics. The latter indicates a non-trivial topological order, which is automatically robust against sufficiently weak perturbations even if they spoil the symmetries of the ideal Rashba-coupled Hamiltonian.

Let us focus on spin $S=\frac{1}{2}$ particles. If $S^z$ were conserved, we would expect the spinor configurations
\begin{equation}\label{T0}
\xi({\bf r}) =
  \left(\begin{array}{c} C_\uparrow e^{i\varphi_\uparrow({\bf r})} \\ C_\downarrow e^{i\varphi_\downarrow({\bf r})} \end{array}\right)
= e^{i \left(\varphi_c({\bf r}) + \frac{1}{2} \varphi_s({\bf r}) \sigma^z \right)}
  \left(\begin{array}{c} C_\uparrow \\ C_\downarrow \end{array}\right)
\end{equation}
to be relevant at low energies when the density fluctuations are small in comparison to the average densities $|C_\uparrow|^2$ and $|C_\downarrow|^2$. This applies to quantum Hall and spin-Hall liquids, and leads to our earlier constructions of the Abelian CS theories, albeit in the language of vortices rather than particles. The $S^z$ conservation here defines the two eigenmodes that carry the opposite $S^z$ spin projections, and the independent fluctuations of these modes at fixed densities are captured by the above spinor configurations.

In contrast, the translation symmetry and the dynamical symmetry of the Rashba spin-orbit coupling create particle modes that carry a conserved momentum $\bf p$ and a conserved ``helical'' spin projection $(\hat{\bf z} \times \hat{\bf p}){\bf S}$. Generic excitations are superpositions of these modes that we can describe by the following spinor field configurations:
\begin{equation}\label{T1}
\xi_{\pm}({\bf p}) = \frac{f_{\pm}({\bf p})}{\sqrt{2}}
  \left(\begin{array}{c} 1 \\ \pm e^{i\phi_{\bf p}}\end{array}\right)
= f_{\pm}({\bf p}) e^{\pm\frac{i\pi}{4}\hat{{\bf p}}\boldsymbol{\sigma}}
  \left(\begin{array}{c} 1 \\ 0 \end{array}\right) \ ,
\end{equation}
where $\phi_{{\bf p}}=\textrm{arg}(-p_{y}+ip_{x})$ is the angle of the in-plane direction perpendicular to $\bf p$ according to the right-hand rule, and $\boldsymbol{\sigma}$ is the vector of Pauli matrices. The subscript $\pm$ indicates the sign of the helical spin projection. The real-space field configurations $\xi({\bf r})$
\begin{equation}\label{T2}
{\bf \xi}_{\pm}({\bf r})=\int\frac{d^{2}p}{(2\pi)^{2}}e^{i{\bf pr}}\xi_{\pm}({\bf p})
\end{equation}\label{T3}
are characterized by the densities
\begin{eqnarray}
\!\!\!\!\!\!\!\!\!\! |\xi_{\pm}({\bf r})|^{2} &=& \int\frac{d^{2}p}{(2\pi)^{2}}\frac{d^{2}p'}{(2\pi)^{2}}
  e^{i({\bf p}-{\bf p}'){\bf r}} \\
&& \times \frac{1+e^{i(\phi_{\bf p}-\phi_{{\bf p}'})}}{2} f_{\pm}^{\phantom{*}}({\bf p})f_{\pm}^{*}({\bf p}') \nonumber \ .
\end{eqnarray}
An incompressible quantum liquid that conserves the helical spin projection will have constant uniform densities $|\xi_\pm({\bf r})|^2 = |C_{\pm}|^{2}$ in both $\pm$ sectors, so that
\begin{equation}\label{T4}
\int\frac{d^{2}q}{(2\pi)^{2}}\,\frac{1+e^{i(\phi_{\bf q}-\phi_{{\bf q}-{\bf p}})}}{2}
  f_{\pm}^{\phantom{*}}({\bf q})f_{\pm}^{*}({\bf q}-{\bf p})=|2\pi C_{\pm}|^{2}\delta({\bf p}) \ .
\end{equation}
This represents one constraint on each of the two complex amplitudes $f_\pm$ per point in space-time. Therefore, the allowed configurations of $\xi_\pm({\bf r})$ are determined by one real function $\varphi_\pm({\bf r})$ in each $\pm$ sector. The complexity of (\ref{T4}) will unfortunately prevent us from identifying the independent variables $\varphi_\pm({\bf r})$, but we will still gain some insight about the quantum states shaped by their fluctuations.

The TR-invariance and other symmetries impose no restrictions on the incompressible densities $C_+$ and $C_-$. The field configurations $\xi_+$ and $\xi_-$ are generally not degenerate, and the dynamics could freeze the fluctuations of one of them. Let us consider such a special case, where for example only $\xi_+$ is important. Having a fixed density $\xi_+^\dagger \xi_+^{\phantom{\dagger}}$ in an incompressible quantum liquid allows us to generate the low-energy fluctuations by applying a unitary operator to a fixed spinor:
\begin{equation}\label{T5}
\xi_+({\bf r}) = \hat{U}_+\lbrack{\bf r};\varphi_+({\bf r})\rbrack \left(\begin{array}{c} C_+ \\ 0 \end{array}\right) \ ,
\end{equation}
The CS gauge field (\ref{CSB}) is given by $\mathcal{Z}_\mu = -i (\partial_\mu \hat{U}_+^{\phantom{\dagger}}) \hat{U}_+^\dagger$ as a function of the independently fluctuating variables $\varphi_+({\bf r})$. All aspects of topological order in an incompressible quantum liquid are then captured by the effective CS Lagrangian (\ref{NACS}) expressed in terms of this gauge field, and by the knowledge of low-energy particle modes that restrict the gauge field configurations. The effective theory also contains $\xi \xi^\dagger$, but the TR-invariance ensures that it can be replaced by $\langle \xi_+^{\phantom{\dagger}}\xi_+^\dagger \rangle = \frac{1}{2} |C_+|^2 \times 1$ after coarse-graining. The winding numbers of vortex excitations determine the quantization of $|C_+|^2$ that appears in the CS theory, and hence the fractional amounts of charge and spin. The ensuing topological orders can be classified by a Laughlin-like sequence of states and hierarchical constructions. However, the statistics of quasiparticles and the ground state degeneracy on Riemann surfaces are expected to differ from the ordinary Laughlin and hierarchical quantum Hall states. The statistics of excitations can be found by considering the adiabatic evolution of a quasiparticle's spinor as it moves on a closed path around another quasiparticle. It depends on the CS gauge field in the vicinity of a vortex, which in the present case gives rise to a non-Abelian Aharonov-Bohm effect and a non-Abelian fractional statistics once $|C_+|^2$ is fractionally quantized. Unfortunately, we cannot reach more detailed conclusions at this time. The characterization of non-Abelian topological orders is an open problem, since the steps outlined here cannot be carried out without a better understanding of the non-Abelian duality.

Nevertheless, we can at least get a glimpse of a $\xi_+$ vortex and appreciate why the quantum liquids of such vortices produce TR-invariant non-Abelian topological orders. We do not know how $\hat{U}_+({\bf r})$ in (\ref{T5}) depends on $\varphi_+({\bf r})$, but we can establish that  it is locally a linear combination of the unit-matrix, $\sigma^x$ and $\sigma^y$, which does not include $\sigma^z$. This follows from the fact that the Pauli matrices are introduced in (\ref{T1}) and (\ref{T2}) only via the products ${\bf p} \boldsymbol{\sigma}$ with a two-dimensional vector ${\bf p}$. Any unitary operator of this kind can be written as
\begin{equation}\label{T6}
\hat{U}_+ = e^{i(\varphi_0 + \varphi_x \sigma^x + \varphi_y \sigma^y)} \ .
\end{equation}
In our case, $\varphi_0({\bf r})$, $\varphi_x({\bf r})$ and $\varphi_y({\bf r})$ are mutually-dependent functions derived non-locally from a single independent function $\varphi_+({\bf r})$. After isolating out the trivial charge fluctuations $\varphi_0$, the rest of this operator is the same as the one in (\ref{Psi1}), so the traceless part of the corresponding CS gauge field $\partial_\mu \xi_+ = i \mathcal{Z}_\mu \xi_+$ takes the form (\ref{Psi1B}) in terms of the vector $\boldsymbol{\varphi} = (\varphi_x, \varphi_y, 0)$:
\begin{eqnarray}\label{T7}
\mathcal{Z}_\mu &=& \partial_\mu \varphi_0 + \left\lbrack 1-\frac{\sin(2|\varphi|)}{2|\varphi|}\right\rbrack
  \frac{(\boldsymbol{\varphi}\partial_{\mu}\boldsymbol{\varphi})(\boldsymbol{\varphi\sigma})}{|\varphi|^{2}} \\
&& +\frac{\sin(2|\varphi|)}{2|\varphi|}\boldsymbol{\sigma}\partial_{\mu}\boldsymbol{\varphi}
   -\left(\frac{\sin|\varphi|}{|\varphi|}\right)^{2}
    (\boldsymbol{\varphi}\times\partial_{\mu}\boldsymbol{\varphi})\boldsymbol{\sigma} \ . \nonumber
\end{eqnarray}

An elementary $\xi_+$ excitation that carries momentum $\bf p$ has an amplitude $f_+({\bf p}') \propto \delta({\bf p}'-{\bf p})$, so its field configuration is:
\begin{equation}\label{T8}
\xi_1({\bf r}) = \frac{1}{\sqrt{2}}\left(\begin{array}{c} 1 \\ e^{i\phi_{\bf p}}\end{array}\right) e^{i{\bf pr}} \ .
\end{equation}
This mode carries both a charge and spin current. Charge currents must be absent in a TR-invariant ground state, so let us consider a superposition of two elementary $\xi_+$ modes that carry counter-propagating currents:
\begin{eqnarray}\label{T9}
\xi_2({\bf r}) &=&
     \frac{1}{2}\left(\begin{array}{c} 1 \\ e^{i\phi_{\bf p}}\end{array}\right)e^{i{\bf p}{\bf r}}
    +\frac{1}{2}\left(\begin{array}{c} 1 \\ e^{i\phi_{-{\bf p}}}\end{array}\right)e^{-i{\bf p}{\bf r}} \nonumber \\
&=& e^{i({\bf p}{\bf r})\lbrack(\hat{{\bf z}}\times\hat{{\bf p}})\boldsymbol{\sigma}\rbrack}
    \left(\begin{array}{c} 1 \\ 0\end{array}\right) \ .
\end{eqnarray}
The corresponding CS gauge field $\partial_{i}\xi_2=i\mathcal{Z}_{i}\xi_2$ is:
\begin{equation}\label{T10}
\mathcal{Z}_i = \lbrack (\hat{{\bf z}}\times\hat{{\bf p}})\boldsymbol{\sigma} \rbrack p_i \ ,
\end{equation}
and it can be reproduced by
\begin{equation}\label{T11}
\varphi_{0}=0\quad,\quad\boldsymbol{\varphi}=({\bf p}{\bf r})(\hat{{\bf z}}\times\hat{{\bf p}})
\end{equation}
from the expression (\ref{T7}). It can be seen either from (\ref{pCurrents3}) or (\ref{pCurrents4}) that the charge and spin current densities carried by this excitation are:
\begin{equation}\label{T12}
j_i = 0 \qquad,\qquad J_i^a = \frac{1}{2} p_i \epsilon_{ja} \hat{p}_{j} \ ,
\end{equation}
meaning that only the helical spin projection perpendicular to the momentum ${\bf p}$ is carried in the direction of ${\bf p}$. The charge and spin densities are:
\begin{eqnarray}\label{T13}
j_0 &=& \xi_2^\dagger \xi_2^{\phantom{\dagger}} = 1 \\
J_0^a \hat{\bf x}^a &=& \frac{1}{2} \xi_2^\dagger \boldsymbol{\sigma} \xi_2^{\phantom{\dagger}}
    = \frac{1}{2} \Bigl\lbrack \hat{{\bf z}}\cos(2{\bf p}{\bf r})-\hat{{\bf p}}\sin(2{\bf p}{\bf r}) \Bigr\rbrack \ . \nonumber
\end{eqnarray}
The number (charge) density is uniform and thus consistent with incompressibility. The spin density appears non-uniform, but it should be noted that it oscillates in both space and time. We have been emphasizing only the spatial dependence, but $\bf pr$ in fact stands for ${\bf pr}-\omega t$, where the frequency $\omega$ is by magnitude equal or greater than the excitation gap $\Delta$ in the incompressible liquid state. These oscillations are removed by coarse-graining over time intervals larger than $\delta t \sim \Delta^{-1}$. The ground state formed by the zero-point quantum fluctuations of many modes such as (\ref{T9}) will not have any spin texture or charge currents.

However, as long as we are allowed to superpose only the $\xi_+$ modes, we cannot eliminate the spin currents. These currents are allowed by the TR symmetry, but must not have an open-ended flow in equilibrium. Therefore, a stable ground state of fluctuating $\xi_+$ must in fact feature spin current loops. Any static arrangement of such loops is an SU(2) vortex lattice, which could be stabilized if the particles were bosonic. The appropriate configurations of $\mathcal{Z}_{\mu}$ in such a vortex lattice would have to involve a vector field $\boldsymbol{\varphi}({\bf r})$ that according to (\ref{T11}) changes along any vortex-centered loop as $\boldsymbol{\varphi}({\bf r})\sim({\bf pr})(\hat{{\bf z}}\times\hat{\bf p})$, where ${\bf p}$ is locally tangential to the loop and quantized to make $\boldsymbol{\varphi}({\bf r})$ single-valued. This inevitably generates a quantized $\sigma^{z}$ flux tube at the vortex center via the $(\boldsymbol{\varphi} \times \partial_{\mu} \boldsymbol{\varphi}) \boldsymbol{\sigma}$ part of (\ref{T7}). The presence of an external SU(2) flux $\Phi_0 \propto \sigma^z$ can naturally generate this type of vortex configurations. A topologically ordered and highly entangled SU(2) quantum liquid state is obtained upon the quantum melting of this vortex lattice.

This qualitative picture hints that the duality transformation involving non-Abelian fields has a lot in common with the Abelian case. However, an elementary vortex cannot be captured by a purely Abelian $\mathcal{Z}_{\mu}\sim\sigma^{z}$ configuration according to (\ref{T7}), so it cannot have an Abelian statistics. As an example, consider a single vortex with concentric circular spin supercurrents. The spatial gauge field component $\mathcal{Z}_\perp$ taken in the direction perpendicular to the local spin-current flow ($\hat{\boldsymbol{\mu}} \perp \hat{\bf p}$) is locally proportional to $\boldsymbol{\varphi\sigma}$ because $\boldsymbol{\varphi}$ always points in the radial direction. Hence, $\mathcal{Z}_\perp$ is a linear combination of only $\sigma^x$ and $\sigma^y$. On the other hand, the component $\mathcal{Z}_\parallel$ in the direction parallel to the current flow is a linear combination of all $\sigma^a$ in the presence of the current flow curvatures ($\partial_\mu \boldsymbol{\varphi}$ is not parallel to $\boldsymbol{\varphi}$ for $\hat{\boldsymbol{\mu}} \parallel \hat{\bf p}$). Therefore, $\mathcal{Z}_\perp$ and $\mathcal{Z}_\parallel$ do not commute. One might say that a dynamical SU(2) gauge field can emit its own ``gauge boson'' quanta and provide a flux-feedback to the externally imposed flux.

The kind of a TR-invariant topological order explored here is different than any one that can be described using the standard CS theories with proper gauge fields. Clearly, the types of many-body entanglement discussed in this section can be extended to higher spin representations, hierarchical states, inhomogeneous quantum fluids, etc. Fluctuations can be generated by virtually any unitary operator that depends on distributed parameters and acts on the $n$-component spinors of fixed amplitudes. Each parameter can be distributed over any complete set of generalized coordinates and coupled to any coordinate-dependent generator of transformations. The Lagrangians (\ref{TopLG}) and (\ref{TopLGp}) provide the formalism to describe all of these possibilities for topological states of matter, at least in principle.

\section{Conclusions and outlook}\label{secConcl}

In this paper we constructed a topological field theory of particles in two spatial dimensions whose charge and spin couple to the external electromagnetic and spin-orbit fields. This theory provides a universal description of both conventional and topological phases, being an extension of the Landau-Ginzburg Lagrangian that implements a state-dependent quantization of classical dynamics via a topological term. The added topological term is innocuous in phases where particles are well localized (e.g. Mott insulators) or very mobile (superconductors and Fermi liquids). However, quantum insulators in which the external magnetic field or spin-orbit coupling frustrate particle motion are qualitatively affected by the topological term and acquire quasiparticle excitations with fractional quantum numbers and exchange statistics. These incompressible quantum liquids are generalized quantum Hall states with topological order (many-body quantum entanglement extending over large distances).

The topological field theory was written in two mutually dual and physically equivalent forms: (\ref{TopLGp}) describes the physical particles directly, while (\ref{TopLG}) describes vortices, the topological defects of particle field configurations. The particle Lagrangian (\ref{TopLGp}) must transparently capture the known dynamics of particles in external magnetic and spin-orbit fields, but cannot alter their exchange statistics in smooth field configurations that dominate the path integral of conventional phases. Therefore, any statistics-altering topological term may couple only to topological defects. Unfortunately, no experimental evidence of vortex dynamics in the materials of interest is available to guide the construction of such topological terms. Instead, we had to resort to the dual theory of vortices (\ref{TopLG}) and construct its topological term first. Topological defects of the dual theory correspond to physical particles whose dynamics is measurable. We constructed the dual topological term by requiring that the dual action be stationary when its field configurations reproduce the classical equations of motion for constant drift currents of particles in external magnetic and spin-orbit fields. The appropriate equations of motion were deduced from the generic model Hamiltonian of two-dimensional topological insulators that exhibit the quantum spin-Hall effect. The Rashba spin-orbit-coupled electrons in realistic quantum wells are modeled by exactly the same type of Hamiltonian, with the same symmetries and qualitative structure. Therefore, by using symmetries we achieved an indirect experimental justification of the topological field theory. The full picture of particle dynamics and topological kinematics is obtained only by establishing the field-theoretical duality between the particle (\ref{TopLGp}) and vortex (\ref{TopLG}) Lagrangians, which constrains their relative form.

The vortex Lagrangian (\ref{TopLG}) is actually a generalization of (dynamically enhanced) Chern-Simons gauge theories. Whenever the dynamics of a quantum state can be approximately captured by an XY model derived from (\ref{TopLG}), its topological properties are similarly captured by an effective theory derived from (\ref{TopLG}) that has a Chern-Simons structure. A Landau-Ginzburg theory is already an abstraction that gives up the ``irrelevant'' microscopic details of the system in order to focus on its universal properties based on symmetries. The true usefulness of a Chern-Simons theory is achieved when even the symmetries are regarded as ``irrelevant'' in order to focus on topological orders. A Chern-Simons theory can be written only after assuming some symmetry, such as U(1)$^n$, but the topological order that it captures is robust even when perturbations remove this symmetry. The topological term of the Lagrangian (\ref{TopLG}) is more versatile. It does not assume any emergent symmetry, but rather lets the Landau-Ginzburg part determine the symmetries and densities in the ground state, which in turn dictate the fractional statistics of quasiparticle excitations.

This paper accomplished several goals. First, we explained the construction of the new topological field theory and its dual. Several known results from the literature were re-derived and adapted to the specific features of the present theory in order to show that it is consistent with the dynamics of systems we wish to model, and that it adequately generalizes the Chern-Simons theory in the cases of standard U(1) quantum Hall liquids.

Second, we introduced a formalism based on the SU(2) gauge symmetry that can describe any spin-orbit coupling and view it as the origin of a generalized quantum Hall effect. The SU(2) ``Hall effect'' has a richer phenomenology that its U(1) counterpart and need not lead to the quantization of spin-Hall conductivity. In particular, we showed that the Rashba spin-orbit coupling has a dynamical symmetry that differs from the conventional U(1) symmetry of the standard quantum Hall effect, but nevertheless creates fractional ground states with topological degeneracy on a torus when interactions stabilize an incompressible quantum liquid. The simplest such topological orders are of the Laughlin kind and do not depend on whether $S^z$ is conserved or not. We further analyzed the quantum numbers of quasiparticle excitations, and identified the nature of charge and spin fractionalization in relation to the symmetries of the ground state. We did this for any combination of external magnetic and spin-orbit fields acting on particles with arbitrary spin.

Third, we demonstrated the ability of the proposed topological field theory to handle a broad spectrum of incompressible quantum liquids with distinct robust topological orders. We elucidated the construction of hierarchical incompressible liquids, mostly by focusing on the Abelian quantum Hall states classified by the U(1) Chern-Simons theories. The analogous SU(2) descriptions of hierarchical Abelian spin-Hall states were obtained in a straight-forward fashion. Then, we considered a few examples of low-energy fluctuations that produce non-Abelian fractional statistics in incompressible quantum liquids. For instance, if such a quantum liquid features a triplet of low-energy modes with SU(2) symmetry, its topological order is described by an effective SU(2) Chern-Simons gauge theory that can be derived from the topological term of (\ref{TopLG}). In general, the effective theory superficially has a Chern-Simons form, but its non-Abelian gauge fields can be constrained to a subspace of all possible configurations through their dependence on a small number of parameters that generate the low-energy fluctuations. We specifically constructed an effective field theory of a highly entangled non-Abelian incompressible quantum liquid that respects the time-reversal symmetry and takes advantage of the Rashba spin-orbit coupling to lower its energy.

The proposed topological field theory is much more general than the Chern-Simons theory, and its clean structure based on representations and symmetries might provide a broad classification scheme of topological orders. The present Lagrangian can be extended to any representation of arbitrary emergent U(1)$^{n_1} \times$SU(2)$^{n_2} \times$SU(3)$^{n_3} \dots$ symmetry groups in which particles or vortices are coupled to an external (non-Abelian) ``electromagnetic'' flux of arbitrary direction in the space spanned by the symmetry generators. The symmetry determines the character of topological orders, the representation is related to the available low-energy particle/vortex modes and the (non)conservation of their quantum numbers, while the flux direction determines the mutual statistics of modes. There are many possibilities that will be explored in future work, together with a systematic analysis of observable phenomena such as the properties of protected boundary states.

Certain symmetry-protected aspects of topological order can be observed in curved geometries because the total angular momentum of quasiparticles couples to the curvature of their two-dimensional plane \cite{Wen1992a, Wen1992c}. Similar ``geometric'' coupling, but of dynamical origin, was noticed in anisotropic fractional quantum Hall states \cite{Haldane2011}. We restricted the analysis in this paper to the simplest continuum systems, and hence did not consider these geometric properties of incompressible quantum liquids. Nevertheless, the spin-geometry coupling can be readily implemented in the presented theory, and will be scrutinized in the future.

The topological term constructed in this paper could be viewed as the lowest-order member of a sequence of topological terms that contain higher powers of flux, (co-variant) derivatives and field operators. This sequence must be restricted by symmetries and its higher order members might lead to additional topological orders if they are macroscopically relevant. Further generalizations to higher spatial dimensions and other types of matter fields would involve constructing topological terms that reflect all possible topological defects and their symmetry-allowed couplings to external gauge fields.

Finally, the proposed theory also describes the dynamics of particles or vortices via its Landau-Ginzburg part. We demonstrated that the topological term is state-independent, but the topological order that it creates depends on the dynamically stable density of particles and vortices. With this level of description, we can ask whether the written theory could be used to chart universal phase diagrams that contain both conventional and topological states of matter. This question is very much worth exploring because no method is currently available to solve such problems, other than the numerical exact diagonalization of systems with a few particles.

\section{Acknowledgements}

I am very grateful to Michael Levin and Zlatko Tesanovic for insightful discussions. This research was supported by the Office of Naval Research (grant N00014-09-1-1025A), the National Institute of Standards and Technology (grant 70NANB7H6138, Am 001), and the U.S. Department of Energy, Office of Basic Energy Sciences, Division of Materials Sciences and Engineering under Award DE-FG02-08ER46544 (summer 2011). This work was supported in part by the National Science Foundation under Grant No. PHYS-1066293 and the hospitality of the Aspen Center for Physics.


\begin{thebibliography}{10}

\bibitem{Tsui1982}
D.~C. Tsui, H.~L. Stormer, and A.~C. Gossard, Physical Review Letters {\bf 48},
   1559  (1982).

\bibitem{Stormer1983}
H.~L. Stormer, A. Chang, D.~C. Tsui, J.~C.~M. Hwang, A.~C. Gossard, and W.
  Wiegmann, Physical Review Letters {\bf 50},  1953  (1983).

\bibitem{Goldman1995}
V.~J. Goldman and B. Su, Science {\bf 267},  1010  (1995).

\bibitem{Saminadayar1997}
L. Saminadayar, D.~C. Glattli, Y. Jin, and B. Etienne, Physical Review Letters
  {\bf 79},  2526  (1997).

\bibitem{Goldman2001}
V.~J. Goldman, I. Karakurt, J. Liu, and A. Zaslavsky, Physical Review B {\bf
  64},  085319  (2001).

\bibitem{Camino2005}
F.~E. Camino, W. Zhou, and V.~J. Goldman, Physical Review B {\bf 72},  075342
  (2005).

\bibitem{Camino2007}
F.~E. Camino, W. Zhou, and V.~J. Goldman, Physical Review Letters {\bf 98},
  076805  (2007).

\bibitem{de-Picciotto1997}
R. de~Picciotto, M. Reznikov, M. Heiblum, V. Umansky, G. Bunin, and D. Mahalu,
  Nature {\bf 389},  162  (1997).

\bibitem{Venkatachalam2011}
V. Venkatachalam, A. Yacoby, L. Pfeiffer, and K. West, Nature {\bf 469},  185
  (2011).

\bibitem{Hasan2010}
M.~Z. Hasan and C.~L. Kane, Reviews of Modern Physics {\bf 82},  3045  (2010).

\bibitem{Qi2010a}
X.-L. Qi and S.-C. Zhang, Reviews of Modern Physics {\bf 83},  1057  (2011).

\bibitem{Moore2010}
J.~E. Moore, Nature {\bf 464},  194  (2010).

\bibitem{Konig2007}
M. K\"{o}nig, S. Wiedmann, C. Br\"{u}ne, A. Roth, H. Buhmann, L.~W. Molenkamp,
  X.-L. Qi, and S.-C. Zhang, Science {\bf 318},  766  (2007).

\bibitem{Kane2005a}
C.~L. Kane and E.~J. Mele, Physical Review Letters {\bf 95},  146802  (2005).

\bibitem{Levin2009}
M. Levin and A. Stern, Physical Review Letters {\bf 103},  196803  (2009).

\bibitem{Karch2010}
A. Karch, J. Maciejko, and T. Takayanagi, Physical Review D {\bf 82},  126003
  (2010).

\bibitem{Cho2010}
G.~Y. Cho and J.~E. Moore, Annals of Physics {\bf 326},  1515  (2011).

\bibitem{Maciejko2010}
J. Maciejko, X.-L. Qi, A. Karch, and S.-C. Zhang, Physical Review Letters {\bf
  105},  246809  (2010).

\bibitem{Swingle2011}
B. Swingle, M. Barkeshli, J. McGreevy, and T. Senthil, Physical Review B {\bf
  83},  195139  (2011).

\bibitem{Neupert2011}
T. Neupert, L. Santos, S. Ryu, C. Chamon, and C. Mudry, Physical Review B {\bf
  84},  165107  (2011).

\bibitem{Santos2011}
L. Santos, T. Neupert, S. Ryu, C. Chamon, and C. Mudry, Physical Review B {\bf
  84},  165138  (2011).

\bibitem{Nikolic2011}
P. Nikolic, Journal of Physics: Condensed Matter {\bf 25},  025602  (2013).

\bibitem{Levin2012}
M. Levin and A. Stern, Physical Review B {\bf 86},  115131  (2012).

\bibitem{Kitaev2000}
A. Kitaev, Physics-Uspekhi {\bf 44 Supplement},  131  (2000).

\bibitem{Kitaev2003}
A. Kitaev, Annals of Physics {\bf 303},  2  (2003).

\bibitem{Nayak2008}
C. Nayak, S.~H. Simon, A. Stern, M. Freedman, and S.~D. Sarma, Reviews of
  Modern Physics {\bf 80},  1083  (2008).

\bibitem{Bonderson2010}
P. Bonderson, S.~D. Sarma, M. Freedman, and C. Nayak,   (2010),
  arXiv:1003.2856.

\bibitem{Pesin2010}
D. Pesin and L. Balents, Nature Physics {\bf 6},  376  (2010).

\bibitem{Rachel2010}
S. Rachel and K.~L. Hur, Physical Review B {\bf 82},  075106  (2010).

\bibitem{Krempa2010}
W. Witczak-Krempa, T.~P. Choy, and Y.~B. Kim, Physical Review B {\bf 82},
  165122  (2010).

\bibitem{Young2008}
M.~W. Young, S.-S. Lee, and C. Kallin, Physical Review B {\bf 78},  125316
  (2008).

\bibitem{Sun2011}
K. Sun, Z. Gu, H. Katsura, and S.~D. Sarma, Physical Review Letters {\bf 106},
  236803  (2011).

\bibitem{Sheng2011}
D.~N. Sheng, Z.-C. Gu, K. Sun, and L. Sheng, Nature Communications {\bf 2},
  389  (2011).

\bibitem{Wang2011}
F. Wang and Y. Ran, Physical Review B {\bf 84},  241103(R)  (2011).

\bibitem{Venderbos2011}
J.~W. Venderbos, M. Daghofer, and J. van~den Brink, Physical Review Letters
  {\bf 107},  116401  (2011).

\bibitem{Murthy2011}
G. Murthy and R. Shankar,   (2011), arXiv:1108.5501.

\bibitem{Goerbig2011}
M.~O. Goerbig, European Physical Journal B {\bf 85},  15  (2011).

\bibitem{Tang2011}
E. Tang, J.-W. Mei, and X.-G. Wen, Physical Review Letters {\bf 106},  236802
  (2011).

\bibitem{Neupert2011a}
T. Neupert, L. Santos, C. Chamon, and C. Mudry, Physical Review Letters {\bf
  106},  236804  (2011).

\bibitem{Bernevig2006a}
B.~A. Bernevig and S.-C. Zhang, Physical Review Letters {\bf 96},  106802
  (2006).

\bibitem{Xiao2011}
D. Xiao, W. Zhu, Y. Ran, N. Nagaosa, and S. Okamoto, Nature Communications {\bf
  2},  596  (2011).

\bibitem{Ghaemi2011}
P. Ghaemi, J. Cayssol, D.~N. Sheng, and A. Vishwanath, Physical Review Letters
  {\bf 108},  266801  (2012).

\bibitem{Papic2011a}
Z. Papi\'{c}, R. Thomale, and D.~A. Abanin, Physical Review Letters {\bf 107},
  176602  (2011).

\bibitem{Abanin2012}
D.~A. Abanin and D.~A. Pesin, Physical Review Letters {\bf 109},  066802
  (2012).

\bibitem{Papic2012}
Z. Papi\'{c}, D.~A. Abanin, Y. Barlas, and R.~N. Bhatt, Journal of Physics:
  Conference Series {\bf 402},  012020  (2012).

\bibitem{Nikolic2010b}
P. Nikolic and Z. Tesanovic, Physical Review B {\bf 83},  064501  (2011).

\bibitem{Nikolic2011a}
P. Nikolic, T. Duric, and Z. Tesanovic, Physical Review Letters {\bf 110},
  176804  (2013).

\bibitem{Seradjeh2009}
B. Seradjeh, J.~E. Moore, and M. Franz, Physical Review Letters {\bf 103},
  066402  (2009).

\bibitem{Greiner2002}
M. Greiner, O. Mandel, T. Esslinger, T.~W. H\"{a}nsch, and I. Bloch, Nature
  {\bf 415},  39  (2002).

\bibitem{Lin2011}
Y.-J. Lin, K. Jim\'{e}nez-Garc\'{i}a, and I.~B. Spielman, Nature {\bf 471},  83
   (2011).

\bibitem{Campbell2011}
D.~L. Campbell, G. Juzeli\={u}nas, and I.~B. Spielman, Physical Review A {\bf
  84},  025602  (2011).

\bibitem{Cooper2011}
N.~R. Cooper, Physical Review Letters {\bf 106},  175301  (2011).

\bibitem{Qi2008b}
X.-L. Qi, T.~L. Hughes, and S.-C. Zhang, Physical Review B {\bf 78},  195424
  (2008).

\bibitem{Lu2012}
Y.-M. Lu and A. Vishwanath, Physical Review B {\bf 86},  125119  (2012).

\bibitem{Frohlich1992}
J. Fr\"{o}hlich and U.~M. Studer, Communications in Mathematical Physics {\bf
  148},  553  (1992).

\bibitem{Dasgupta1981}
C. Dasgupta and B.~I. Halperin, Physical Review Letters {\bf 47},  1556
  (1981).

\bibitem{Fisher1989}
M.~P.~A. Fisher and D.~H. Lee, Physical Review B {\bf 39},  2756  (1989).

\bibitem{Sachdev1990}
S. Sachdev and R. Jalabert, Modern Physics Letters {\bf 4},  1043  (1990).

\bibitem{Sachdev2004}
S. Sachdev, arXiv:cond-mat/0401041  (2004), chapter 9 in "Quantum magnetism",
  U. Schollwock, J. Richter, D. J. J. Farnell and R. A. Bishop eds, Lecture
  Notes in Physics, Springer, Berlin (2004).

\bibitem{WenQFT2004}
X.-G. Wen, {\em Quantum Field Theory of Many-Body Systems} (Oxford University
  Press, New York, 2004).

\bibitem{Wen1990a}
X.~G. Wen and Q. Niu, Physical Review B {\bf 41},  9377  (1990).

\bibitem{Balatsky1991}
A. Balatsky and E. Fradkin, Physical Review B {\bf 43},  10622  (1991).

\bibitem{Lopez1995}
A. Lopez and E. Fradkin, Physical Review B {\bf 51},  4347  (1995).

\bibitem{Fradkin1998}
E. Fradkin, C. Nayak, A. Tsvelik, and F. Wilczek, Nuclear Physics B {\bf 516},
  704  (1998).

\bibitem{Fradkin1999}
E. Fradkin, C. Nayak, and K. Schoutens, Nuclear Physics B {\bf 546},  711
  (1999).

\bibitem{Fradkin2001}
E. Fradkin, M. Huerta, and G.~R. Zemba, Nuclear Physics B {\bf 601},  591
  (2001).

\bibitem{Zhang1989}
S.~C. Zhang, T.~H. Hansson, and S.~A. Kivelson, Physical Review Letters {\bf
  62},  82  (1989).

\bibitem{Zhang1989a}
S.~C. Zhang, T.~H. Hansson, and S. Kivelson, Physical Review Letters {\bf 62},
  980  (1989).

\bibitem{Zhang1992}
S.-C. Zhang, International Journal of Modern Physics B {\bf 6},  25  (1992).

\bibitem{Murthy2003}
G. Murthy and R. Shankar, Reviews of Modern Physics {\bf 75},  1101  (2003).

\bibitem{Murthy2012}
G. Murthy and R. Shankar, Physical Review B {\bf 86},  195146  (2012).

\bibitem{Laughlin1983}
R.~B. Laughlin, Physical Review Letters {\bf 50},  1395  (1983).

\bibitem{Jain1989}
J.~K. Jain, Physical Review Letters {\bf 63},  199  (1989).

\bibitem{Moore1991}
G. Moore and N. Read, Nuclear Physics B {\bf 360},  362  (1991).

\bibitem{Bernevig2006}
B.~A. Bernevig, T.~L. Hughes, and S.-C. Zhang, Science {\bf 314},  1757
  (2006).

\bibitem{Hsieh2009}
D. Hsieh, Y. Xia, D. Qian, L. Wray, J.~H. Dil, F. Meier, J. Osterwalder, L.
  Patthey, J.~G. Checkelsky, N.~P. Ong, A.~V. Fedorov, H. Lin, A. Bansil, D.
  Grauer, Y.~S. Hor, R.~J. Cava, and M.~Z. Hasan, Nature {\bf 460},  1101
  (2009).

\bibitem{Hsieh2009a}
D. Hsieh, Y. Xia, D. Qian, L. Wray, F. Meier, J.~H. Dil, J. Osterwalder, L.
  Patthey, A.~V. Fedorov, H. Lin, A. Bansil, D. Grauer, Y.~S. Hor, R.~J. Cava,
  and M.~Z. Hasan, Physical Review Letters {\bf 103},  146401  (2009).

\bibitem{Xia2009}
Y. Xia, D. Qian, D. Hsieh, L. Wray, A. Pal, H. Lin, A. Bansil, D. Grauer, Y.~S.
  Hor, R.~J. Cava, and M.~Z. Hasan, Nature Physics {\bf 5},  398  (2009).

\bibitem{Zhang2010}
Y. Zhang, K. He, C.-Z. Chang, C.-L. Song, L.-L. Wang, X. Chen, J.-F. Jia, Z.
  Fang, X. Dai, W.-Y. Shan, S.-Q. Shen, Q. Niu, X.-L. Qi, S.-C. Zhang, X.-C.
  Ma, and Q.-K. Xue, Nature Physics {\bf 6},  584  (2010).

\bibitem{Hofstadter1976}
D.~R. Hofstadter, Physical Review B {\bf 14},  2239  (1976).

\bibitem{Fisher1989a}
M.~P.~A. Fisher, P.~B. Weichman, G. Grinstein, and D.~S. Fisher, Physical
  Review B {\bf 40},  546  (1989).

\bibitem{herbut95}
I.~F. Herbut and Z. Tesanovic, Physica C {\bf 255},  324  (1995).

\bibitem{Franz2000}
M. Franz and Z. Tesanovic, Physical Review Letters {\bf 84},  554  (2000).

\bibitem{Vafek2001}
O. Vafek, A. Melikyan, M. Franz, and Z. Tesanovic, Physical Review B {\bf 63},
  134509  (2001).

\bibitem{tesanovic04}
Z. Tesanovic, Physical Review Letters {\bf 93},  217004  (2004).

\bibitem{balents05}
L. Balents, L. Bartosch, A. Burkov, S. Sachdev, and K. Sengupta, Physical
  Review B {\bf 71},  144508  (2005).

\bibitem{Witten1984}
E. Witten, Communications in Mathematical Physics {\bf 92},  455  (1984).

\bibitem{Ardonne2004}
E. Ardonne, P. Fendley, and E. Fradkin, Annals of Physics {\bf 310},  493
  (2004).

\bibitem{Wen1992a}
X.~G. Wen and A. Zee, Physical Review Letters {\bf 69},  953  (1992).

\bibitem{Wen1992c}
X.~G. Wen and A. Zee, Physical Review Letters {\bf 69},  3000  (1992).

\bibitem{Haldane2011}
F.~D.~M. Haldane, Physical Review Letters {\bf 107},  116801  (2011).

\end{thebibliography}



\end{document}